\documentclass[preprint]{aastex}
\usepackage{geometry}   % Changes the page dimensions
\usepackage{graphicx}   % Support the \includegraphics command and options
\usepackage{amsmath}
\usepackage[percent]{overpic}
\usepackage{array}      % For better arrays (eg matrices) in maths
\usepackage{verbatim}   % Adds better verbatim
\usepackage{bm}
\usepackage[export]{adjustbox}
\usepackage{rotating}
\usepackage{enumerate}
\usepackage[table]{xcolor} 
\usepackage{color}
\usepackage{epsfig}     % Allows inclusion of images without specifying extention (reads eps for latex, pdf for pdftex).
\usepackage{url}        % Nice URL formatting.
\usepackage{subfigure}  % Letter labeled subfigures.
\usepackage{tikz}  
\usetikzlibrary{shapes,arrows}
\tikzstyle{flow} = [rectangle, rounded corners, minimum width=6.75in, minimum height=1.0in,text centered, draw=black, fill=white, ultra thick]
\definecolor{darkgreen}{RGB}{0,100,0}
\definecolor{darkred}{RGB}{175,0,0}

\def\mean#1{{\langle #1\rangle}}
\usepackage{hyperref}
\hypersetup{pdfborder={0 0 0}, colorlinks=true, citecolor=blue, linkcolor=blue}

\begin{document}
\title{Synchronic coronal hole mapping using multi-instrument EUV images: Data preparation and detection method.}
\author{R. M. Caplan, C. Downs, and J. A. Linker}
\affil{Predictive Science Inc.\footnote{{\tt URL:} http://www.predsci.com}}
\affil{9990 Mesa Rim Road Suite 170 San Diego, California 92121, USA}
\email{caplanr@predsci.com}
\date{\today}
%\maketitle
\begin{abstract}
We describe a method for the automatic mapping of coronal holes (CH) using simultaneous multi-instrument EUV imaging data. Synchronized EUV images from STEREO/EUVI A\&B 195\AA\ and SDO/AIA 193\AA\ are preprocessed, including PSF deconvolution and the application of data-derived intensity corrections that account for center-to-limb variations (limb brightening) and inter-instrument intensity normalization. We systematically derive a robust limb-brightening correction that takes advantage of unbiased long-term averages of data and respects the physical nature of the problem. The new preprocessing greatly assists in CH detection, allowing for the use of a simplified variable-connectivity two-threshold region growing image segmentation algorithm to obtain consistent detection results.  We generate synchronic EUV and CH maps, and show a preliminary analysis of CH evolution.

Several data and code products are made available to the community ({\tt www.predsci.com/chd}): For the period of this study  (06/10/2010 to 08/18/14) we provide synchronic EUV and coronal hole map data at 6-hour cadence, data-derived limb-brightening corrections for STEREO/EUVI A\&B 195\AA\ and SDO/AIA 193\AA, and inter-instrument correction factors to equate their intensities. We also provide the coronal hole image segmentation code module ({\tt ezseg}) implemented in both FORTRAN OpenMP and GPU-accelerated C-CUDA.  A complete implementation of our coronal hole detection pipeline in the form of a ready-to-use MATLAB driver script  {\tt euv2chm} utilizing {\tt ezseg} is also made available.
\end{abstract}

%%%%%%%%%%%%%%%%%%%%%%%%%%%%%%%%%%%%%%%%%%%%%%%%%%%%%%%%%%%%%%%%%%%%%%%%%%%%%%%%%%%%%%%%%%%%%%%%%%%%%%%%%%%%%%%%%%%%%

\section{Introduction}
\label{sec_intro}
Coronal holes (CHs) manifest themselves as regions of relatively low intensity in EUV and X-Ray images (or correspondingly bright regions in He I 10830 absorption) \citep{Bohlin1977}.  They have long been recognized to be associated with open magnetic fields \citep{Altschuler_etal1972}, sources of fast solar wind streams \citep{Krieger_etal1973} and recurrent geomagnetic activity \citep{Neupert_Pizzo1974}.  The boundaries of coronal holes are recognized to play an important role in the origin of the slow solar wind.  In theories that invoke a quasi-static origin,  the slow wind arises from regions of large expansion factor near the boundaries \citep{Wang_Sheeley1990,Cranmer_etal2007}.  Recently, it was shown that distance to CH boundaries is in fact a better predictor of solar wind speed than expansion factor \citep{Riley_etal2015}. Theories that assert a dynamic origin of the slow wind require reconnection between open and closed field regions \citep{Fisk_etal1998, Antiochos_etal2011}; the imprints of such reconnection should be seen near coronal hole boundaries.  Explanations for the  quasi-rigid rotation of some extended coronal holes also require reconnection there \citep{Wang_etal1996,Fisk_etal1998,Lionello_etal2005}.  Understanding the evolution of coronal hole boundaries is thus important not only for investigating coronal dynamics but also for deducing the genesis of the solar wind.

The recent profile of the STEREO and SDO missions give us an opportunity to study CH evolution over medium time periods (days to weeks).  However, a meaningful assessment, as well as comparison with models, requires an objective method for identifying CH boundaries in a consistent manner over many days.

It is difficult to systematically identify CH boundaries within EUV images because they are not always clear-cut, and the choice of boundary may be somewhat subjective (indeed they can even differ based on arbitrary factors such as the chosen color table of the image). Even with a strict definition for a coronal hole, and a means to reject false positives (such as darker quiet sun regions and filaments), factors such as systematic instrumental effects, line-of-sight changes, and temporal evolution all may cause variations in CH intensity and/or appearance in EUV imaging data. Together these factors can make it difficult to automatically detect CH regions in a consistent manner as they evolve in space and time.

All of these issues have therefore yielded many algorithms and methods for identifying CHs in the literature.  Historically, CH detection was accomplished by manually tracing the perceived boundaries of the CHs by hand, frame by frame \citep{2002_Harvey_Polar_Holes}. Modern coronal hole detection methods are typically built around adaptive image segmentation algorithms that use the properties of individual (or several) images to separate the image into distinct regions such as CHs or Active Regions (ARs). Image segmentation has been the subject of research for over 30 years.  We refer the reader to the surveys of \citet{1975_Zucker_Survey_reg_grow} and \citet{2002_Freixenet_img_seg_survay_w_exp} for an overview.  

Many of the algorithms that have been developed for image segmentation are difficult to implement or generally not suitable for use in CH detection due to the noisy nature of EUV images, the ill-defined boundaries of CHs, as well as other unique features of EUV images such as the similar intensities of dark quiet sun regions.  For this reason, many applications of image segmentation to solar EUV images focus on the simplest techniques that yield acceptable results for the specific goal of each study; see \citet{2010_Aaschwanden_ImageProc_FeatureSeg_SolarImages_Review} for a review.  

The most basic segmentation method, that of intensity thresholding, is the basis for many algorithms.  Histograms can be used to show different intensity value ranges in various structures with thresholds chosen to best separate these structures; for example, \citet{1998_Gallagher_QSnet-hist-method} used this method to identify the structure of the quiet-sun network.  Methods using global and local histograms to select and apply thresholding segmentations have been used for CH detection \citep{2008_Scholl_CHD_auto-hist-improc,2009_Scholl_CHD_auto-hist-improc,2009_Krista_CHD_autothresh} and utilized in studying short-term local CH evolution \citep{2011_Krista_CHD_evolve_velcalc,2013_Krista_CHD_local_evolve_hist}.  A previous study that used AIA and EUVI data for synchronic CH mapping (as is done in this paper) was performed in \citet{2014_Lowder_CHD_AIA_EUVI}.  They used a single-valued thresholding, where the threshold value is automatically selected by finding the local minimum in the histograms of subsections of a given image, and then using the average of all subsection values to obtain the global threshold for segmentation. 

A different approach to image segmentation is taken by the SPoCA code \citep{2014_Verbeeck_SPoCA}, which currently supplies the Heliophysics Event Knowledgebase\footnote{http://www.lmsal.com/hek/index.html} (HEK) with maps of coronal holes visible by SDO/AIA. The SPoCA algorithm uses a slightly modified version of the fuzzy C-means (FCM) algorithm applied to the image histograms to clump similar areas together.  Other authors use different methods of classifiers that utilize EUV images from multiple wavelengths to detect structures including coronal holes \citep{2006_DeWit_FastSolarSeg_MultBand,2013_Suresh_PhD_image_seg_euv}. 

A core element in all of these techniques is the concept of identifying CHs using adaptive thresholds or adaptive segmentation logic that is based on the structural features presented in an image at a single time or in groups of images within a relatively short period \citep[an exception is][where the authors used 10-year averages of hand-drawn CHs from He 1083nm images to form their threshold values]{2005_Henny_Auto_CH}. Thus, by design, the definition of a coronal hole can shift or adapt depending on the relative distribution of structures in the current image(s), which may vary with time as the sun rotates and evolves morphologically. In this work, we take a different philosophical approach to the problem.  Here we attempt to construct an image processing pipeline and detection algorithm that will allow us to employ an \emph{inherently fixed} definition of a coronal hole, one that ideally remains consistent as coronal holes evolve in time and space.

This idea is motivated by a goal to construct high-quality full-sun coronal hole maps for a given time and to track the temporal evolution of such features in the simplest and most consistent manner as possible. Specifically, we aim to develop a technique that can: a) account for systematic changes in brightness and contrast as coronal holes rotate throughout the field of view of one instrument (e.g. center to limb changes),  b) remain relatively insensitive to variations in the appearance and structure of the corona during rotation and the solar cycle (histogram variation), and c) consistently detect the same features from heterogeneous, multi-spacecraft imaging data as features rotate from one viewpoint to another. Satisfying these goals opens the door to study not only the long term properties of coronal holes on the sun \citep[e.g.,][]{2014_Lowder_CHD_AIA_EUVI} or the short term evolution of transient holes \citep[e.g.,][]{2013_Krista_CHD_local_evolve_hist} but also the medium term evolution of a coronal hole as it rotates around the sun. This is particularly relevant during the 2010--2014 time period that we cover in this study, because the favorable orientation of the STEREO and SDO spacecraft allows for the instantaneous mapping of nearly the entire EUV corona (also known as a synchronic map), meaning that the evolution of a single coronal hole can be studied continuously for weeks at a time.

Our approach to achieve these goals is divided into two main elements. The first is the development of a data-driven, physically motivated, preprocessing pipeline to prepare EUV images for CH detection. In addition to using PSF deconvolution to naturally enhance the coronal hole contrast, we present a technique to derive intensity dependent limb-brightening corrections from tailored year-long averages of imaging data, which enables the flattening of EUV intensity variation from disk center to the limb in a manner that respects the inherent geometry and physics of the problem. We also use a similar technique to derive inter-instrument transformations to relate intensity values across instruments, allowing the use of a single set of CH detection parameters for all three spacecraft. This preprocessing pipeline is very general and has a number of additional applications where the time-dependent mapping of EUV data is of interest.

The second element is the detection algorithm itself. Our code module, named {\tt ezseg}, utilizes a two-threshold simplified implementation of a region-growing segmentation algorithm. It relies on seeding points that have a high probability of belonging to coronal holes, and subsequently growing the seeded region to intensities that are deemed to generally correspond to the boundary of CHs. A variation of this algorithm has been employed previously in other contexts, such as the popular SWAMIS code \citep{2007_DeForest_SWAMIS} for magnetic feature tracking.  Our implementation differs from SWAMIS in its algorithmic computation method, using manually selected threshold values, and an added variable connectivity requirement (see Sec.~\ref{sec_alg}).  The {\tt ezseg} algorithm is implemented as a modular CUDA-C routine (for running on Nvidia GPUs), as well as an OpenMP-enabled FORTRAN routine, both of which can be called from any custom-made driver code, including those written in scripting/data analysis languages (e.g. MATLAB, IDL, Octave, SciLab, R, etc.).  For convenience and to provide a ready-to-use code, we make available a sample driver implementation in  MATLAB in addition to the {\tt ezseg} codes themselves.

This paper, which focuses on how we build and apply the above preprocessing and coronal hole detection techniques to EUV imaging data from 2010--2014, is outlined as follows:  Sec.~\ref{sec_datagrab} describes the process of data acquisition and issues involved.  Sec.~\ref{sec_preproc} describes the preprocessing steps of PSF deconvolution, limb-brightening correction, and inter-instrument transformations.  Validation of the preprocessing is also discussed. Sec.~\ref{sec_alg} describes our detection algorithm and mapping procedures.  Sample results of the generated CH maps spanning the 4 year period and a preliminary analysis of CH evolution are given in Sec.~\ref{sec_results}.   We conclude in Sec.~\ref{sec_conclusion} where we give the locations of the provided data products and codes, and comment on future applications of this work.

\begin{figure}[tbp]
\begin{tikzpicture}[node distance=1.5in]
\node (data) [flow] {
$\begin{array}{l}%
    \parbox{0.8\textwidth}{ 
    {\bf \Large 1) Data Acquisition} (Secs.~\ref{sec_datagrab},\ref{sec_preproc}) [{\tt euv2chm\_data\_prep}]
    }
    \\
 \begin{array}{lr}%
  \parbox{0.5\textwidth}{ 
    \begin{itemize} \itemsep0.5pt \parskip0pt \parsep0pt
      \item Obtain level-1 preprocessed EUV disk images using SSW in IDL.
      \item Sample data formatting script made available.
    \end{itemize}
  }
  &
  \begin{minipage}[c]{0.4\textwidth}%
      \hfill \includegraphics[height=0.8in]{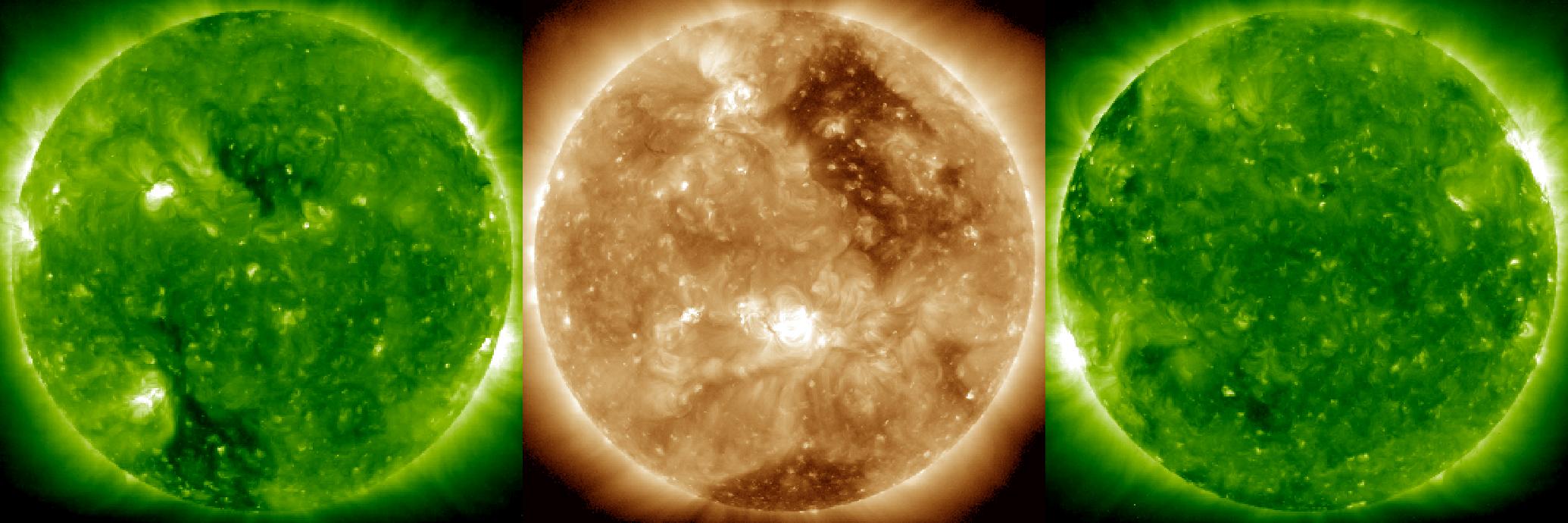}
  \end{minipage}
\end{array}
\end{array}$
};
\node (psf) [flow, below of=data] {
$\begin{array}{l}%
    \parbox{0.8\textwidth}{ 
    {\bf \Large 2) PSF Deconvolution} (Sec.~\ref{sec_psf})
    }
    \\
 \begin{array}{lr}%
  \parbox{0.5\textwidth}{ 
    \begin{itemize} \itemsep0.5pt \parskip0pt \parsep0pt
      \item Use publicly available PSFs and deconvolution algorithms to remove scattered light.
    \end{itemize}
  }
  &
  \begin{minipage}[c]{0.4\textwidth}%
      \hfill \includegraphics[height=0.8in]{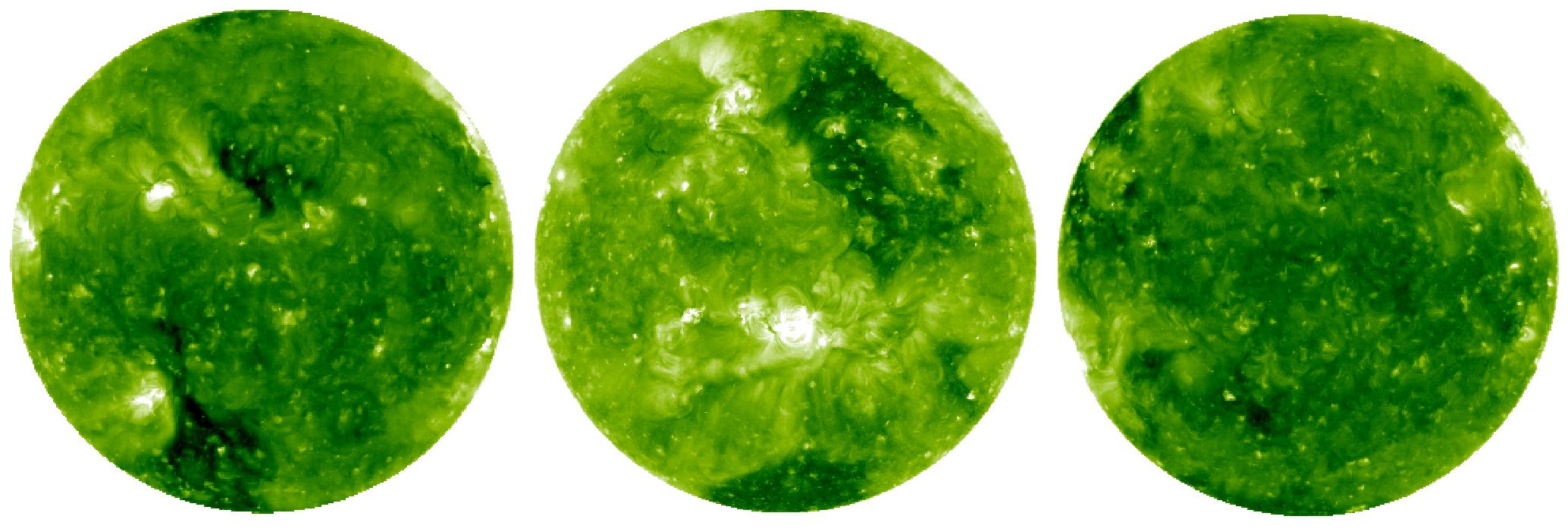}
  \end{minipage}
\end{array}
\end{array}$
};
\node (lbc) [flow, below of=psf] {
$\begin{array}{l}%
    \parbox{0.8\textwidth}{ 
    {\bf \Large 3) Limb-brightening Correction} (Sec.~\ref{sec_lbcc}) [{\tt euv2chm}]
    }
    \\
 \begin{array}{lr}%
  \parbox{0.5\textwidth}{ 
    \begin{itemize} \itemsep0.5pt \parskip0pt \parsep0pt
      \item Flatten natural center-to-limb variation using a data-driven approach.
      \item Correction data made available.
    \end{itemize}
  }
  &
  \begin{minipage}[c]{0.4\textwidth}%
      \hfill \includegraphics[height=0.8in]{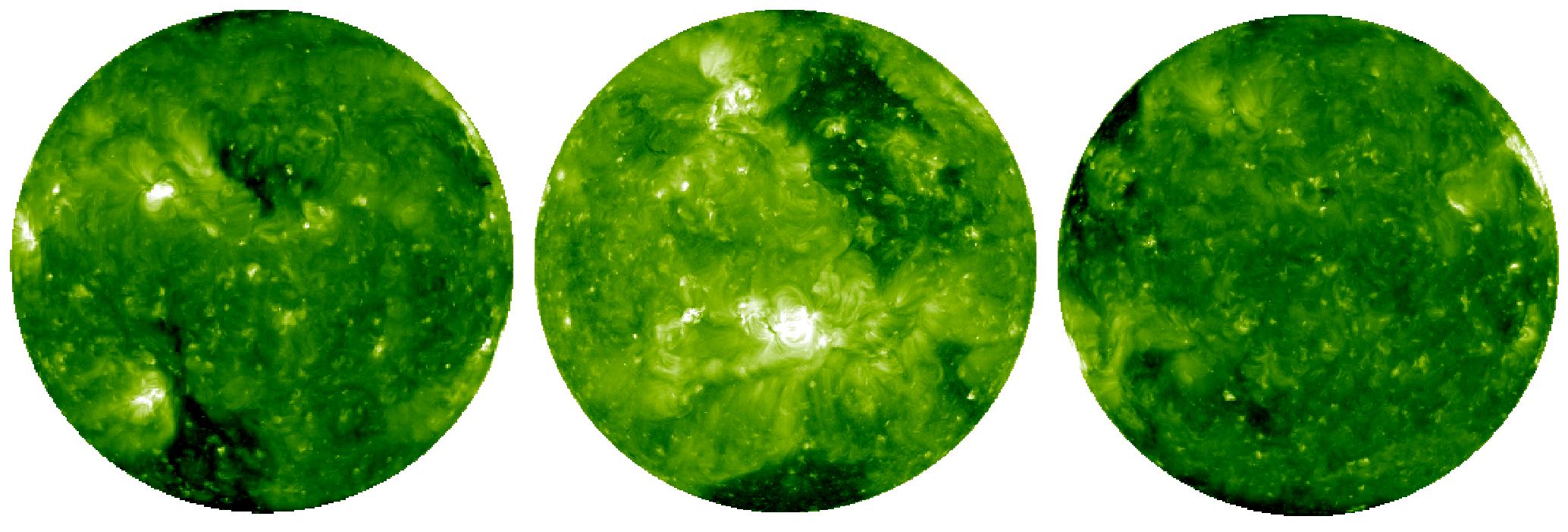}
  \end{minipage}
\end{array}
\end{array}$
};
\node (iit) [flow, below of=lbc] {
$\begin{array}{l}%
    \parbox{0.8\textwidth}{ 
    {\bf \Large 4) Inter-instrument Transformation} (Sec.~\ref{sec_iit}) [{\tt euv2chm}]
    }
    \\
 \begin{array}{lr}%
  \parbox{0.5\textwidth}{ 
    \begin{itemize} \itemsep0.5pt \parskip0pt \parsep0pt
      \item Normalize intensities between the 3 instruments using a data-driven approach.
      \item Transformation data made available.
    \end{itemize}
  }
  &
  \begin{minipage}[c]{0.4\textwidth}%
      \hfill \includegraphics[height=0.8in]{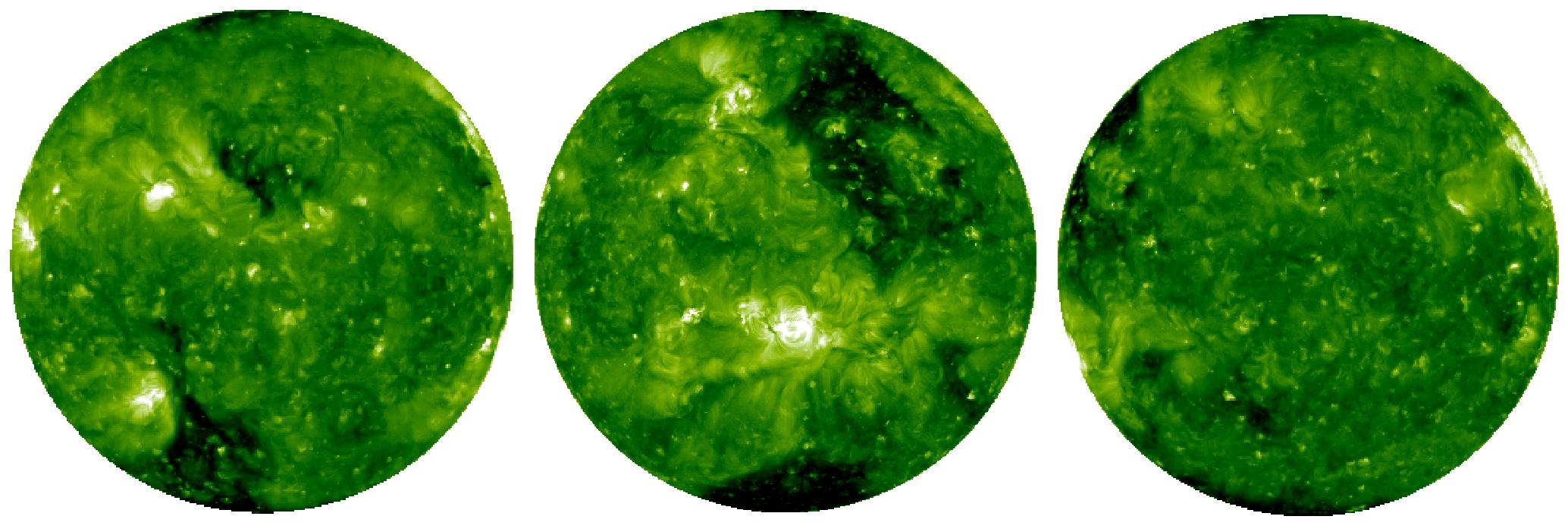}
  \end{minipage}
\end{array}
\end{array}$
};
\node (chd) [flow, below of=iit] {
$\begin{array}{l}%
    \parbox{0.8\textwidth}{ 
    {\bf \Large 5) Coronal Hole Detection} (Sec.~\ref{sec_ezseg}) [{\tt ezseg},{\tt euv2chm}]
    }
    \\
 \begin{array}{lr}%
  \parbox{0.5\textwidth}{ 
    \begin{itemize} \itemsep0.5pt \parskip0pt \parsep0pt
      \item Detect the CHs using a region growing algorithm. 
      \item Software ({\tt EZSEG}) made available.
    \end{itemize}
  }
  &
  \begin{minipage}[c]{0.4\textwidth}%
      \hfill \includegraphics[height=0.8in]{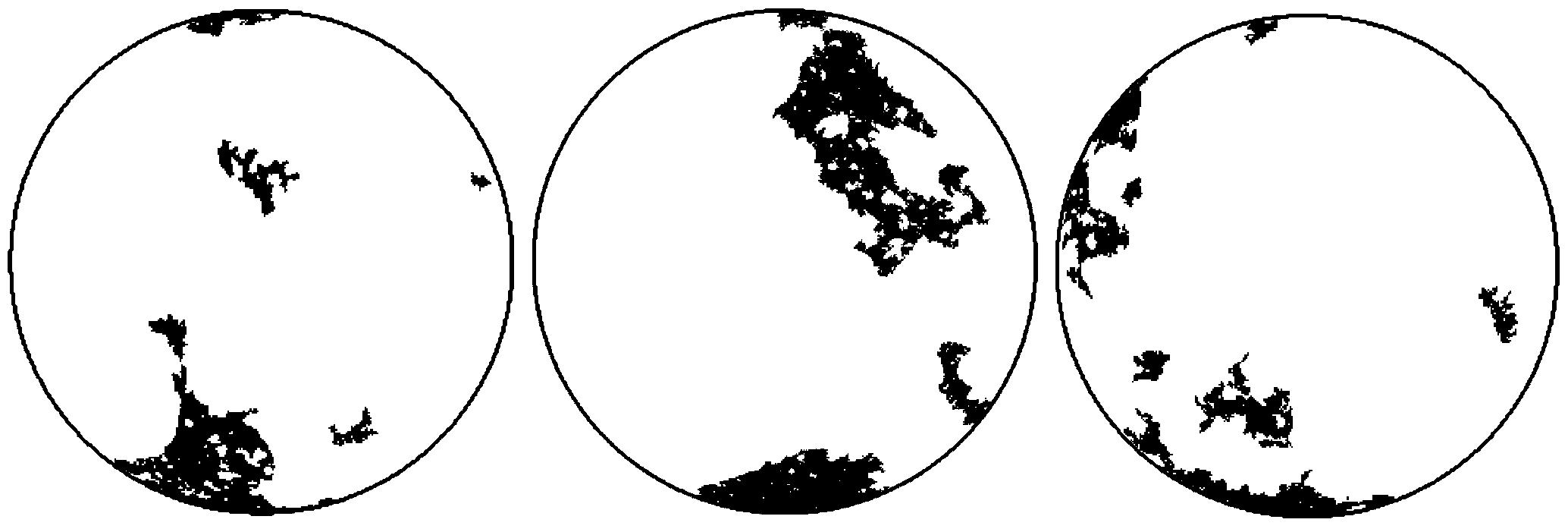}
  \end{minipage}
\end{array}
\end{array}$
};
\node (map) [flow, below of=chd] {
$\begin{array}{l}%
    \parbox{0.8\textwidth}{ 
    {\bf \Large 6) Mapping} (Sec.~\ref{sec_mapping}) [{\tt euv2chm}]
    }
    \\
 \begin{array}{lr}%
  \parbox{0.5\textwidth}{ 
    \begin{itemize} \itemsep0.5pt \parskip0pt \parsep0pt
      \item Merge the 3 results to generate synchronic EUV and CH maps.
      \item Final maps for 4 years at 6-hour cadence are made available.
    \end{itemize}
  }
  &
  \begin{minipage}[c]{0.4\textwidth}%
      \hfill \includegraphics[height=0.75in]{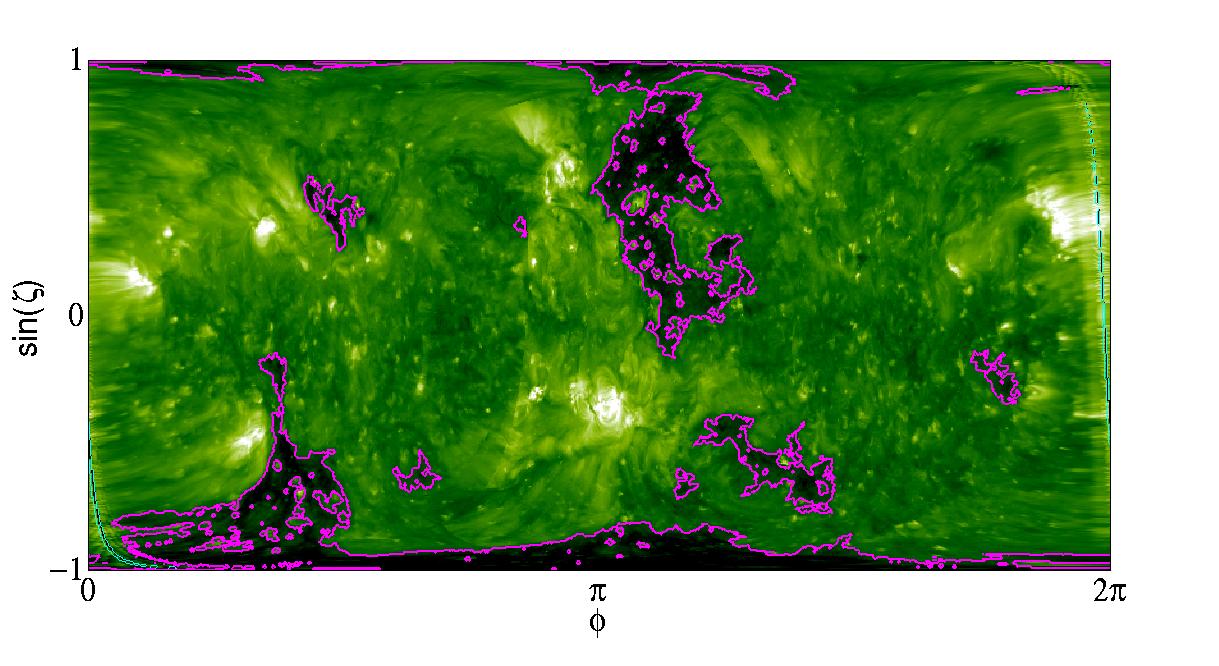}%
      \includegraphics[height=0.75in]{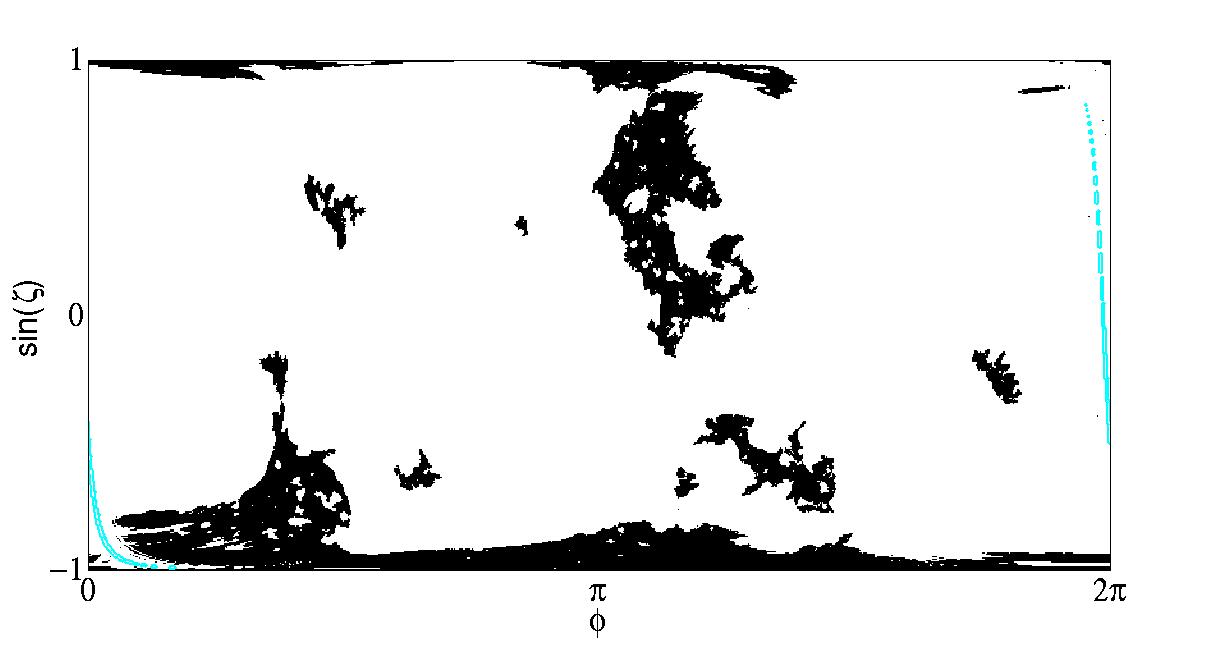}
  \end{minipage}
\end{array}
\end{array}$
};
\draw [-stealth', ultra thick] (data) -- (psf);
\draw [-stealth', ultra thick] (psf) -- (lbc);
\draw [-stealth', ultra thick] (lbc) -- (iit);
\draw [-stealth', ultra thick] (iit) -- (chd);
\draw [-stealth', ultra thick] (chd) -- (map);
\end{tikzpicture}
\caption{Workflow to generate synchronic EUV and CH maps (examples are from 02/03/2011).  The section describing the step and the name of the provided code example are indicated.\label{fig_workflow}}
\end{figure}
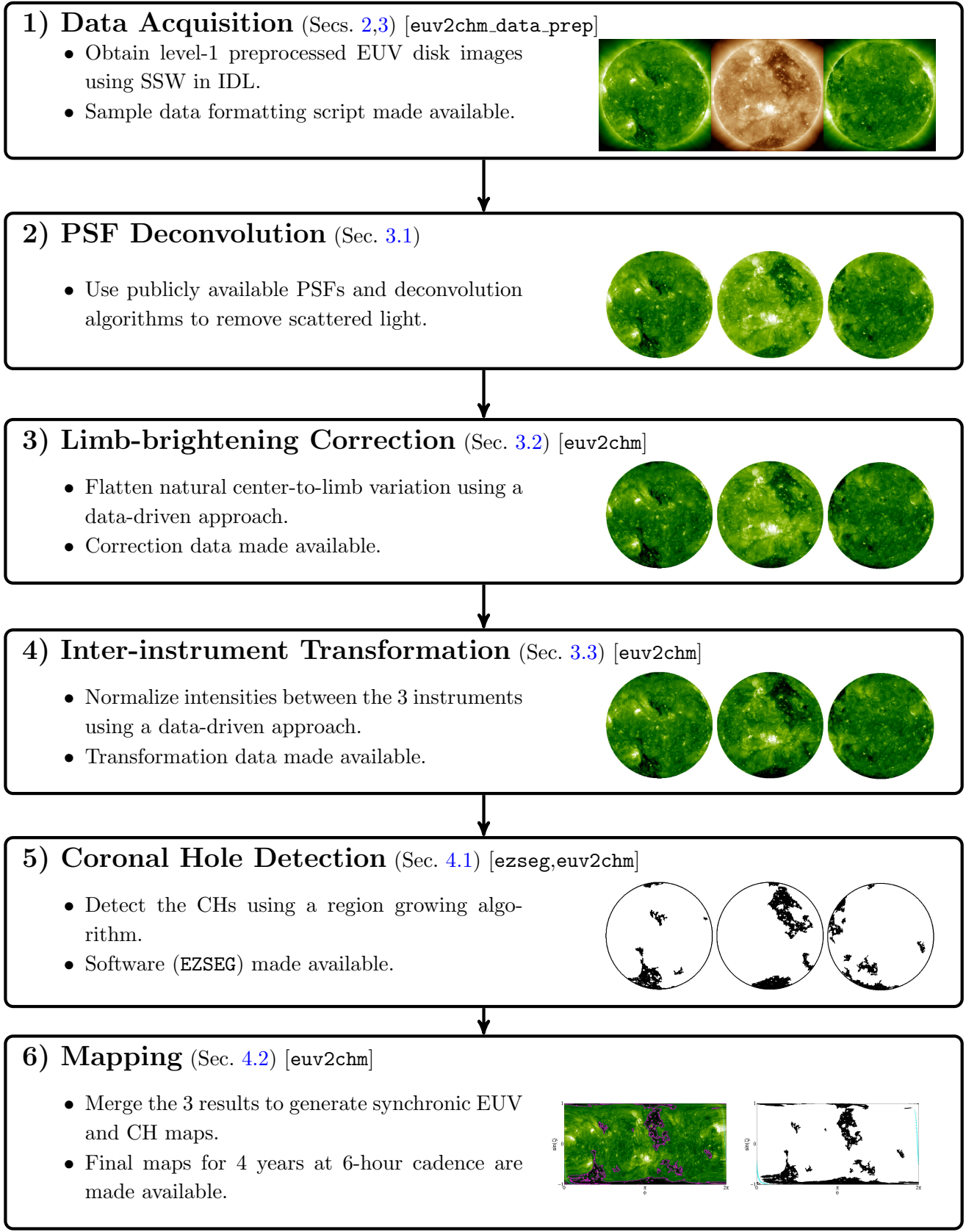

To aid the reader through the described data and processing pipeline, we provide an outline of each of the steps in Fig.~\ref{fig_workflow}, noting the section number where each step is described in detail, and (where applicable) the name of the provided software examples that perform each step.

%%%%%%%%%%%%%%%%%%%%%%%%%%%%%%%%%%%%%%%%%%%%%%%%%%%%%%%%%%%%%%%%%%%%%%%%%%%%%%%%%%%%%%%%%%%%%%%%%%%%%%%%%%%%%%%%%%%%%

\section{Observational Data}
\label{sec_datagrab}
We use images from the Fe XII 195\AA\ and 193\AA\ channels from STEREO-EUVI and SDO-AIA instruments respectively for our study.  Fe XII imaging observations have historically been used to detect CHs because the combination of high contrast and line strength (high count rates) make them ideal for CH detection. For example, standard Differential Emission Measure (DEM) models employed by \citet{2010_Dwyer_AIA_responses} predict around a ten-fold difference in AIA 193\AA\ photon count between typical CH and quiet sun regions. Although there are subtle differences in the temperature response functions of the AIA 193\AA\ and EUVI 195\AA\ channels, they are similar enough to be used together to generate synchronic maps with proper preprocessing (see Sec.~\ref{sec_preproc}).  To simplify notation, throughout the paper we will use STA, STB, and AIA to refer to STEREO-A EUVI 195\AA, STEREO-B EUVI 195\AA, and SDO AIA 193\AA\;respectively.

We obtain the EUV images from AIA and EUVI using SolarSoftWare (SSW) \citep{1998_Freeland_SSW} in IDL. For a given date range and cadence, the Virtual Solar Observatory (VSO) and Joint Science Operation Center (JSOC) are queried to find sets of STA, STB, and AIA science images whose observation times are within a maximum time interval of 30 minutes (typically within 30 seconds) of each other. For this study we obtain image sets at a 6-hour cadence from June 2010 to October 2014, which yields a total of 6051 frames for each instrument.

Where possible, bad data frames are detected and replaced by nearby frames by checking header data.  However, some bad EUVI data frames get through this process, so they must be manually replaced, and/or properly handled by the processing codes. Most bad images can be detected automatically, but some (such as the rare out-of-focus images) are missed, requiring a visual inspection to identify them.  These frames are then skipped in our processing and detection steps.

%%%%%%%%%%%%%%%%%%%%%%%%%%%%%%%%%%%%%%%%%%%%%%%%%%%%%%%%%%%%%%%%%%%%%%%%%%%%%%%%%%%%%%%%%%%%%%%%%%%%%%%%%%%%%%%%%%%%%

\section{Preprocessing}
\label{sec_preproc}
Since coronal holes are identified as low-intensity regions in EUV images, it is very important to minimize intensity variations across the disk, both instrumental (noise, scattered light, scaling, etc.) and physical (limb brightening, line-of-sight obscuration, etc).

We start by calling the standard image reduction and processing routines in SSW. EUVI images are reduced using {\tt secchi\_prep.pro}, which includes exposure time correction and image registration, flat field correction, unit conversion, and bias. The AIA images, which are only available in a calibrated Level 1 format, are registered and exposure normalized using {\tt aia\_prep.pro} and binned to match the EUVI resolution ($2048\times 2048$).  We refer to the results of this preliminary processing as the ``original'' line-of-sight disk images. The default units in SSW for EUVI data is photons/s, while those for AIA are DN/s.  These units are similar enough to display the images using the same color scale (especially after applying our inter-instrument transformations (see Sec.~{\ref{sec_iit}})).  Therefore, throughout this paper, we display the images (in $\mbox{log}_{10}$) using the standard EIT green color table spanning image values between $0$ and $3$.

For images that undergo deconvolution (see below), the deconvolution step must be applied after calibration but before rotation to align the image with solar north. The images then receive the limb-brightening correction, and are scaled to each other as described in the remainder of this section.  We note that the application of noise-reduction techniques to the images, such as the ongoing work by \citet{2014_Kirk_AGU_Noise_Talk} for AIA, would also be a potentially beneficial preprocessing step. However, applications of sophisticated noise-reduction techniques to solar EUV imaging data are not currently available, and developing them is beyond the scope of this study.  We also note that the image data from STA and STB is known to suffer from ICER compression artifacts, especially in dark regions off the limb.  Although we have observed some small artifacts in the darker regions of coronal holes, their presence does not appear to have any noticeable effect on our data pipeline or coronal hole detection.  

The end goal of this section is to transform/flatten the EUV images such that every disk pixel appears (as much as possible) as if it were observed on the sun at disk center with a single idealized lens.  This flattening of the images allows us to create synchronic EUV maps and greatly assist our CH detection algorithm in Sec.~\ref{sec_ezseg}.

%%%%%%%%%%%%%%%%%%%%%%%%%%%%%%%%%%%%%%%%%%%%%%%%%%%%%%%%%%%%%%%%%%%%%%%%%%%%%%%%%%%%%%%%%%%%%%%%%%%%%%%%%%%%%%%%%%%%%

\subsection{PSF Deconvolution}
\label{sec_psf}
Although many instrumental effects are removed from the images through the standard Level 1 preprocessing, optical effects such as scattered light, diffraction, and others remain.  These can cause many coronal hole detection algorithms to be less effective.  For example, unwanted stray light from brighter regions can decrease the relative intensity contrast between CHs and nearby regions of the quiet sun, and cause CHs to be less uniformly dark.  Some of these optical effects can be removed by deconvolving the images with the instrument's point spread function (PSF).  Such PSFs can only be approximated and the deconvolution process is an ill-defined inverse problem.  However, deconvolving the images with an approximate PSF can still greatly improve the image in the ways we desire for CH detection.

A full discussion for in-flight PSF estimation and correction using lunar transients for the EUVI and AIA instruments can be found in \citet{2012_Shearer_PSFmoon} \& \citet{2013_Poduval_newAIA_PSFs} respectively. For our purposes, the most relevant correction comes from the estimation of the stray light profile of the PSF. For EUVI we obtain estimated PSFs tailored to each spacecraft (A\&B) derived by \citet{2013_Shearer_thesis} from within the \texttt{euvi\_deconvolve.pro} package accessible in SSW.  For AIA, we use the freely available PSFs derived by \citet{2013_Poduval_newAIA_PSFs}.  All of these PSFs contain an extended profile away from the PSF core that helps correct stray light.  Research in the formulation of PSFs for EUVI and AIA is ongoing \citep[for example, see][]{2015_Gonzales_AIA+EUVI_PSFModeling_Preprint} and these efforts have the potential to improve the correction of the images further.

Once the estimated PSFs are obtained, the EUV images must be deconvolved with them. Due to the large number of images to be processed we require a fast and efficient algorithm to reduce the overall processing time. To this end we use the SGP deconvolution algorithm from \citet{2009_Bonettini_SGP}\footnote{\label{ftnt_idlsgp} See \citet{2012_Prato_Deconvolution_IDLGPU} for the description of a freely available IDL implementation of the SGP algorithm (with support for the third-party IDL package GPULib: \url{http://www.txcorp.com/home/gpulib}).}, which offers a speed improvement with similar results when compared to the popular Richardson-Lucy deconvolution algorithm \citep{1974_Lucy_RLdeconvolve}.   A standalone C and CUDA-C version of the SGP implementation is under development by the authors of \citet{2012_Prato_Deconvolution_IDLGPU}, and binaries of the new codes were graciously provided to us, allowing us to utilize Nvidia GPUs for our processing (independent of IDL or other third-party software).  For example, deconvolving a single 4096x4096 image takes $\approx\!7.5$ minutes using the standard Richardson-Lucy routine in IDL SSW, $\approx\!3$ minutes using the SGP IDL scripts\textsuperscript{\ref{ftnt_idlsgp}}, $\approx\!1$ minute using the serial stand-alone C SGP code, and only $\approx\!4$ seconds using the stand-alone GPU SGP code\footnote{The CPU used for the timing was a Xeon W3680 3.3GHz, while the GPU used was a GTX970}.  Using single precision drops the GPU compute time down further to only  $\approx\!2$ seconds per image (without noticeable loss in image quality).  We also note that a new GPU code for computing Richardson-Lucy deconvolution for AIA is in development \citep{2015_Cheung_AIA_PSF_GPU} that can process an AIA image in less than a second and will be incorporated into the standard AIA data pipeline when completed.  This will eliminate the need to separately deconvolve the AIA images as done here.

\begin{figure}[tbp]
\centering
\includegraphics[width=7.0in]{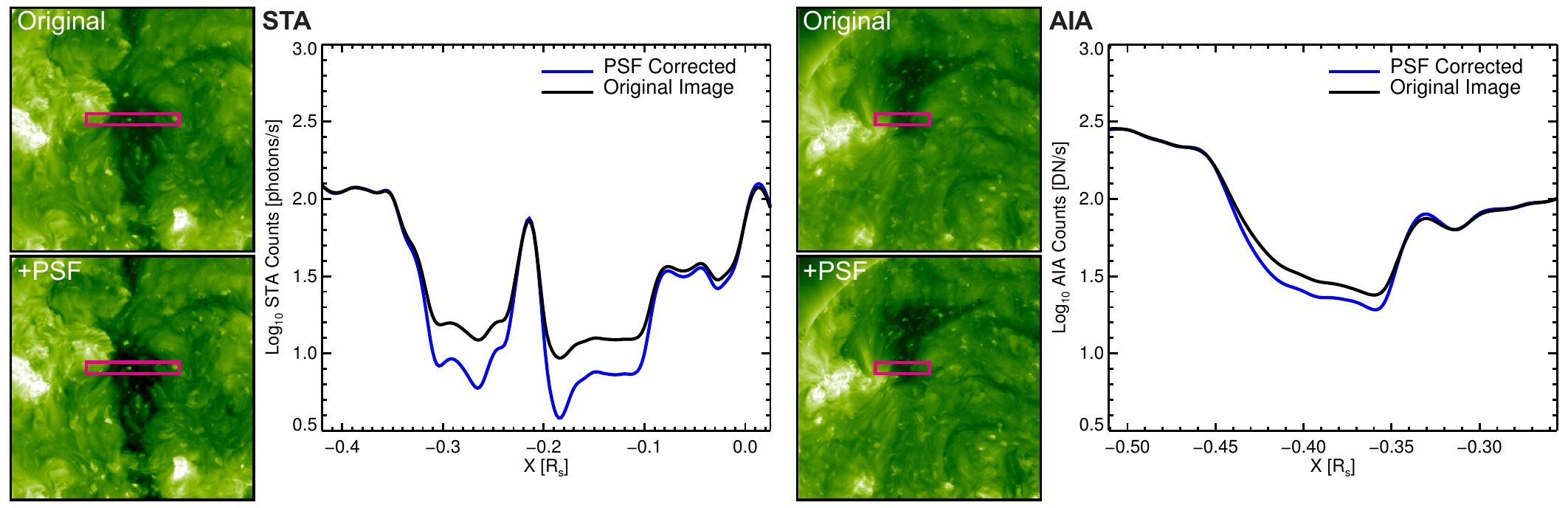}
\caption{Example of the PSF correction applied to STA (left) and AIA (right) images. The context images show the emission before and after PSF deconvolution. The line plots show the averaged emission for a strip across the respective coronal hole features (magenta boxes).\label{fig_psf}}
\end{figure}

The PSF deconvolution is applied to the calibrated images before rotation. The advantage of the deconvolution step as a contrast enhancement for coronal hole detection is illustrated in Fig.~\ref{fig_psf}. Here we show context images and slices through selected coronal holes for all three instruments. While the nominal emission values for the quiet-sun and nearby brighter regions remains similar, the stray light correction removes contaminating emission from the darkest portion of coronal holes. This increases the relative contrast ratio between coronal holes and neighboring features. Note that because the base optical properties between the AIA and EUVI telescopes differ considerably, the relative contrast enhancement gained from PSF deconvolution is not expected to be the same.

%%%%%%%%%%%%%%%%%%%%%%%%%%%%%%%%%%%%%%%%%%%%%%%%%%%%%%%%%%%%%%%%%%%%%%%%%%%%%%%%%%%%%%%%%%%%%%%%%%%%%%%%%%%%%%%%%%%%%

\subsection{Limb-brightening correction}
\label{sec_lbcc}
Because the extended EUV emitting portion of the corona is (mostly) optically thin, solar EUV images are affected by limb brightening due to the increasing line-of-sight path length towards the limb.  Thus, the same solar feature observed near disk center of the image will tend to be brighter in intensity near the limb.  Since many coronal hole detection algorithms make use of intensity thresholds (including ours), such variation in intensity can be problematic, particularly since we intend to track the evolution of coronal holes as they rotate across the solar disk and from one instrument viewpoint to the other.

In order to equalize (or flatten) the intensities across the disk, we seek limb-brightening correction curves (LBCCs) that will convert intensities observed at any position on the disk to their disk center equivalent.  We choose to define the LBCCs as a function of $\mu$, where 
\[
\mu=\cos\theta, \qquad \theta\in[0,\pi/2],
\] 
where $\theta$ is the angle from disk center ($\mu=1$) to the limb ($\mu=0$).  We note that the ``limb'' here is defined as the disk edge in the coronal EUV images that, due to the height of the chromosphere and transition region, differs slightly from the photospheric limb (see Sec.~\ref{sec_mapping} where this difference affects how we map the images).  Throughout this paper we choose a value of the base coronal radius of $R_0=1.01\,R_{\odot}$. We also choose to describe the LBCCs in terms of the $\mbox{log}_{10}$ of image intensity because EUV images are typically viewed in log-space.

Assuming that the rate of the increase in intensity from disk center to the limb is independent of the disk center intensity ($I_0$) itself (for example, in the case of a spherically-symmetric solar atmosphere with constant temperature) leads to LBCCs of the form
\begin{equation}
\label{eq_lbcc_def}
L(\mu) = \mbox{log}_{10}\,\left[\frac{F(\mu)}{F_0}\right] = I(\mu) - I_0,
\end{equation}
where $F_0 = F(\mu=1)$ is the image intensity at disk center, and $I = \mbox{log}_{10}\,F$.   
Given an image of $\log_{10}$ intensities $I$, applying the LBCC is simply described as
\[
I^{\prime}(\mu) = I(\mu) - L(\mu),
\]
where $I^{\prime}(\mu)$ is the corrected image.  The $I_0$-independent assumption of the LBCC of Eq.~(\ref{eq_lbcc_def}) can, in some cases, be an adequate approximate limb-brightening correction but we use it here to motivate and illustrate our data-derived approach for obtaining an $I_0$-dependent correction for limb brightening.   

In the following sections, we examine analytic approximations to LBCCs in the case of hydrostatic equilibrium, and then move on to describing the formulation of data-derived LBCCs, starting with our data selection process.  Our final preferred method for limb-brightening correction {---} that of an $I_0$-dependent limb-brightening correction image transformation {---} is then described in detail, which results in a more accurately flattened image and enhanced contrast.

\subsubsection{Theoretical limb-brightening correction curves}
\label{sec_lbcc_theo}
If one assumes an idealized single-temperature hydrostatic atmosphere and a weak dependence of the instrument's temperature response function on density, a theoretical limb-brightening correction curve can be formulated (see Appendix~\ref{sec_lbcc_theory}) as
\begin{equation}
\label{eq_lbcdef1}
L(\mu;T_e) = I(\mu;T_e)-I_0(T_e),
\end{equation}
where
\begin{equation}
\label{eq_lbcdef2}
I(\mu;T_e)=\mbox{log10}\left[\int_{0}^{\infty} \mbox{exp}\left[-2\,\frac{R_0}{r}\,\frac{r-R_0}{\lambda(T_e)}\right]\,dx\right],
\end{equation}
and
\[
\lambda(T_e)=\frac{k_B\,T_e}{g_0\,\tilde{\mu}\,m_p}
\]
is the gravitational scale height; $r=\sqrt{x^2+2\,x\,\mu\,R_0+R_0^2}$ is the distance to solar center, $R_0$ is the approximate radial distance of the base of the corona, $x$ is the distance from the observer to $R_0$, $\tilde{\mu}$ is the mean molecular weight \citep[we use the coronal value of $\tilde{\mu}=0.6$][]{2014_Priest_BOOK}, $k_B$ is the Boltzmann constant, $m_p$ is the proton mass, and $g_0=G\,M_{\odot}/R_{0}^2$.  It is important to note that under these assumptions, this limb-brightening correction curve over $\mu$ is independent of the base density and of the temperature response function, and is only dependent on the temperature itself.

\begin{figure}[tbp]
\centering
\subfigure[]{\includegraphics[width=0.45\textwidth]{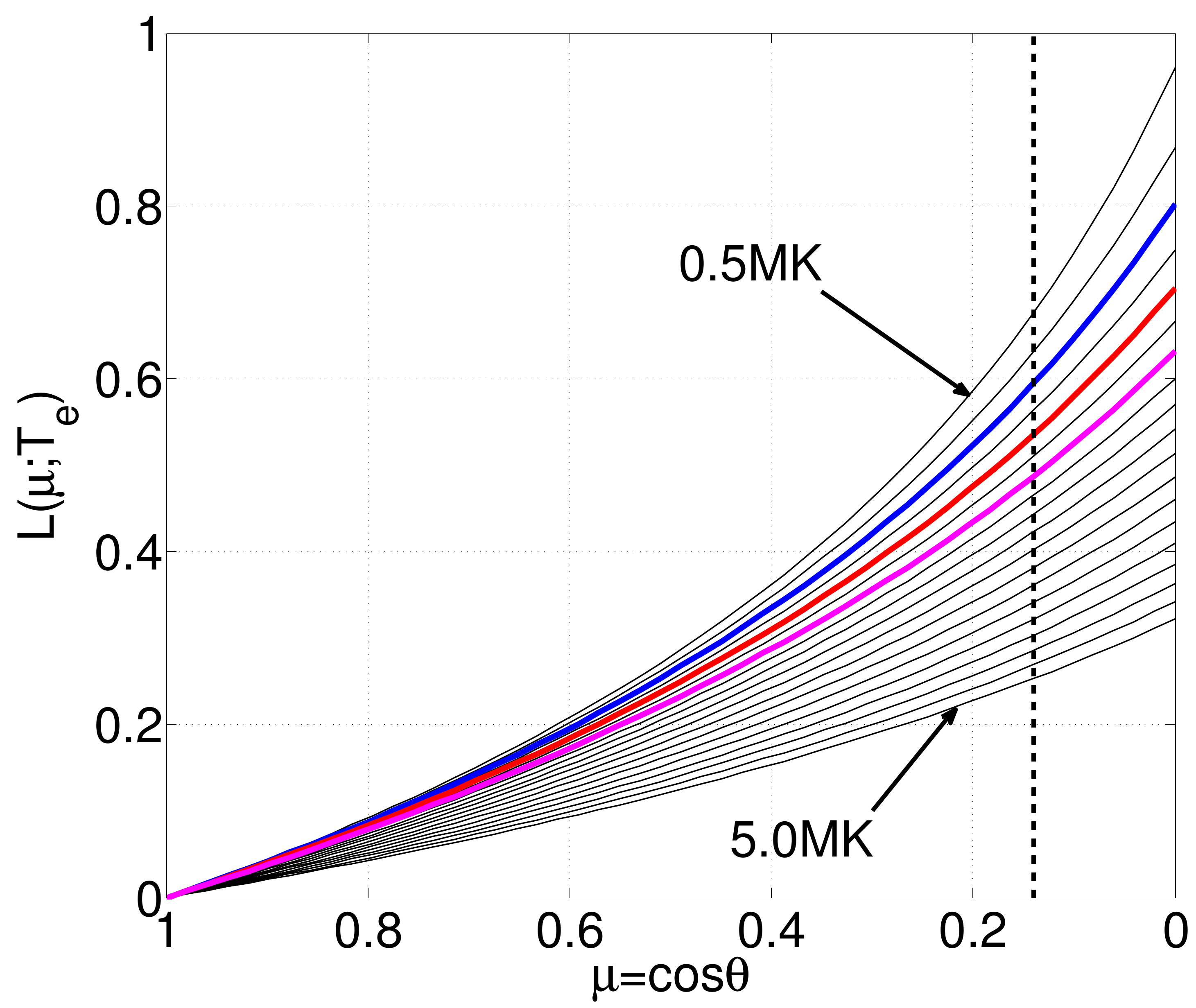}}
\subfigure[]{\includegraphics[width=0.45\textwidth]{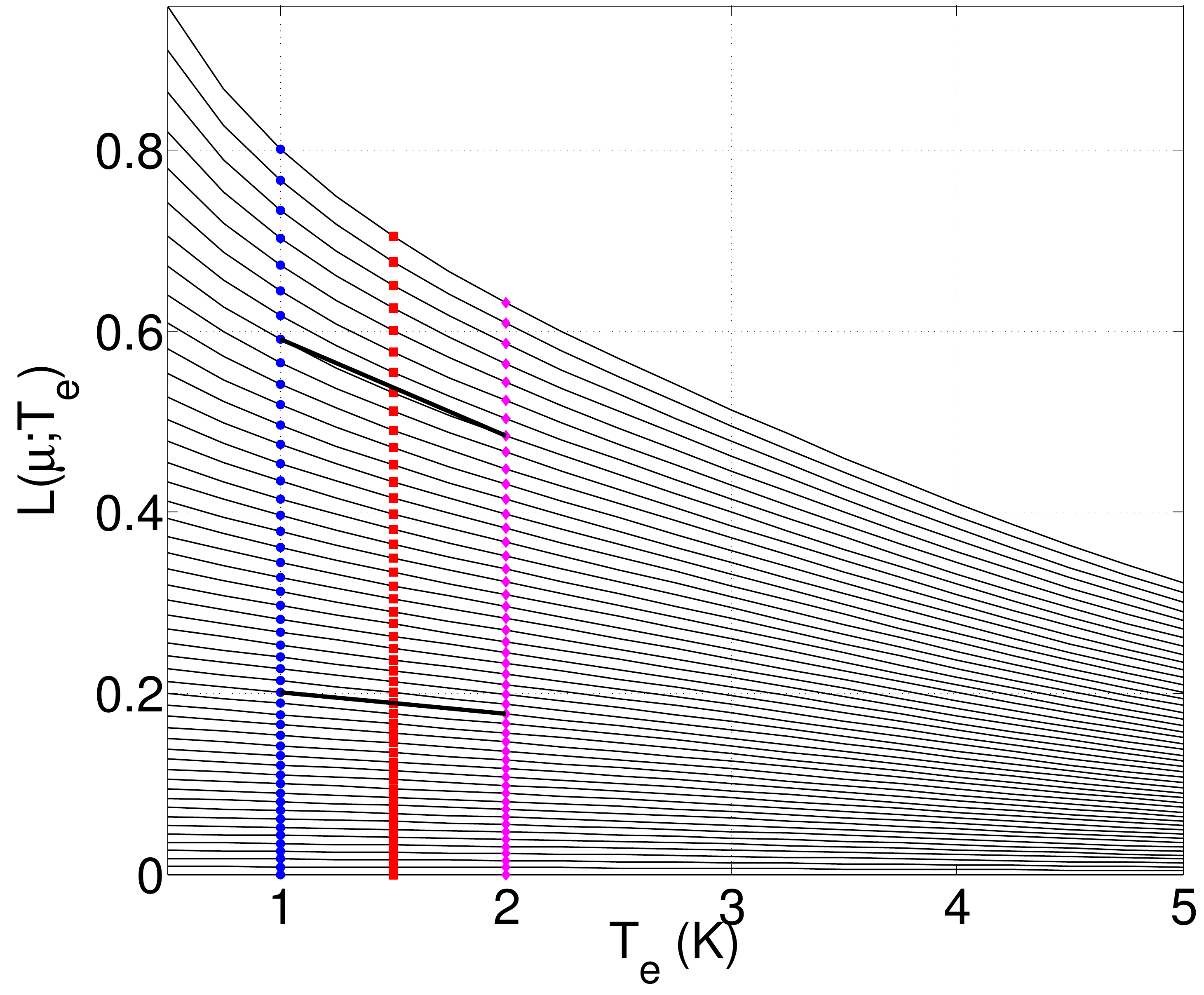}}
\caption{Theoretical limb-brightening correction curves computed from Eq.~(\ref{eq_lbcdef1}).  (a) From top to bottom:  curves for temperatures of $T_e=0.5\,\mbox{MK}$ to $T_e=5\,\mbox{MK}$ in increments of $0.25\,\mbox{MK}$. The vertical dashed line indicates the $\mu$ value at $r={R_\odot}$ given a coronal radius of $R_0=1.01\,R_{\odot}$. (b) From top to bottom: curve values for $\mu=0$ to $\mu=1$ in increments of $0.02$. The curves for $1\,\mbox{MK}$, $1.5\,\mbox{MK}$, and $2\,\mbox{MK}$ are highlighted in blue, red, and magenta respectively.  Line segments in (b) are displayed along curves for $\mu=0.6$ and $\mu(r=R_\odot)$, illustrating the near-linear temperature dependence of the curves.\label{fig_lbcc_theo}} 
\end{figure}

In Fig.~\ref{fig_lbcc_theo} we show the theoretical limb-brightening correction curves computed numerically from Eq.~(\ref{eq_lbcdef1}) for temperatures between $0.5\,\mbox{MK}$ and $5\,\mbox{MK}$ visualized as a function of $\mu$ (left) and $T_e$ (right).
We see that the effect of limb brightening is much larger towards the limb than near disk center and decreases overall with increasing temperature.  It is also noticeable from the right plot that the temperature dependence of the curves can be well approximated by a linear function over small ranges of temperature (see Sec.~\ref{sec_lbcc_data2} where we make use of this approximation).

Using theoretical LBCCs on the EUVI and AIA images is problematic however, because for any given EUV image of the sun, distinct coronal structures will have different temperatures and densities, both of which are unknown. Furthermore, the total emission for a given pixel is a convolution of the Differential Emission Measure distribution (DEM) of the plasma along that line-of-sight with the temperature response function of the given filter. Therefore, a limb-brightening curve of a single temperature will not be representative for the center-to-limb variation of all visible structures. For purposes of coronal hole detection, one could choose a curve corresponding to the estimated average temperature of a coronal hole (say 0.5 to 0.8 MK) but this too would not be optimal because it would over-correct the center-to-limb variation of hotter structures in the image, thus reducing their contrast with coronal holes. Additionally, even for a single temperature, additional effects such as opacity near the limb due to resonant scattering \citep{1994_Schrijver_OpticallyThickEUVLines,2000_Schrijver_EUVopacityQuietSun} are not taken into account in the optically thin model for limb brightening. 

For these reasons, we opt for a more data-driven approach to model the center-to-limb intensity variance guided by properties of the theoretical limb-brightening curves described above.  The data-driven nature of the model is advantageous as it will inherently include effects not captured in the analytical model and can be used for the full range of structures within the EUV image.

\subsubsection{Data selection}
\label{sec_lbcc_data1}
To obtain data-derived limb-brightening corrections, we start by extracting intensity values as a function of $\mu$ from the image.  Due to solar rotation and the inherent inhomogeneity of the corona, one must take care in determining the temporal and spatial selection of the data to be used.  

To mitigate selection bias and to better sample variations of coronal structures, we average the image data over time.  We have found that a one-year average is optimal for the time period we studied. This averaging window was chosen by observing the variations in the computed limb-brightening correction curve values (using the methods described in this section) as the averaging window was increased.  The window size was fixed when the variations in the correction parameters were observed to be converging to a constant amplitude.  

\begin{figure}[tbp]
\centering
\includegraphics[width=0.32\textwidth]{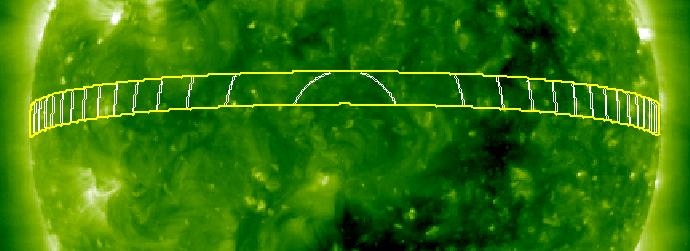}
\includegraphics[width=0.32\textwidth]{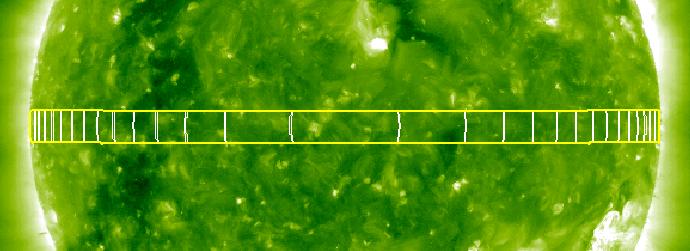}
\includegraphics[width=0.32\textwidth]{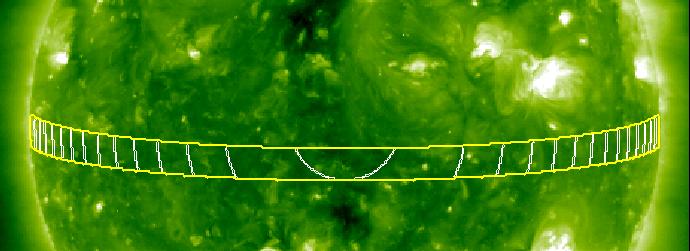}
\caption{Examples of latitude-limited $\mu$-bins used to compute histograms for limb-brightening correction.  Left to right: STB, AIA, and STA for 06/10/2010.  The latitude is limited by $\pm \pi/64\approx 2.8^{\circ}$ from center with a total of $15$ $\mu$-bins with a width of $\Delta \mu\approx 0.06$ outside of the central bin. \label{fig_lbcc_latlim}}
\end{figure}

The discrete nature of the images require the selection of bins of image data to represent the intensity values $I$ over $\mu$.  Using bins of data also helps to avoid effects of noise and small transient structures exhibiting large intensity variation.   A seemingly good choice for the $\mu$-bin locations/sizes is such that an equal area of data is contained in each bin.  This could be done for the full disk for $n$-bins by using $\mu$-bin locations defined by $\mu_i=\sqrt{1-(i-1)/(n-1)}$, $i={1\rightarrow n}$. However, such a choice is not optimal because solar rotation implies that structures visible at high latitudes will not rotate through all possible center-to-limb angles, causing them to contribute unequally to the $\mu$-bins. Instead, we enforce that all $\mu$-bins span the same range in heliographic latitude. In practice, this is done by determining the projected latitudes based on the current $B_0$ angle of the spacecraft and restricting bins to a strip that intersects disk center such that all $\mu$-bins cover the same range of heliographic latitude during the data averaging window. We have found that a latitude limit of $\pm \pi/64\approx2.8^{\circ}$ from center and bin sizes of $\Delta \mu\approx 0.06$ outside the center bin yields acceptable results for the time period discussed in this paper.  It is also important to note that because the emission properties of the image understandably change rapidly past the photospheric edge, we only collect data up to $r=R_{\odot}$ which, for our choice of $R_0=1.01\,R_{\odot}$, is represented by the $\mu$ value of
\[
\mu(r=R_{\odot})=\sqrt{1-R_{\odot}^2/R_0^2}=0.14037.
\]
Fig.~\ref{fig_lbcc_latlim} shows examples of the latitude limited $\mu$-bins used. 

\subsubsection{Data-derived LBCCs}
\label{sec_lbcc_data1b}
\begin{figure}[tbp]
\centering
\includegraphics[width=0.45\textwidth]{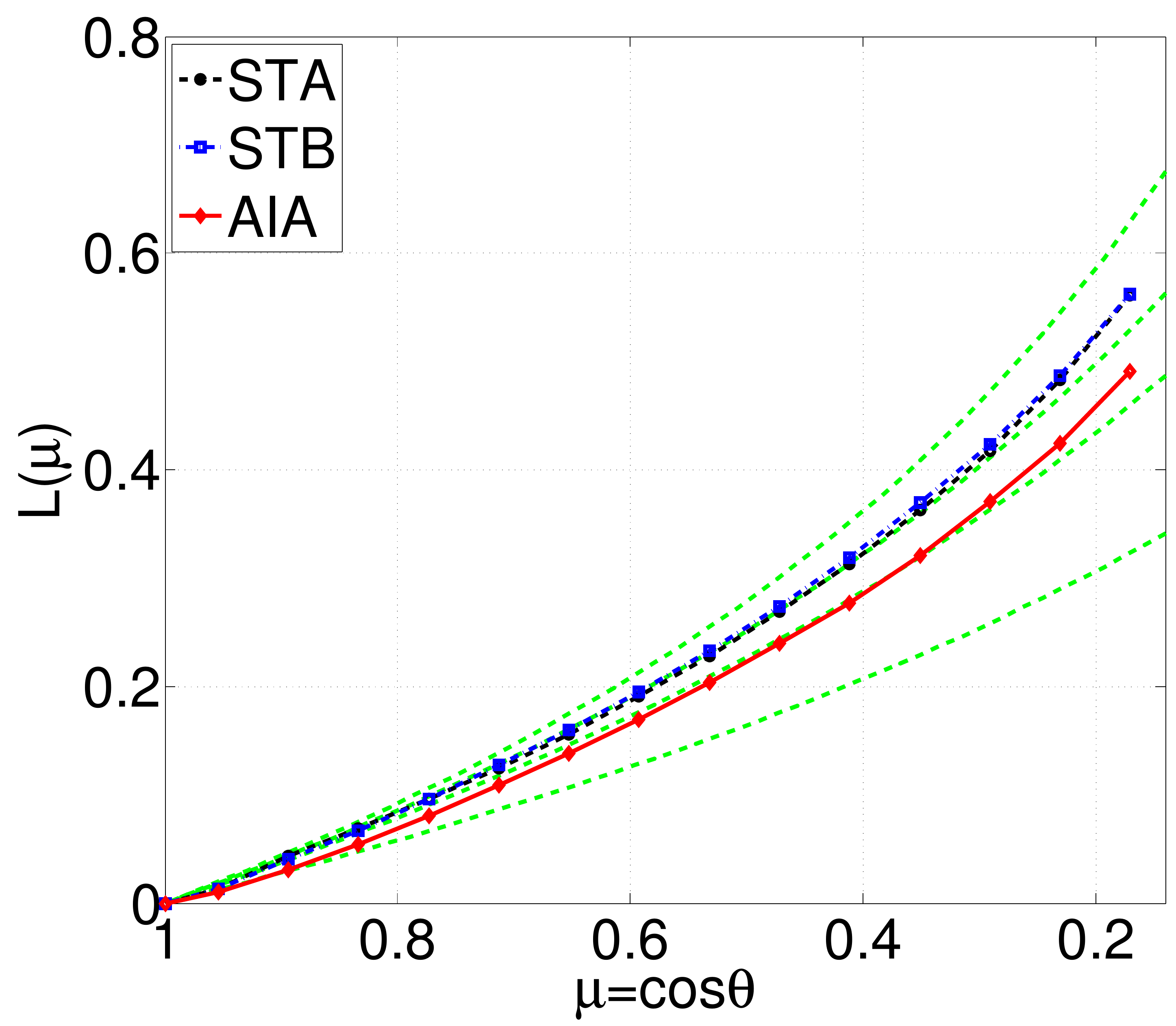}
\includegraphics[width=0.45\textwidth]{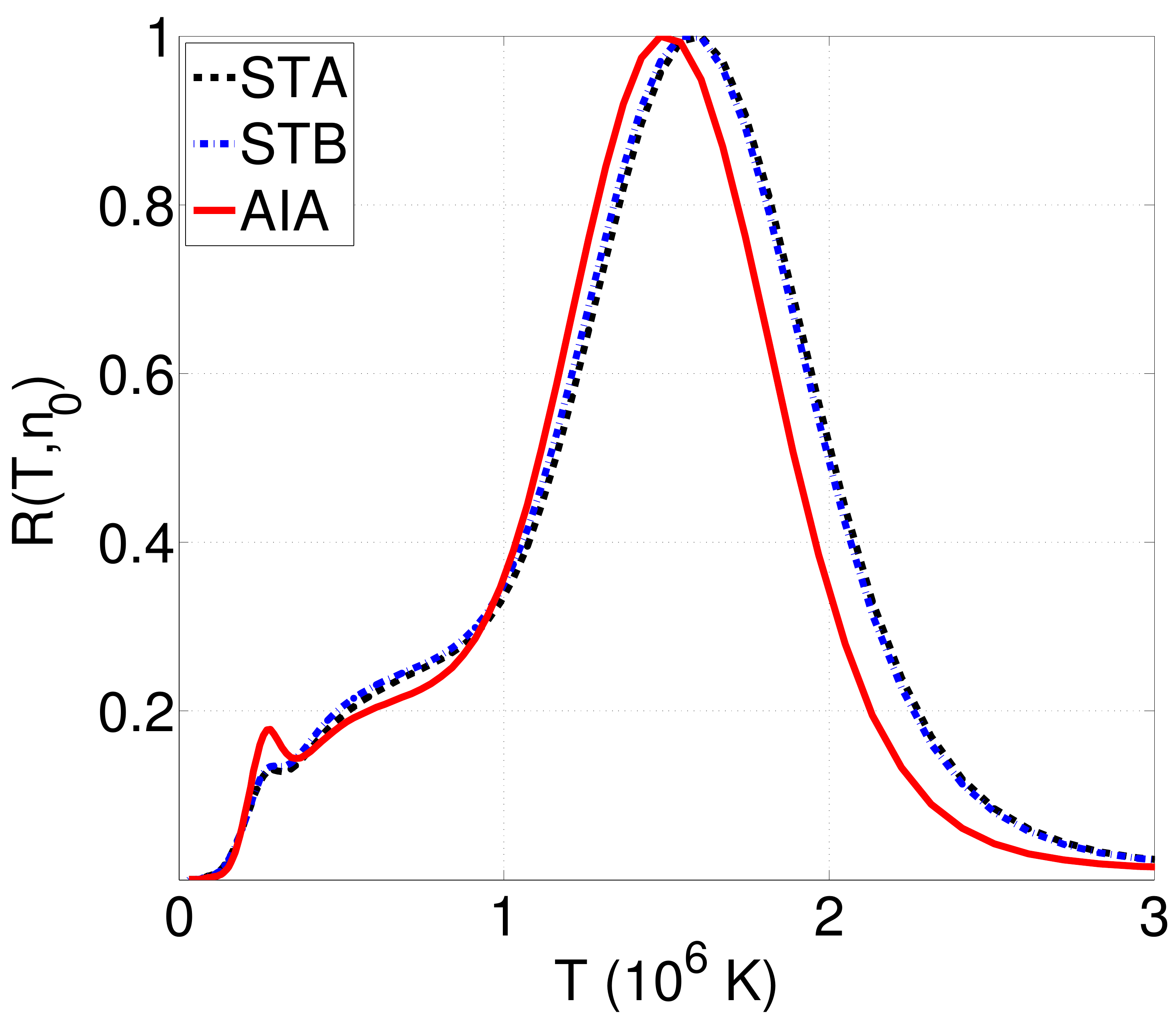}
\caption{Left:  LBCCs for 02/03/2011 for STA (black dashed), STB (blue dot-dashed), and AIA (red solid) computed with average intensities over one year at 6-hour cadence.  The curves were computed with the $\mu$-bin locations and latitude limit level shown in Fig.~\ref{fig_lbcc_latlim}.  The theoretical curves of Eq.~(\ref{eq_lbcdef1}) (green dashed-lines) for $T\in\{0.5,1.25,2,3.75\}\mbox{MK}$ (top to bottom) are shown for comparison.  Right: Normalized response functions for STA/B and AIA at a number density of $n_0=5\times 10^8 \mbox{cm}^{-3}$. The response functions are determined using calibration data available in SSW and synthetic spectra generated from the CHIANTI 7.1 database. \label{fig_lbccs1}}
\end{figure}

The simplest way to obtain a limb-brightening correction curve from the $\mu$-bins obtained in Sec.~\ref{sec_lbcc_data1} is to measure the average intensity inside each of the $\mu$-bins over the time-window ($\bar I(\mu)$), and subtract the averaged intensities at the disk center bin ($\bar I_0$) (i.e. directly applying the definition of the LBCC of Eq.~(\ref{eq_lbcc_def})). In Fig.~\ref{fig_lbccs1}, we display example LBCCs computed in this manner for each of the three instruments.
The curves are seen to be very similar in form to the single-temperature LBCCs of Eq.~(\ref{eq_lbcdef1}).  We also note that the LBCC (at least for the time chosen) of AIA differs noticeably from those of STA and STB.  This is likely due to the differences in their respective response functions (also shown in Fig.~\ref{fig_lbccs1}) as well as differences in the resolution and stray light profiles of the telescopes.

The data-derived LBCCs are still $I_0$-independent, but they are expected to be more accurate than using the theoretical model of Sec.~\ref{sec_lbcc_theo} because they capture effects not included in the analytical model (such as opacity, density variations, etc).  Also, because they represent the mean convolved limb-brightening rate of all multi-thermal structures, they eliminate the need to arbitrarily choose a mean coronal temperature as is required when using Eq.~(\ref{eq_lbcdef1}).  Despite these improvements, we wish to further enhance the technique to treat specific coronal structures independently (i.e. an $I_0$-dependent correction).  This can be achieved using the information present in the full histogram of intensities within each $\mu$-bin.  In the next section, we explore how image transformations can be used to closely match these histograms.  

%As an example, given the histogram of averaged intensity values in the disk center bin and the histogram at another $\mu$-bin, one can shift the histograms by applying the image transformation $I^{\prime}(\mu) = I(\mu) - s(\mu)$ where the shift distance $s(\mu)$ in this case is directly the LBCC value $L(\mu)$ of Eq.~(\ref{eq_lbcc_def}).  However, such a transformation would once again result in an $I_0$-independent LBCC. Therefore, we instead use an alternative histogram matching strategy described in detail in the next Section, one that is designed to allow for the application of $I_0$-dependent limb-brightening corrections to the image.  

\subsubsection{Intensity-dependent limb-brightening correction} 
\label{sec_lbcc_data2}
Up to now, the LBCCs are applied as a transformation of the $\mbox{log}_{10}$ image described as
\begin{equation}
\label{eq_lbcc_L}
I_0 = I(\mu) - L(\mu),
\end{equation}
i.e. given an image pixel at a $\mu$ level, one subtracts the LBCC $L(\mu)$ to get the pixel's approximate value as if it were at disk center.  Thus, every pixel at the same $\mu$ level is corrected with the same $L$ value.  To remove this constraint, we can instead use a nonlinear (linear in log space) image transformation described as
\begin{equation}
\label{eq_lbcc_beta_y}
I_0 = \beta(\mu)\,I(\mu) + y(\mu).
\end{equation}
This transformation is equivalent to using a unique LBCC for each and every possible disk center intensity $I_0$ value ($I_0$-dependent), which can be expressed by combining Eq.~(\ref{eq_lbcc_L}) and Eq.~(\ref{eq_lbcc_beta_y}) to yield
\begin{equation}
\label{eq_lbcc_beta_y_L}
L(\mu,I_0) = \left(\frac{1}{\beta(\mu)} - 1\right)\,I_0 - \frac{y(\mu)}{\beta(\mu)}.
\end{equation}
The form of the transformation of Eq.~(\ref{eq_lbcc_beta_y}) is chosen primarily because it is the lowest-order nonlinearity that can modify both the position and the shape of the $\mu$-bin histograms to match the central histogram.  However, we observed that the temperature dependence of the theoretical LBCCs in Fig.~\ref{fig_lbcc_theo} can be well approximated as a linear function of temperature for each $\mu$ value.  Assuming a correspondence between image intensity ($I_0$) and temperature (see Appendix~\ref{sec_lbcc_theory}), the linearly $I_0$-dependent form of Eq.~(\ref{eq_lbcc_beta_y_L}) can be viewed as a first-order approximation of the temperature dependence of the theoretical LBCCs.

%However, it is interesting to note that $I_0$-dependent LBCCs are a function of both $\mu$ and $I_0$, which is similar to the theoretical LBCCs in Sec.~\ref{sec_lbcc_theo} being functions of $\mu$ and temperature. 

%While a direct correspondence between image intensity and temperature in an EUV image is somewhat suspect, this could be justified (at least on average) by a heuristic physics based argument. 
%Drawing from a long history of idealized coronal heating scaling laws \citep[e.g.][and references therein]{2002_Aschwanden+Schrijver_HeatingScalingLaws}, if one assumes that there is on average some general power law correspondence between number density and temperature (no matter how weak) then the strong dependence of emissivity on density implies that the intensity of a given structure will be positively correlated with temperature (e.g. Eq.~(\ref{eq_lbcc_mu1})).

%Assuming this positive correlation, we can utilize the theoretical LBCC curves of Fig.~\ref{fig_lbcc_theo} as a guide for the model.  
%We observe that the dependence of the theoretical LBCCs over temperature appears nearly linear in the range of temperatures where the response functions of STA\&B and AIA are most prominent (especially away from the limb).
%Therefore, the choice of the linear form of Eq.~(\ref{eq_lbcc_beta_y_L}) as a function of $I_0$ can be seen as a first-order approximation of the temperature dependence of the theoretical LBCCs.

\begin{figure}[tbp]
\centering
\includegraphics[width=0.45\textwidth]{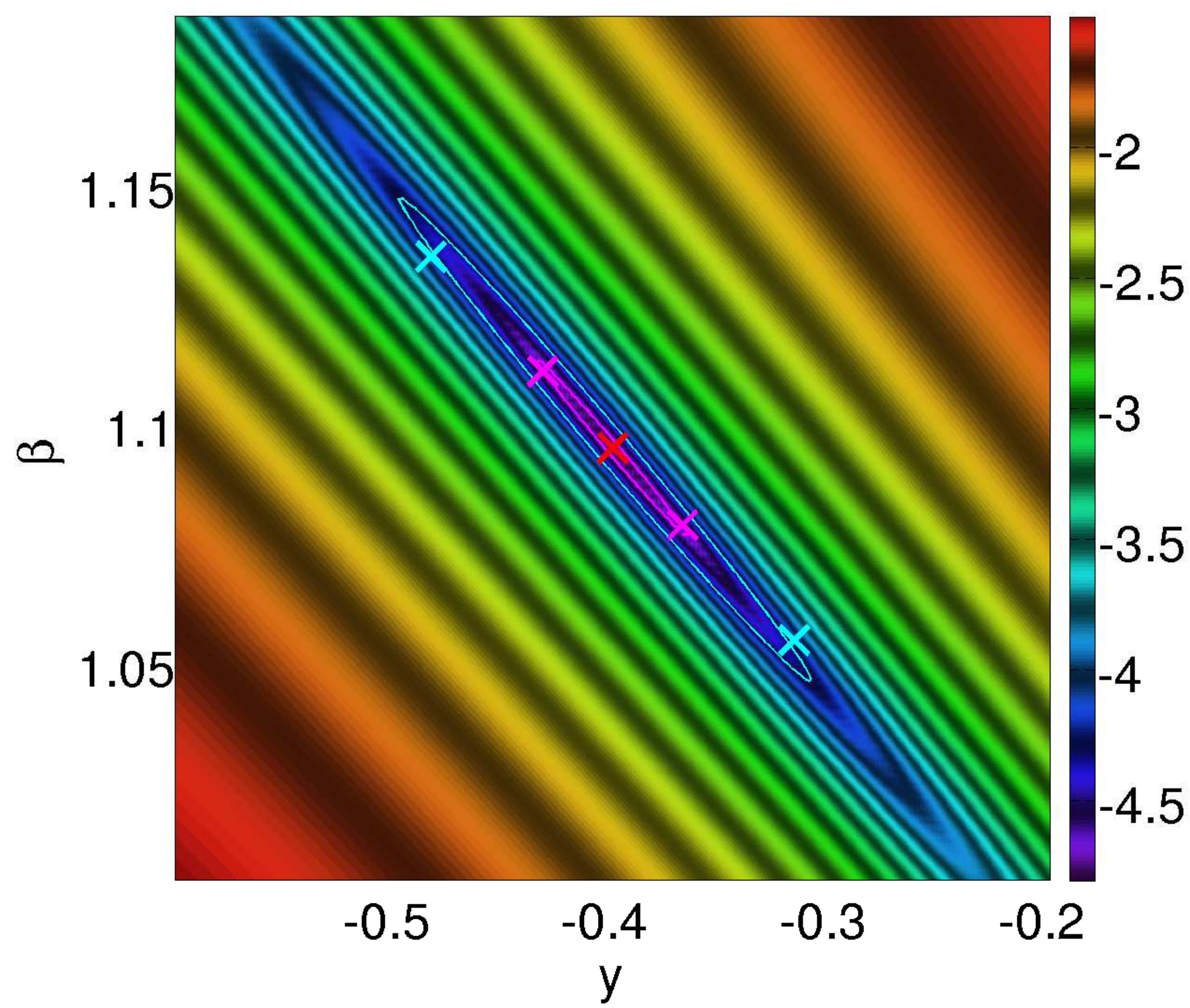}
\includegraphics[width=0.44\textwidth]{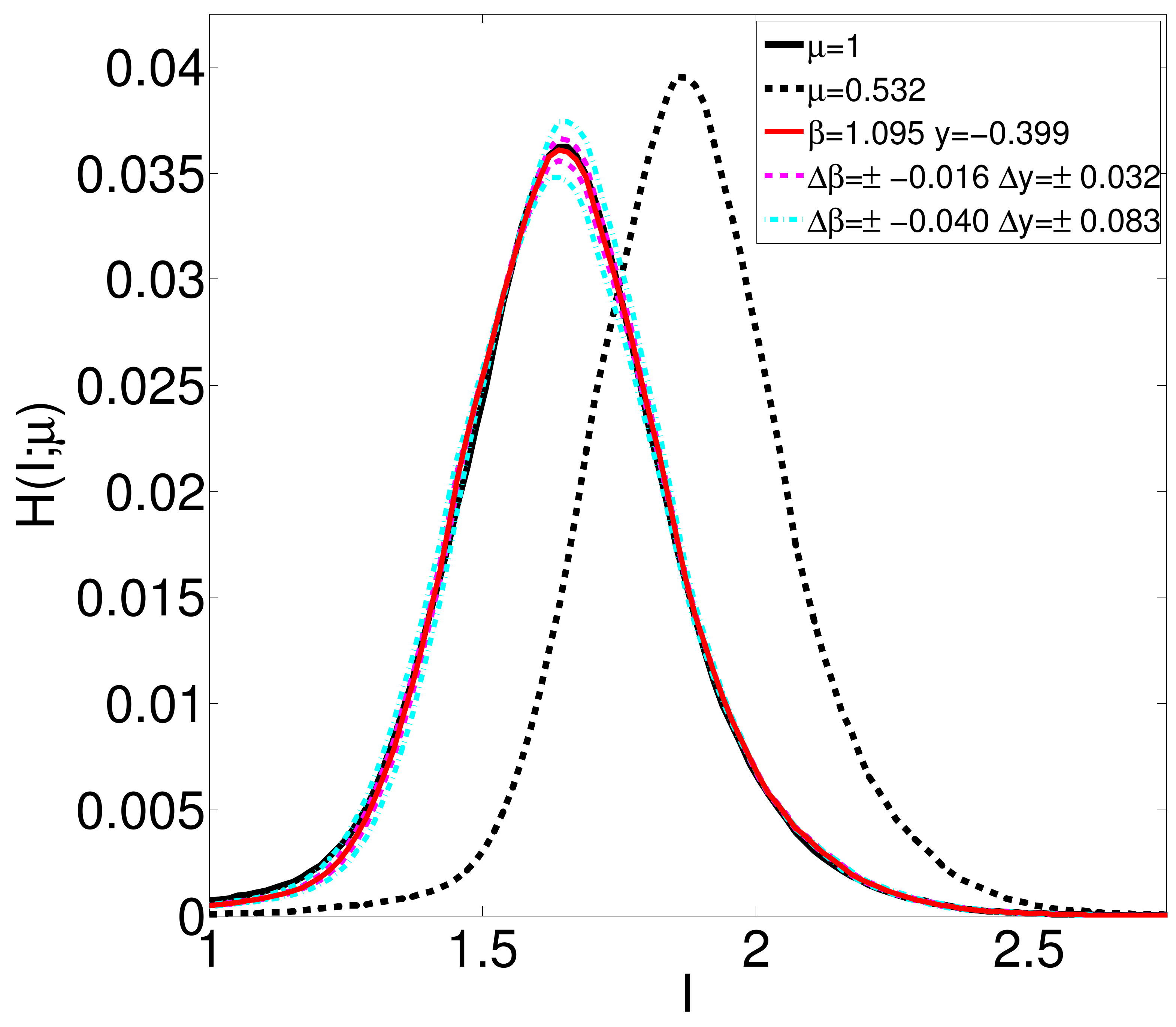}
\caption{Left:  An example of the sum-of-squares error (log) in the $\beta$-$y$ parameter space for the difference between a normalized histogram $H(I;\mu=0.532)$ transformed by Eq.~(\ref{eq_lbcc_beta_y}) and the central normalized histogram at $\mu=1$.  Although there is a unique minimum, there is a large parameter space where the error is small and produces visibly-satisfactory histogram matches.  Right:  Original and transformed histograms for the $\beta$-$y$ values indicated in the left figure.\label{fig_lbcc_sse}}
\end{figure}

In order to find the optimal values of $\beta(\mu)$ and $y(\mu)$ over time, we perform a two-dimensional least-squares optimization between the normalized histograms at each $\mu$ level ($H(I;\mu)$) to the disk center normalized histogram ($H(I_0;\mu)$).  The histograms are normalized by dividing the bin sample value by the total number of samples.  It turns out that this is not a well-behaved optimization because there is a fairly large parameter space in which the histograms match very well.  To illustrate this, we show the sum-of-squares error (SSE) for a typical histogram-matching optimization in Fig.~\ref{fig_lbcc_sse}.   
We see that although there is a unique minimum (in this case a SSE value of $~10^{-5}$), there are many combinations of $\beta$ and $y$ values that still yield very low SSE values corresponding to visibly acceptable histogram matches.  Thus, we expect our computed $\beta(\mu)$ and $y(\mu)$ curves to suffer from some local variations over time and over $\mu$ (as can be seen in Fig.~\ref{fig_lbcc_beta_y}) yet still produce consistent quality transformations.  Due to the discrete and sometimes noisy nature of the histograms, many classic optimization algorithms that require smoothness have convergence problems. Therefore, we use the derivative-free Nelder-Mead Simplex Method, which is shown to have good convergence properties in two-dimensional problems such as ours \citep{1998_Lagarias_NM_Convergence}.

\begin{figure}[tbp]

\centering
$\begin{array}{rccc}
\;
& 
\hbox{\includegraphics[width=2in]{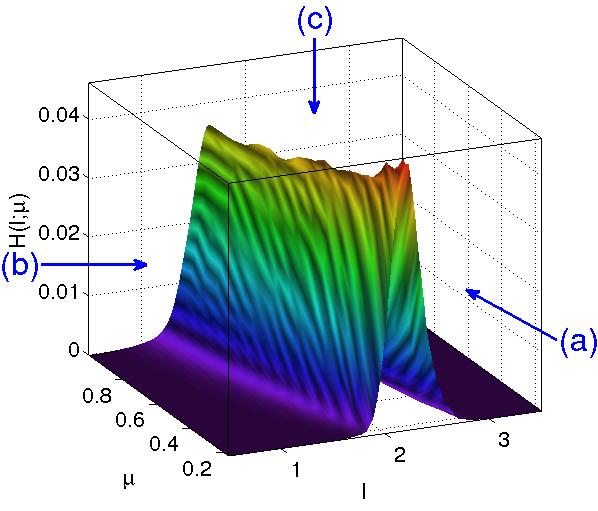}} 
&
\hbox{\includegraphics[width=2in]{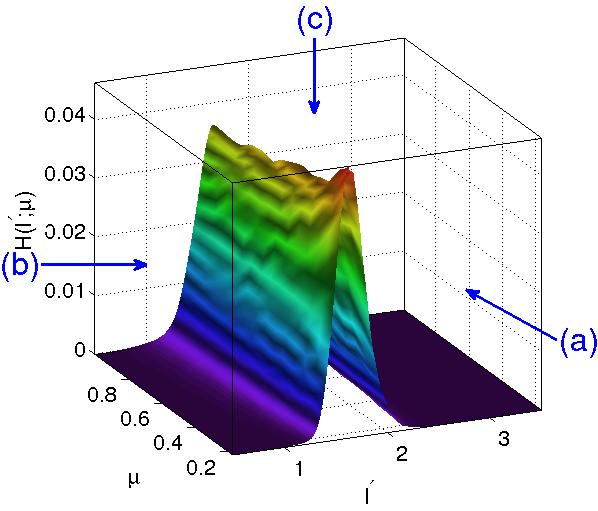}}
&
\hbox{\includegraphics[width=2in]{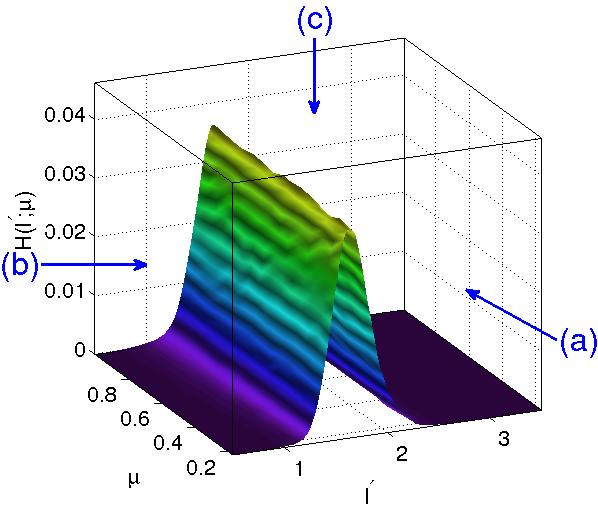}}
\\
\mbox{(a)}
& 
\hbox{\includegraphics[valign=m,width=2.05in]{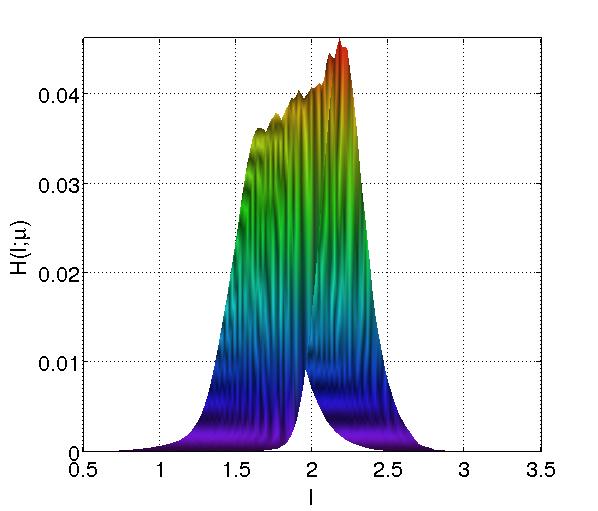}} 
&
\hbox{\includegraphics[valign=m,width=2.05in]{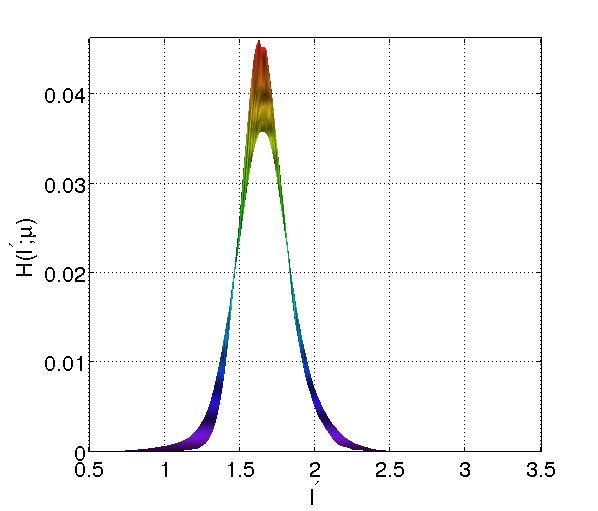}}
&
\hbox{\includegraphics[valign=m,width=2.05in]{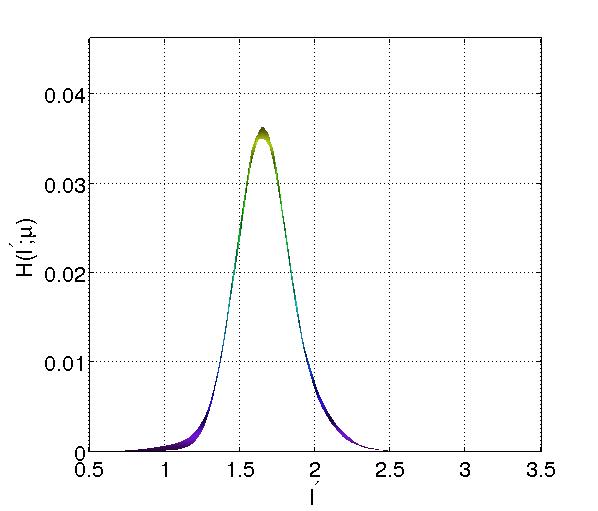}}
\\
\mbox{(b)}
& 
\hbox{\includegraphics[valign=m,width=2.05in]{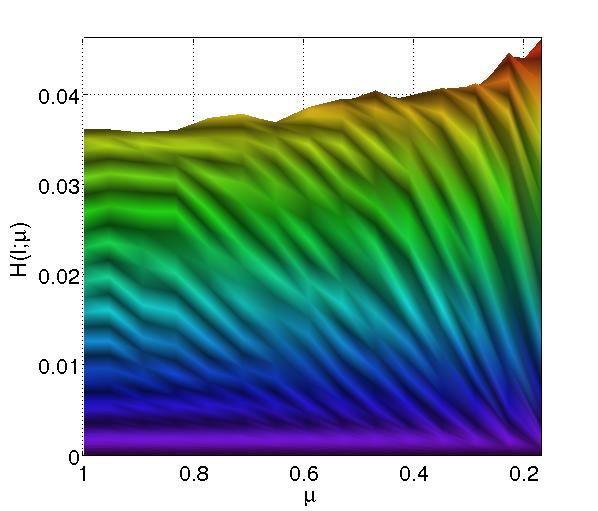}} 
&
\hbox{\includegraphics[valign=m,width=2.05in]{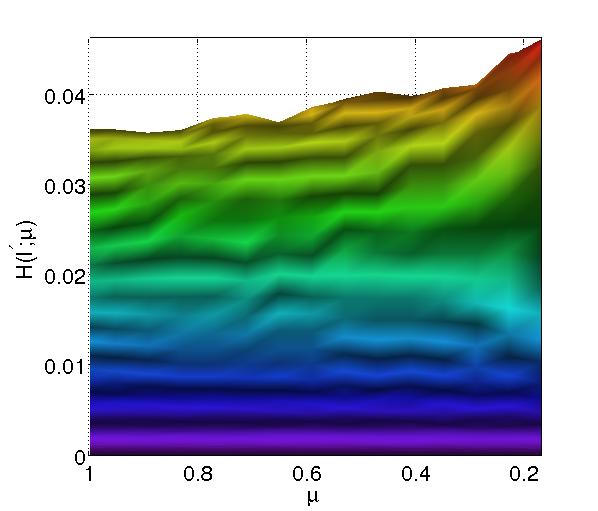}}
&
\hbox{\includegraphics[valign=m,width=2.05in]{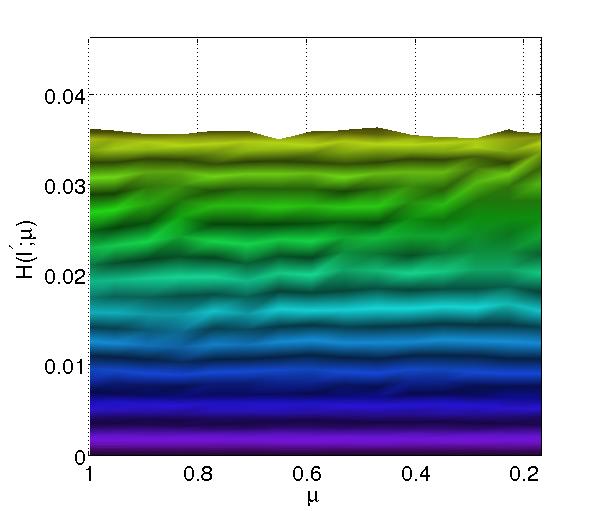}}
\\
\mbox{(c)}
& 
\hbox{\includegraphics[valign=m,width=2.05in]{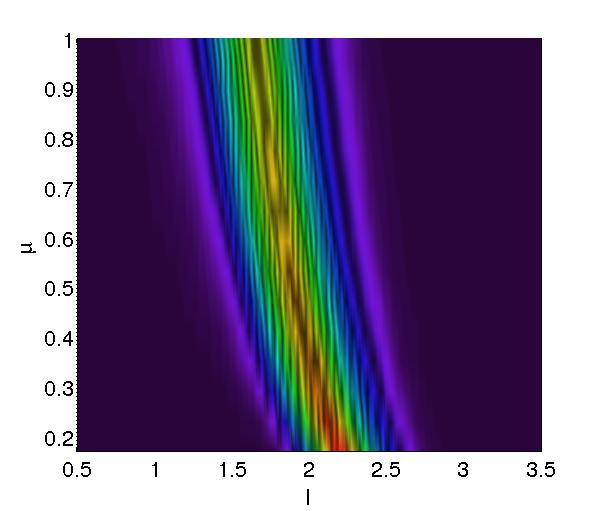}} 
&
\hbox{\includegraphics[valign=m,width=2.05in]{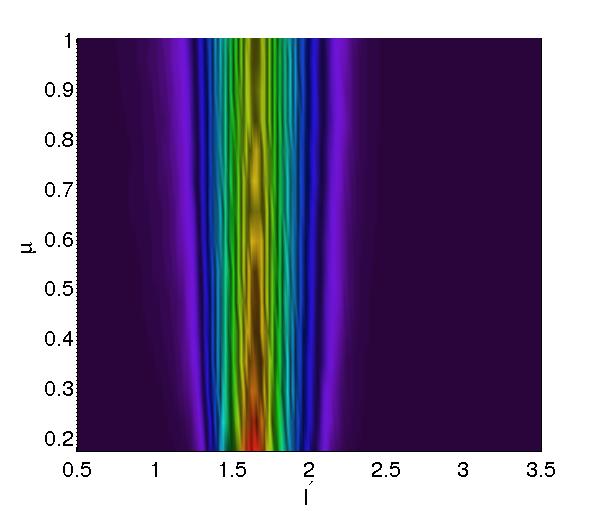}}
&
\hbox{\includegraphics[valign=m,width=2.05in]{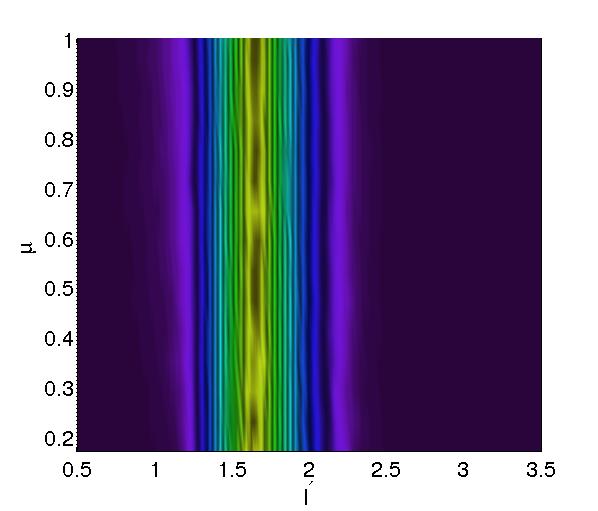}}
\\
\; & \mbox{(1) Original}\,I(\mu)& \mbox{(2)}\,I^{\prime}(\mu)=I(\mu)-L(\mu) & \mbox{(3)}\,I^{\prime}(\mu)=\beta(\mu)\,I(\mu)+y(\mu)
\end{array}$
\caption{Surface plots of one-year averaged (centered at 02/03/2011) normalized intensity histograms ($H(I;\mu)$) for all $\mu$-bins of STA disk image data before and after applying limb-brightening correction transformations.  The surface plots of $H(I;\mu)$ are shown at a variety of viewing angles (top to bottom) as indicated in the top perspective plots.  The viewing angles show the relationships (a) $H(I;\mu)$ versus $I$, (b) $H(I;\mu)$ versus $\mu$, and (c) $H(I;\mu)$ shown on the $I${--}$\mu$ plane.  Left to right: (1) Original histograms, (2) the histograms after applying an optimized $I_0$-independent LBCC (Eq.~(\ref{eq_lbcc_def})), and (3) the histograms after applying the optimized $I_0$-dependent LBCC image transformation of Eq.~(\ref{eq_lbcc_beta_y}).\label{fig_lbcc_hists}}
\end{figure}

In Fig.~\ref{fig_lbcc_hists}, we show surface plots of the normalized one-year averaged intensity histograms $H(I;\mu)$ for STA disk images centered at 02/03/2011 before and after the transformation of Eq.~(\ref{eq_lbcc_beta_y}) over all values of $\mu$.  For comparison, we also show the result of merely shifting the histograms for each bin along the intensity axis to match the position of the center bin histogram (i.e. applying an $I_0$-independent LBCC of Eq.~(\ref{eq_lbcc_def})).
Due to the amplitude and shape differences of the histograms over the $\mu$-bins, the $I_0$-dependent image transformation creates a much more consistent histogram along $\mu$ than that of Eq.~(\ref{eq_lbcc_def}), resulting in a better limb-brightening correction of the images.
 
\subsubsection{Limb brightening results and analysis}
\label{sec:lbcc_analy}
\begin{figure}[tbp]
\centering
\includegraphics[width=0.43\textwidth]{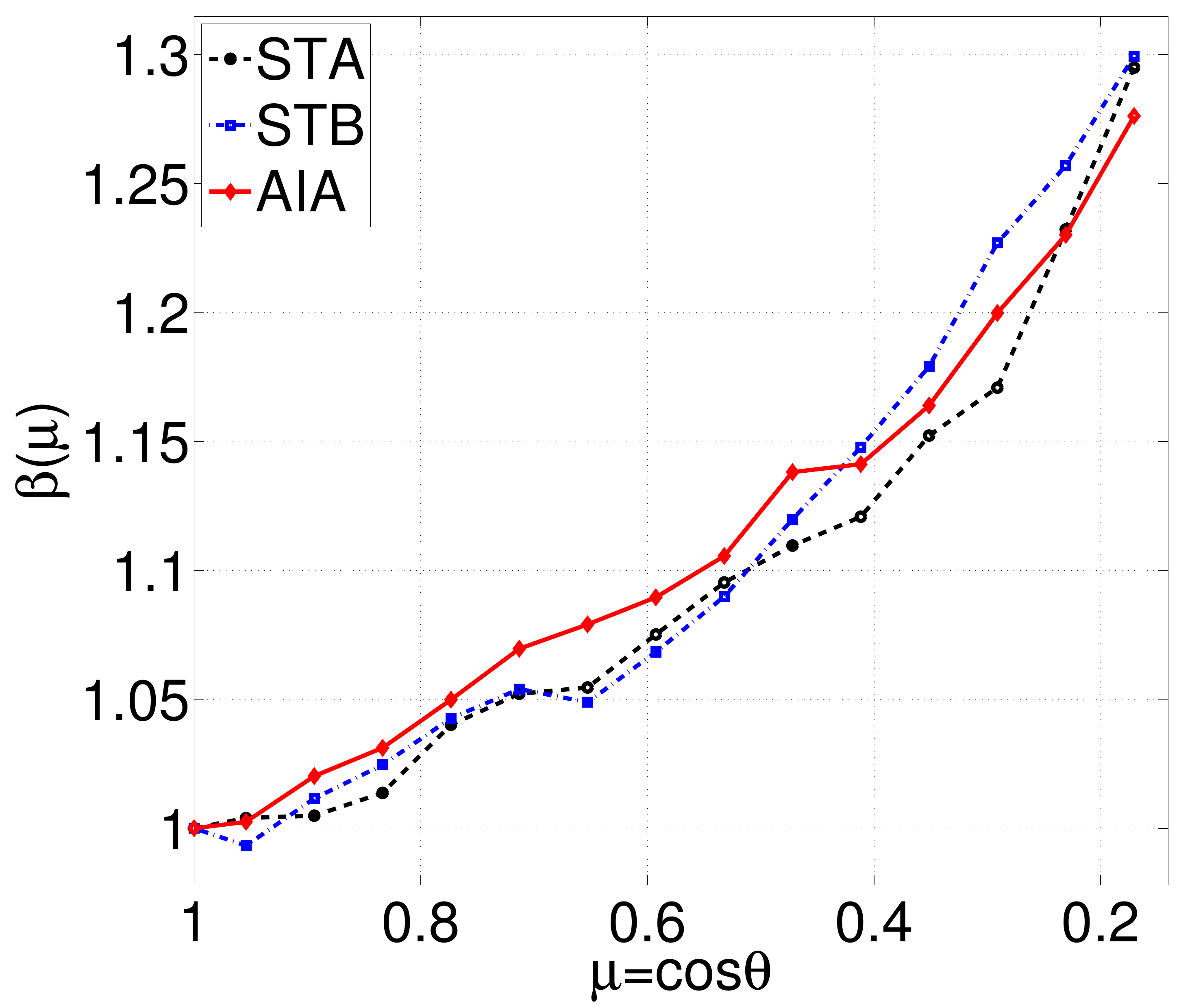}
\includegraphics[width=0.43\textwidth]{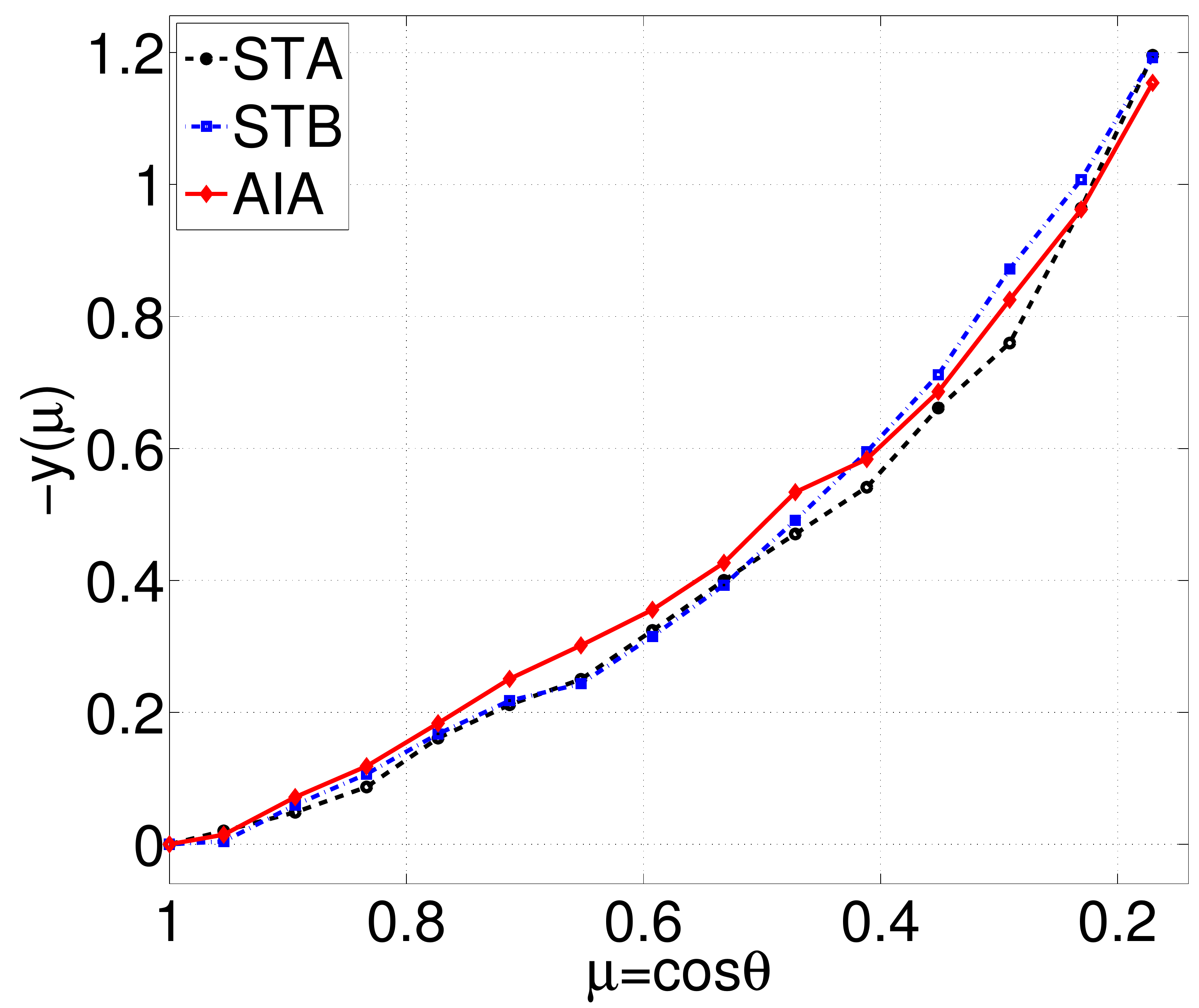}
\caption{Optimal values of $\beta(\mu)$ and $y(\mu)$ computed for 02/03/2011 for STA (black dashed), STB (blue dot-dashed), and AIA (red solid).  We used $300$ intensity value bins for each of the $\mu$-bin histograms.  The parameters for the averaging window, $\mu$-bin locations and sizes, and latitude limit used are the same as in Fig.~\ref{fig_lbccs1}. \label{fig_lbcc_beta_y}}
\end{figure}

\begin{figure}[tbp]
\centering
\includegraphics[width=0.32\textwidth]{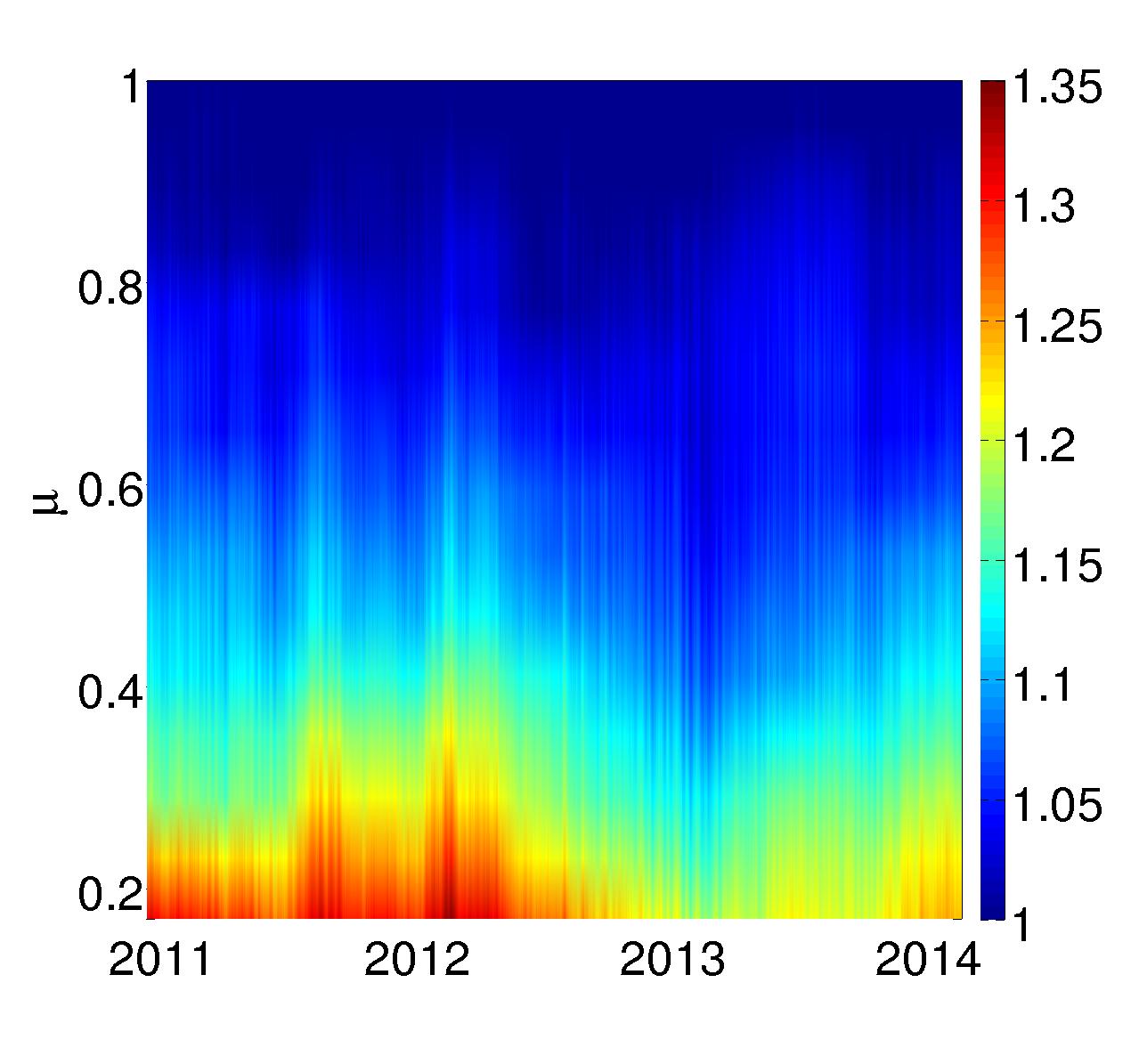}
\includegraphics[width=0.32\textwidth]{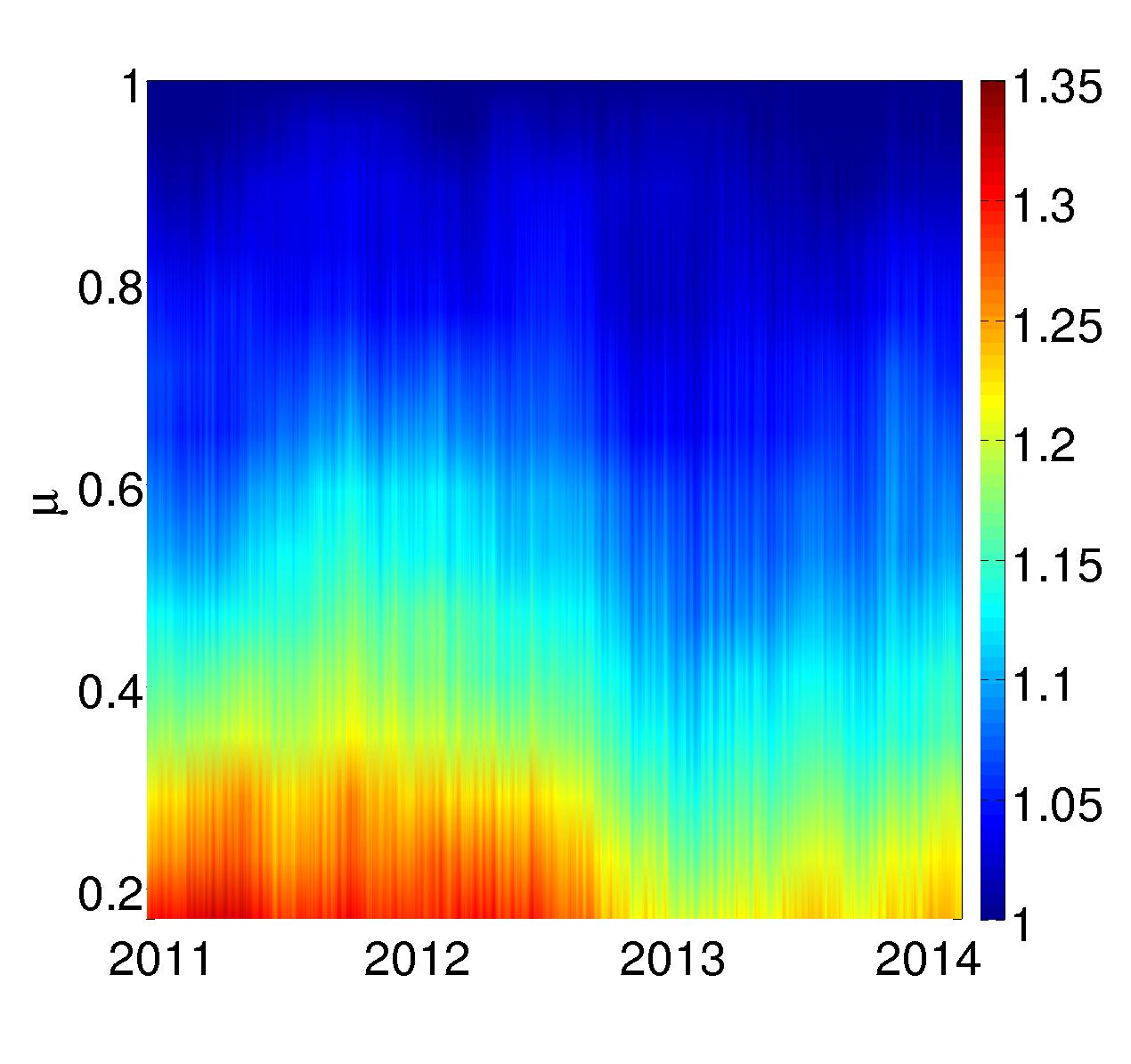}
\includegraphics[width=0.32\textwidth]{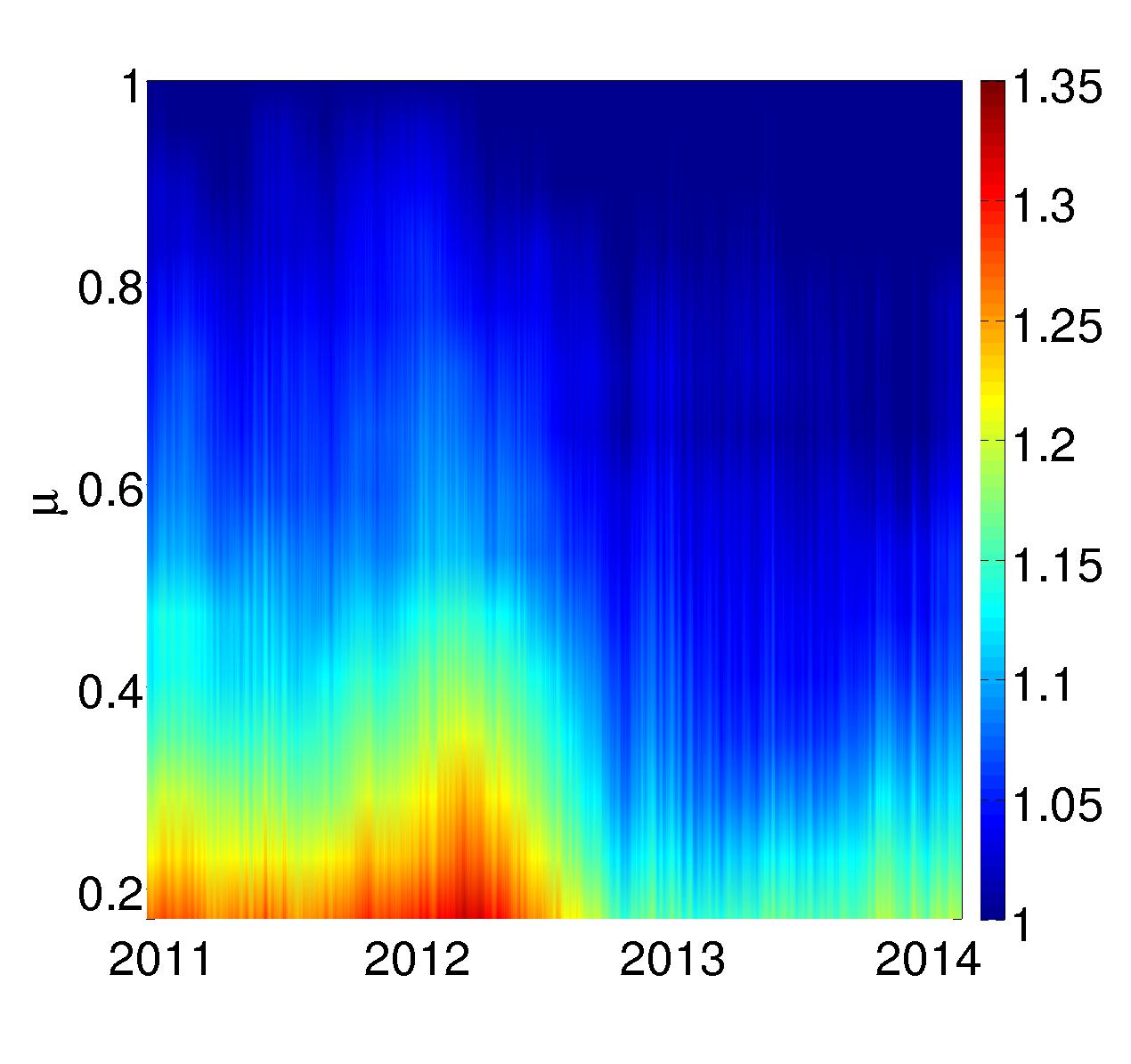}
\\
\includegraphics[width=0.32\textwidth]{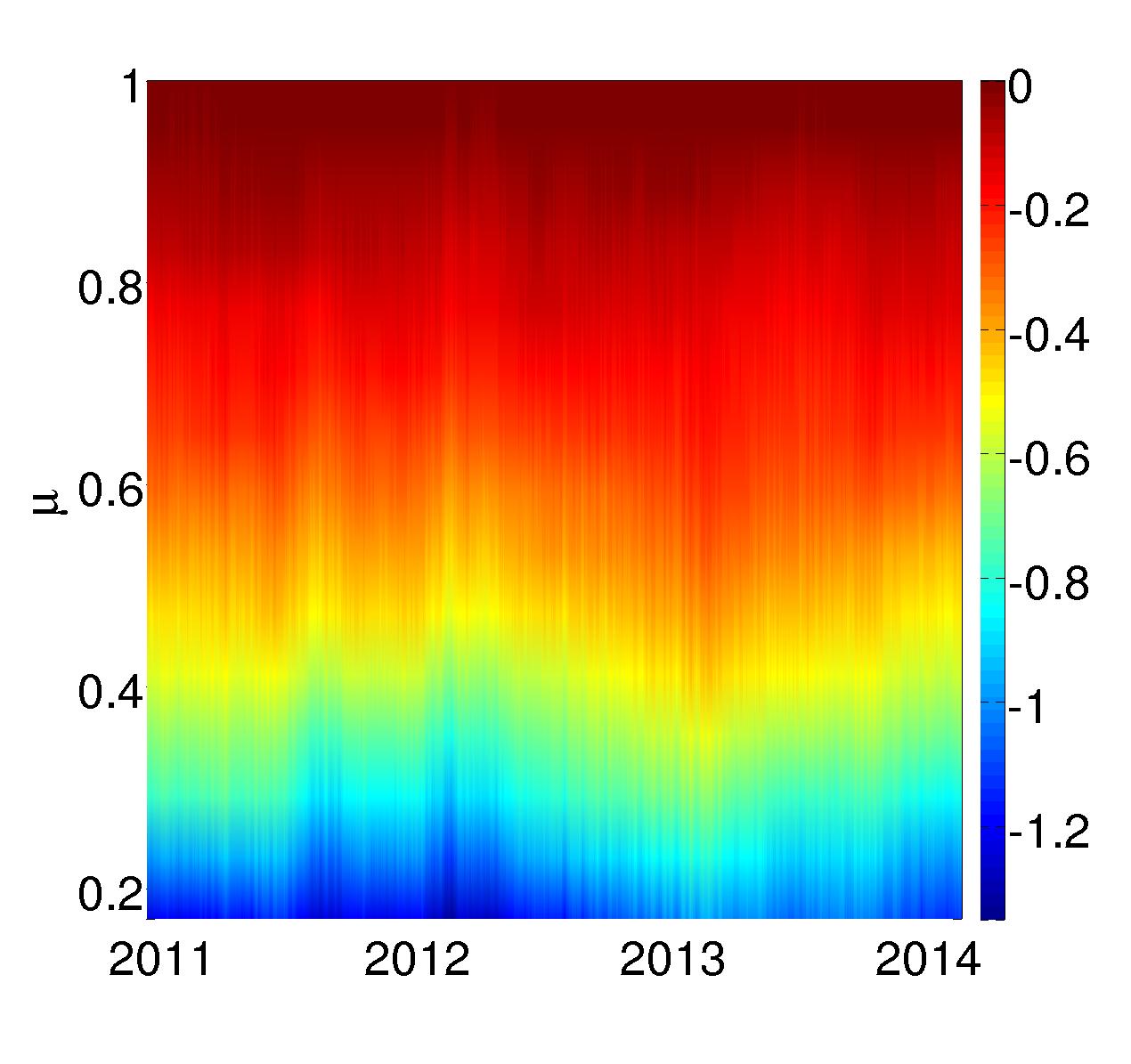}
\includegraphics[width=0.32\textwidth]{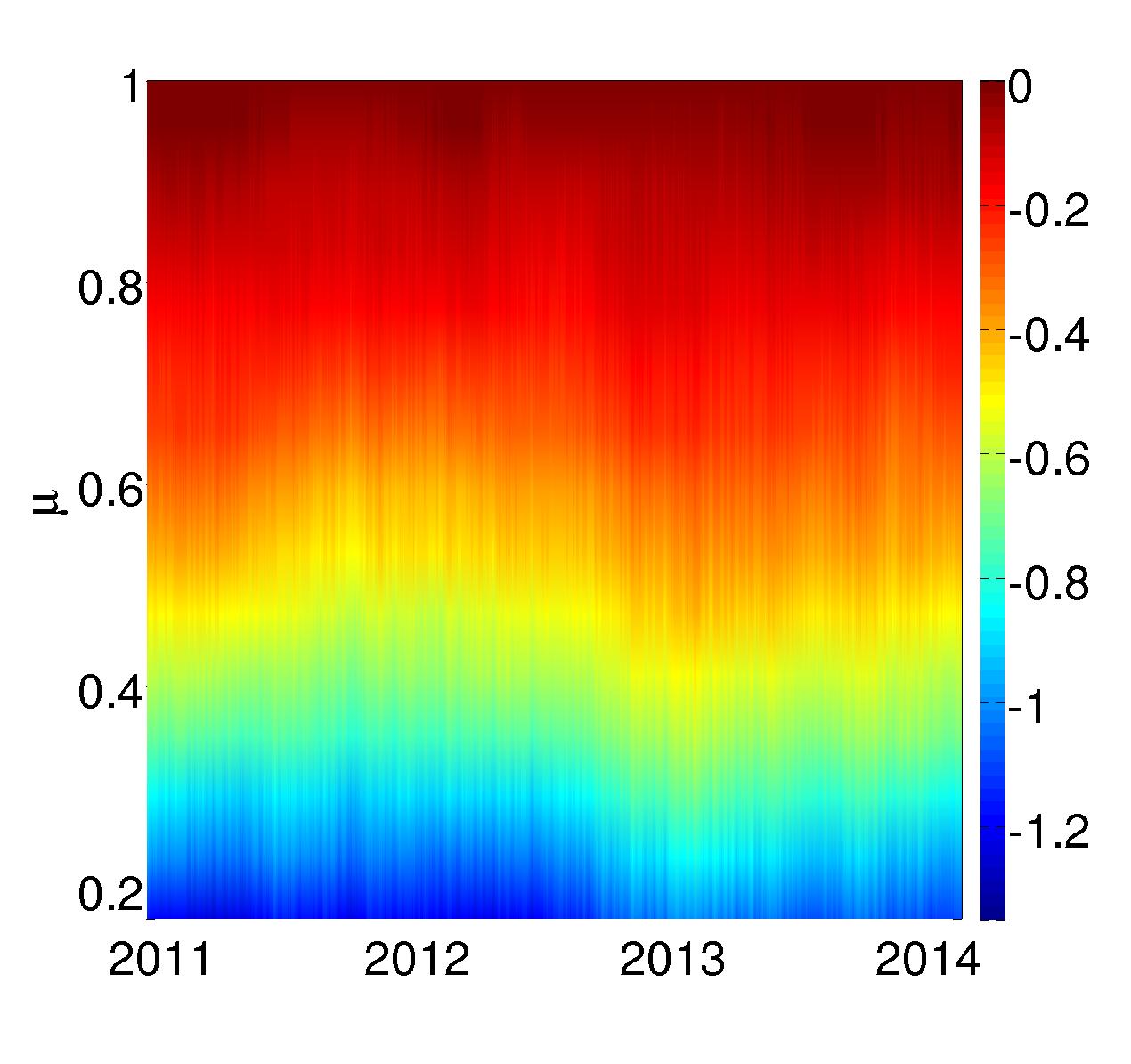}
\includegraphics[width=0.32\textwidth]{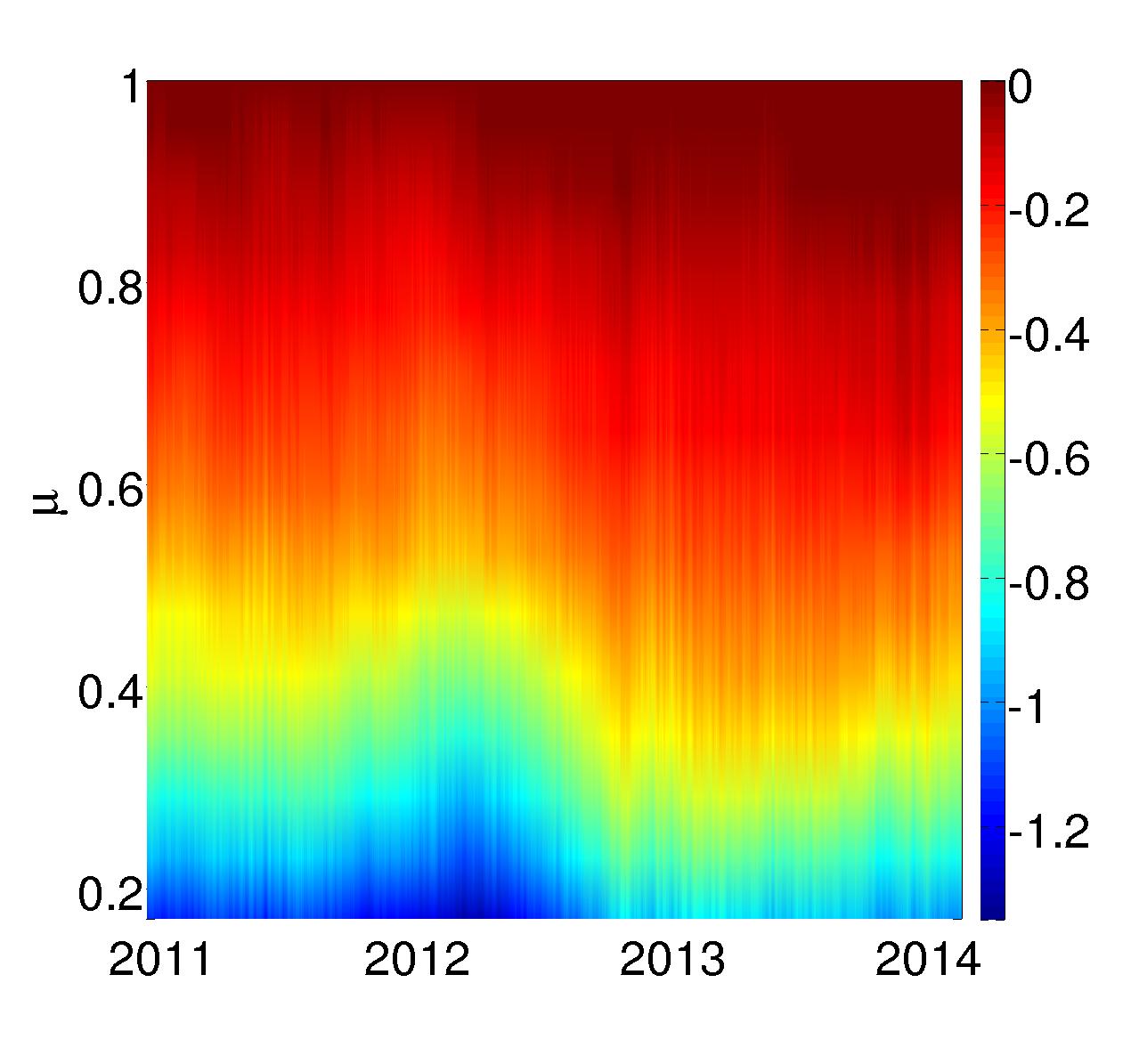}
\caption{Optimal values of the limb-brightening correction image transformation (Eq.~(\ref{eq_lbcc_beta_y})) factors $\beta(\mu)$ (top) and $y(\mu)$ (bottom) for STA, STB, and AIA (left to right) for the time period of 12/10/2010 to 02/17/2014.  The parameters for obtaining the optimal values are the same as in Fig.~\ref{fig_lbcc_beta_y}.\label{fig_lbcc_alldata}}
\end{figure}

We have computed and made available (see Sec.~\ref{sec_results}) the optimal $\beta(\mu)$ and $y(\mu)$ values (based on one year running-average data) for use with the LBCC image transformation of Eq.~(\ref{eq_lbcc_beta_y}) from 12/10/2010 to 02/17/2014 at 6-hour cadence.  An example of the $\beta(\mu)$ and $y(\mu)$ values over $\mu$ is shown for 02/03/2011 in Fig.~\ref{fig_lbcc_beta_y}, while the computed values over the entire time period are displayed in Fig.~\ref{fig_lbcc_alldata}.
Each instrument results in similar values of $\beta(\mu)$ and $y(\mu)$, but exhibits variation over time.  These changes emphasize the need for a data-driven approach in order to obtain optimal corrections that respect changes in the images over time.  Since the LBCC factors are only formulated up to $r=R_{\odot}$, when applying the corrections to the EUV images, we extrapolate the curves to correct the pixels in the small region between $r=R_\odot$ and $r=R_0$.

\begin{figure}[tbp]
\centering
\includegraphics[width=0.32\textwidth]{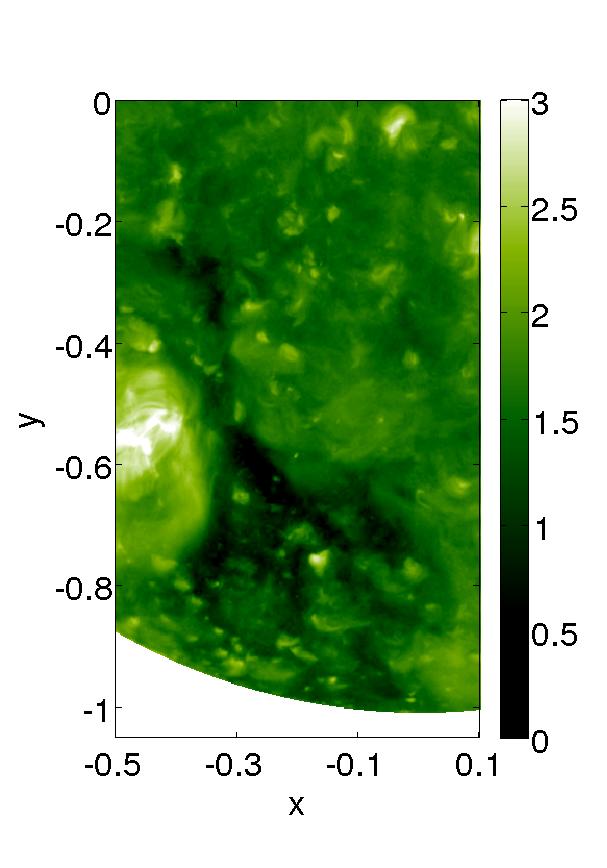}
\includegraphics[width=0.32\textwidth]{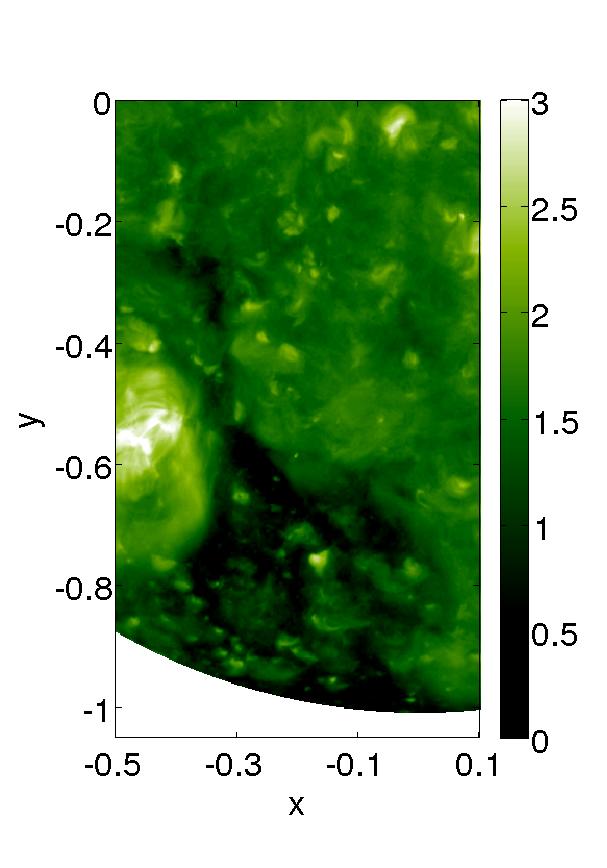}
\includegraphics[width=0.32\textwidth]{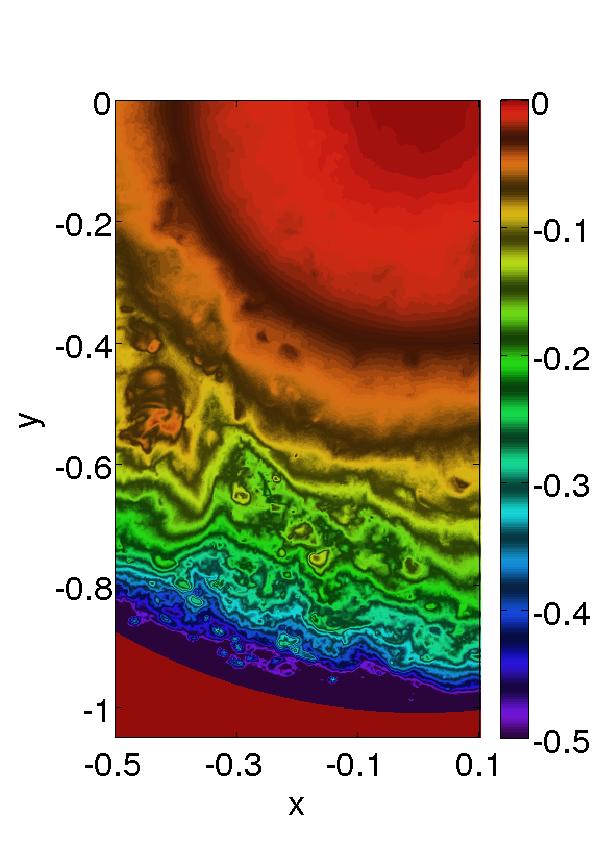}
\caption{Example of the limb-brightening correction of Eq.~(\ref{eq_lbcc_beta_y}) applied to the STB EUV image for 02/03/2011.  Left to right: Original EUV image, EUV image after applying the limb-brightening correction, and the intensity difference values for the correction.
\label{fig_lbcc_euv}}
\end{figure}

An illustration of the effect of applying the transformation of Eq.~(\ref{eq_lbcc_beta_y}) to an EUV image containing an extended coronal hole near the limb is shown in Fig.~\ref{fig_lbcc_euv}.  A color plot of the difference between the original and corrected images (i.e. the effective LBCCs) is also shown. 
We see that the coronal hole in the image has a more consistent intensity level from the limb towards disk center after the correction as desired.  Also, it is apparent that the correction is highly structure/intensity dependent, and that brighter structures are darkened less than darker structures, which increases the overall contrast (a desirable trait for coronal hole detection).  An advantage of this contrast enhancement over those of standard image processing techniques is that it is both unit-preserving and respects the physical nature of the limb-brightening problem.  

%That is, assuming a general correspondence of image intensity to temperature (see Appendix \ref{sec_lbcc_theory}), its functional form approximates the theoretical LBCCs of Sec.~\ref{sec_lbcc_theo}).

\begin{figure}[tbp]
\centering
\includegraphics[width=0.5\textwidth]{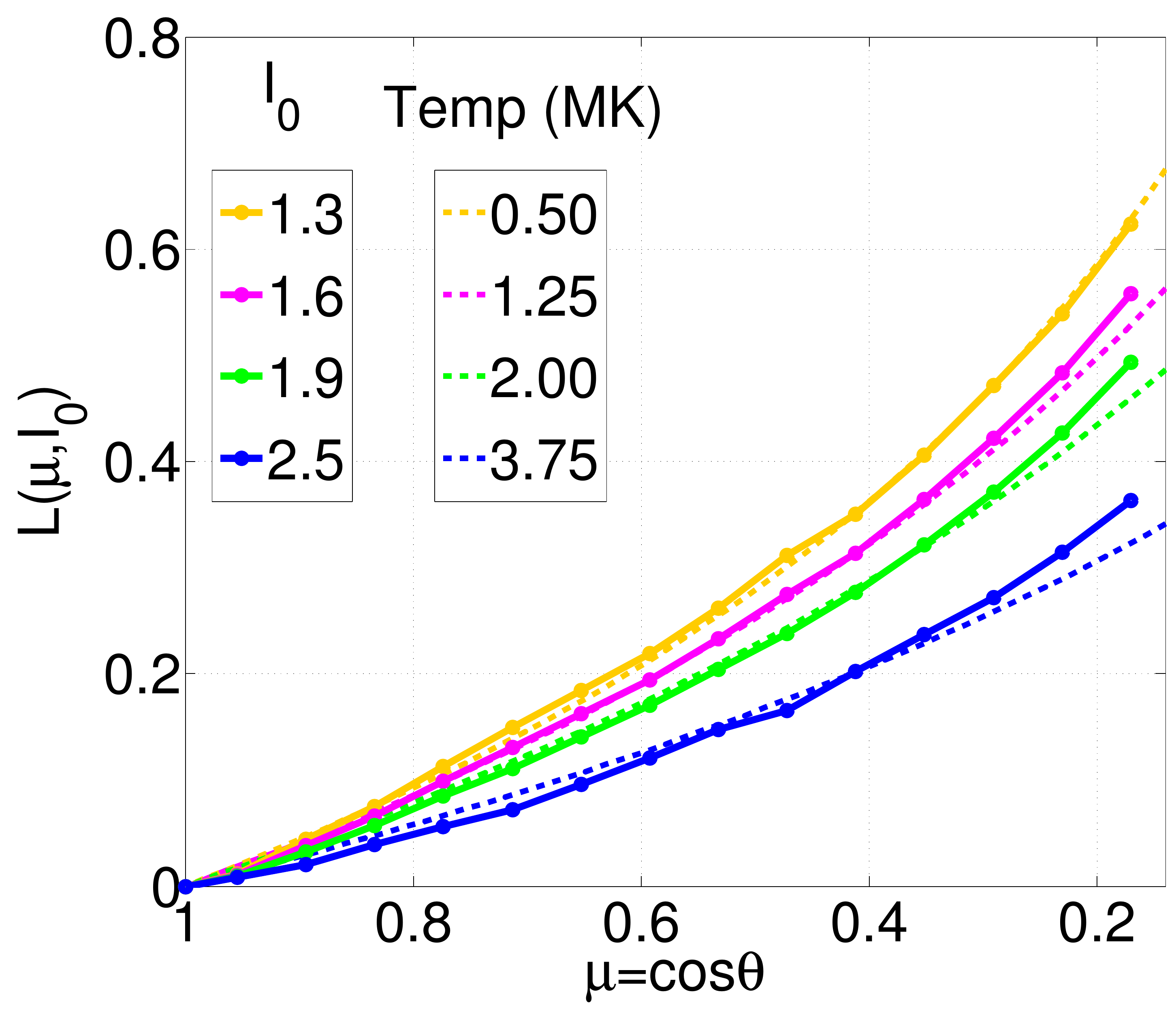}
\includegraphics[width=0.475\textwidth]{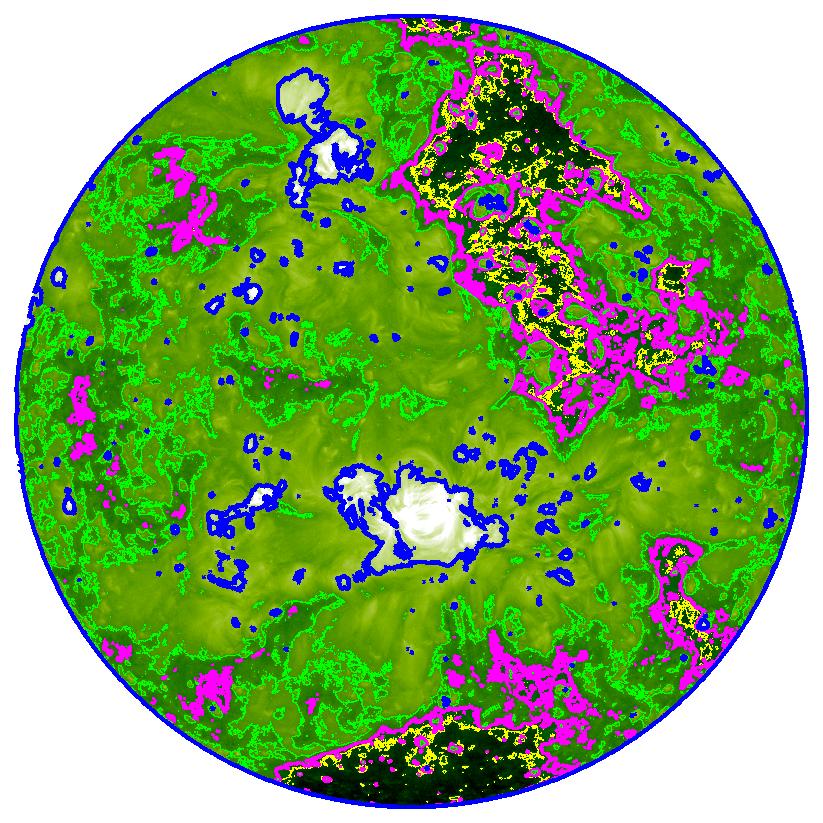}
\caption{Left:  LBCCs for 02/03/2011 for AIA computed using Eq.~(\ref{eq_lbcc_beta_y_L}) with the $\beta(\mu)$ and $y(\mu)$ values from Fig.~\ref{fig_lbcc_alldata} for central intensity values $I_0\in\{1.3,1.6,1.9,2.5\}$.  The same parameters as in Fig.~\ref{fig_lbcc_alldata} were used.  The theoretical LBCCs of Eq.~(\ref{eq_lbcdef1}) for $T\in\{0.5,1.25,2,3.75\}\mbox{MK}$ (top to bottom) are also shown.  Right:  The limb-brightening corrected AIA EUV image for the same date displaying contours at the selected values of $I_0$.\label{fig_lbcc_curves}}
\end{figure}

We now explore the ability of the $I_0$-dependent image transformation to represent LBCCs for various disk center intensities.  In Fig.~\ref{fig_lbcc_curves} we show four generated LBCCs for 02/03/2011 computed with Eq.~(\ref{eq_lbcc_beta_y_L}) from the values of $\beta(\mu)$ and $y(\mu)$ from Fig.~\ref{fig_lbcc_beta_y}.  The curves are generated using chosen $I_0$ values such that they match closely to theoretical LBCCs of selected temperatures.
We see that the LBCCs for increasing values of $I_0$ are smoothly varied in the direction of the theoretical LBCCs computed with Eq.~(\ref{eq_lbcdef1}) for increasing temperature, and that they closely match the shape and range of the curves.  This qualitatively fits with the expectation that structures that vary in EUV intensity levels (low to high), such as coronal holes, quiet-sun, and active regions, are similarly ordered in temperature (low to high).  This idea is shown visually in the right panel of Fig.~\ref{fig_lbcc_curves} where a corrected AIA EUV image is displayed with contours of the chosen $I_0$ values. The general correlation of different temperature curves to different $I_0$ values suggests that one could possibly employ this technique to serve as an indirect statistical estimate of average coronal temperatures based on images at a single wavelength. We plan to explore this concept in the future.

%%%%%%%%%%%%%%%%%%%%%%%%%%%%%%%%%%%%%%%%%%%%%%%%%%%%%%%%%%%%%%%%%%%%%%%%%%%%%%%%%%%%%%%%%%%%%%%%%%%%%%%%%%%%%%%%%%%%%

\subsection{Inter-instrument intensity transformations}
\label{sec_iit}
After completing the steps described in sections \ref{sec_psf}{--}\ref{sec_lbcc}, the preprocessed EUV images are now suitable for performing coronal hole detection. However, in order to make synchronic maps using multiple instruments, it is highly desirable to derive inter-instrument transformations (IITs) that allow us to equate intensities from one instrument to another. In addition to being a useful data product on its own, we find this helpful for two reasons: First, a proper scaling will allow one to employ the same thresholding parameters in the CH detection algorithm, basically eliminating the need to find the separate ``best'' values for all three instruments. This also helps ensure that the algorithm will detect a CH with the same tolerance in each image. Second, scaling the images to each other allows one to create continuous synchronic EUV intensity maps from the three instruments, which helps for tracking time-dependent features as they rotate or propagate from one viewpoint to another. This is very useful when judging the quality of the CH detection, as well as tracking CH evolution across instruments (see Sec.~\ref{sec_results}).

\begin{figure}[tbp]
\centering
\subfigure[ ]
{
\begin{overpic}[width=0.45\textwidth]{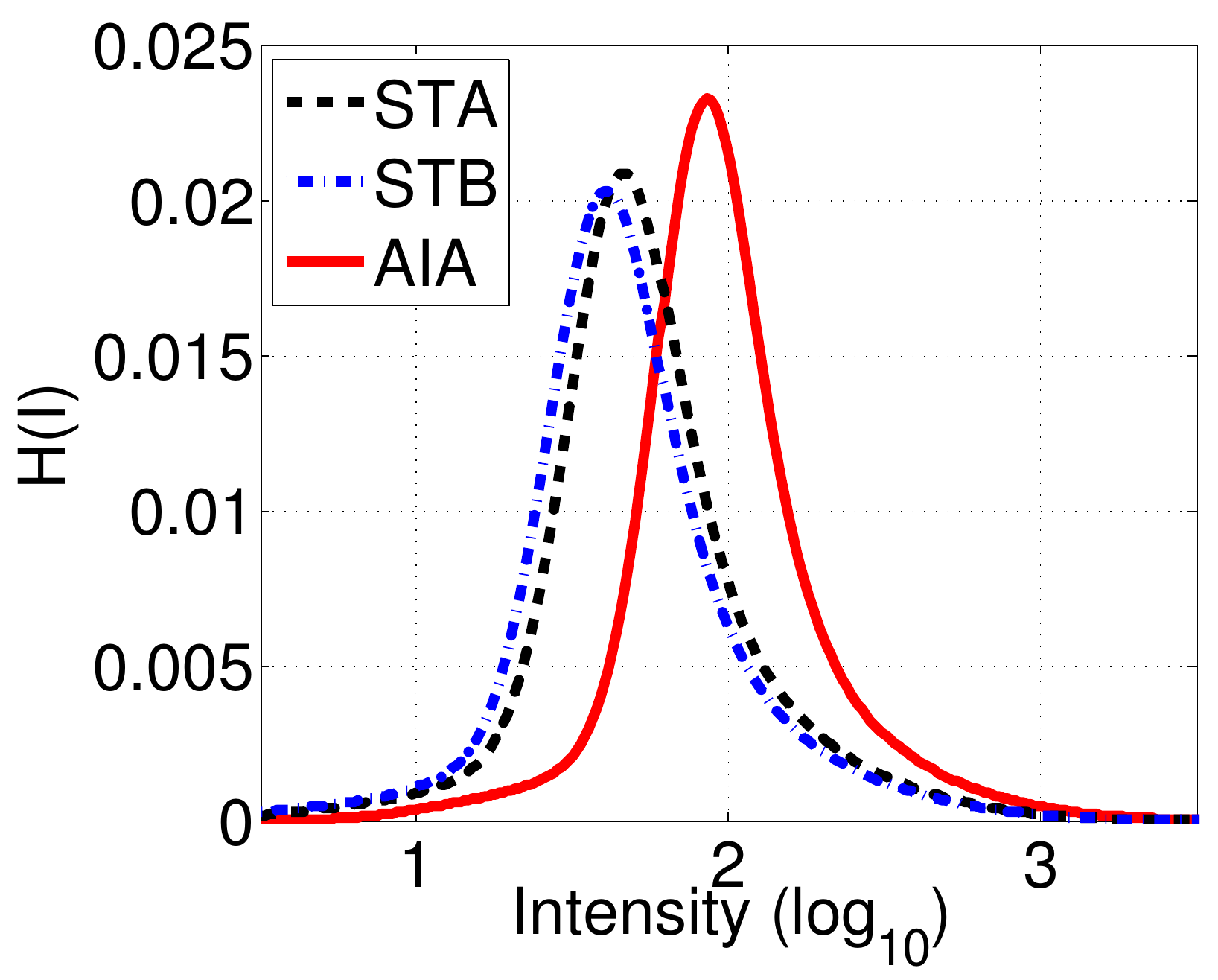}
     \put(72,60){\includegraphics[scale=0.2]{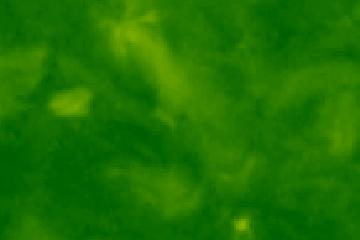}}
     \put(72,45){\includegraphics[scale=0.2]{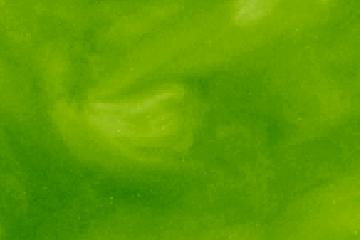}}
     \put(72,30){\includegraphics[scale=0.2]{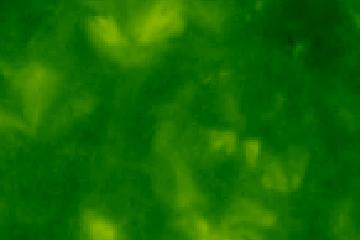}}
\end{overpic}
}
\subfigure[ ]
{
\begin{overpic}[width=0.45\textwidth]{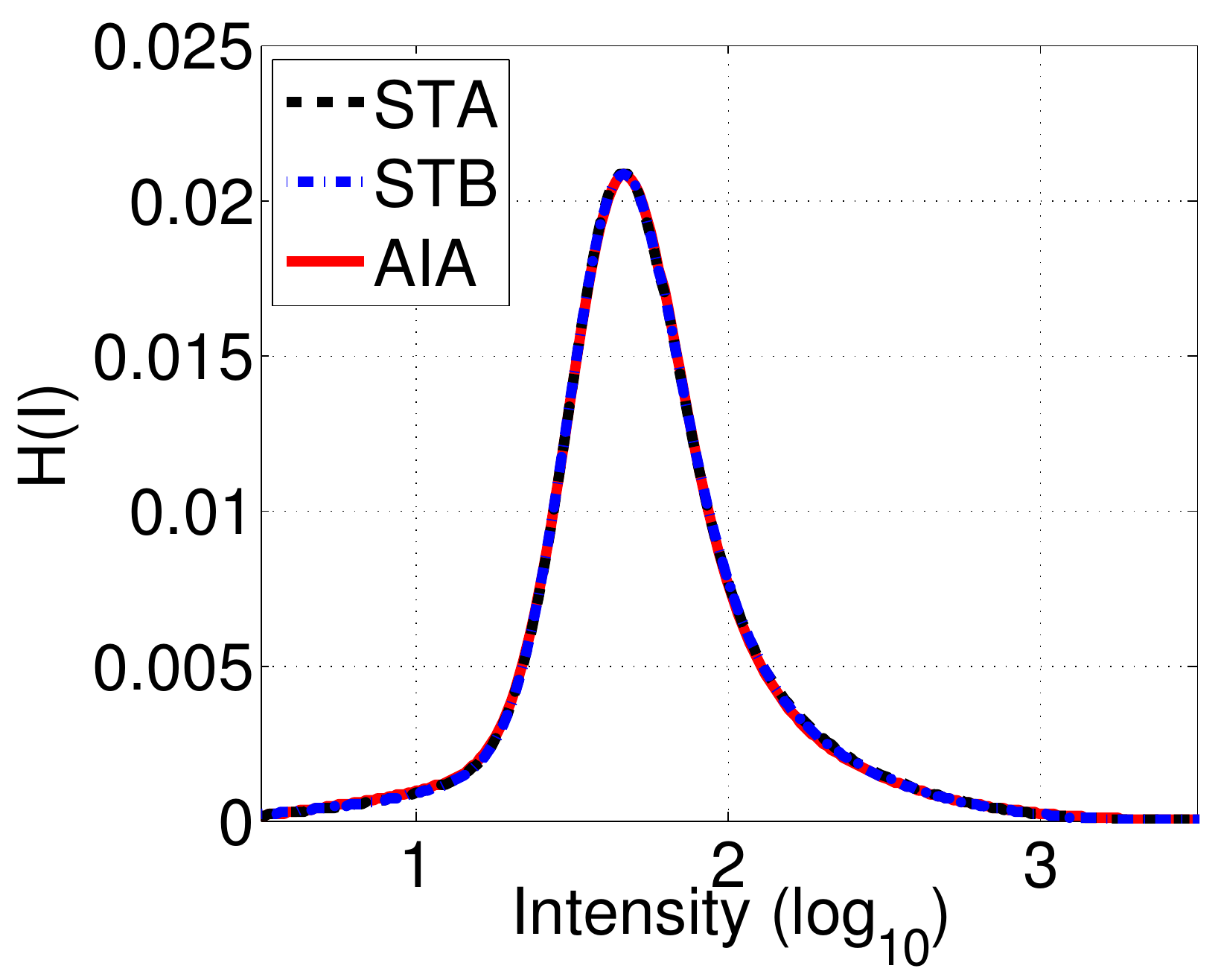}
     \put(72,60){\includegraphics[scale=0.2]{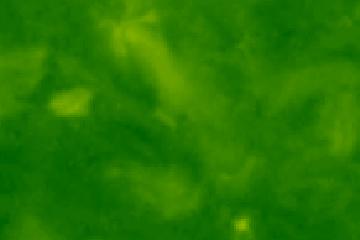}}
     \put(72,45){\includegraphics[scale=0.2]{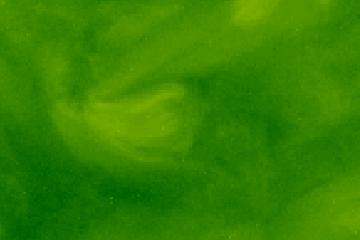}}
     \put(72,30){\includegraphics[scale=0.2]{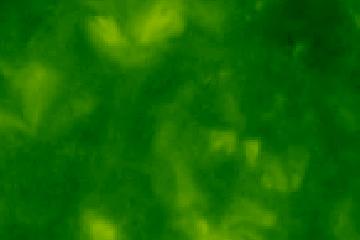}}
\end{overpic}
}
\caption{Example of the normalized histograms of one-year 6-hour cadence averaged data centered about 02/03/2011 for STA, STB, and AIA before (a) and after (b) the correction of Eq.~(\ref{eq_scale_alpha_x}).  The histograms use $400$ intensity bins, a latitude limit of $\pm 75.3^{\circ}$ from center and an optimization relative tolerance of $10^{-4}$.  The inset shows selected quiet sun portions of each instrument's image at the central date before and after applying the correction.\label{fig_iit_hist}}
\end{figure}

Transforming one instrument's EUV image so that it appears as if another instrument has viewed it is very similar to the task of the limb-brightening correction curve calculations in Sec.~\ref{sec_lbcc} performed on each $\mu$ section.  Therefore, the same techniques can be applied and we once again take a data-driven approach.  In this case, there is no a priori reason to assume a nonlinear transformation like Eq.~(\ref{eq_lbcc_beta_y}) is necessary and perhaps a linear histogram shift (or average intensity comparison) is all that is needed.  However, we have found that, similar to the LBCCs, the shapes of the histograms from different instruments do not match reliably and consequently, the results based on linear histogram shifts are not ideal.  We find once again that a transformation of the form
\begin{equation}
\label{eq_scale_alpha_x}
I_{a} = \alpha\,I_{b} + x,
\end{equation}
where the subscripts $a$ and $b$ represent the log10 images of instrument $a$ and instrument $b$, provides a better match between the histograms of the multiple instruments (see Fig.~\ref{fig_iit_hist}).
  
When using linear histogram shifts or average intensity differences, the order of operations of LBCC and inter-instrument scaling is arbitrary and each can be applied separately.  However, due to the combination of the nonlinear nature of the transformations we choose to use in each step, the LBCC and inter-instrument scaling cannot be applied independently.  Therefore, one must apply one correction to the EUV images, and then use the corrected data to compute the second correction factors.  We choose to first apply the limb-brightening corrections computed in Sec.~\ref{sec_lbcc} to the set of EUV images, and then use the resulting images to compute the proper $\alpha$ and $x$ correction factors for the inter-instrument corrections.

In order to ensure that the spacecraft are observing more or less the same portions of the sun during a given averaging window, we shift our averaging window in time corresponding to the separation in Carrington longitudes of the spacecraft's position.  Additionally, we limit the latitude of the disk data based on the $B_0$ angles of each instrument such that each instrument sees the same area.  For our data set, the limit needed is $\pm 75.3^{\circ}$ from center.

Fig.~\ref{fig_iit_hist} shows an example of the histograms of one-year averaged data centered about 02/03/2011 for STA, STB, and AIA both before and after the correction of Eq.~(\ref{eq_scale_alpha_x}) is applied.  Also shown are selected quiet sun portions of each image at the same date.
We see that the correction factors match the averaged histograms nearly exactly, and when applied to the EUV image, cause the three instruments' image intensities, on average, to match very closely.
 
\begin{figure}[tbp]
\centering
\includegraphics[width=0.75\textwidth]{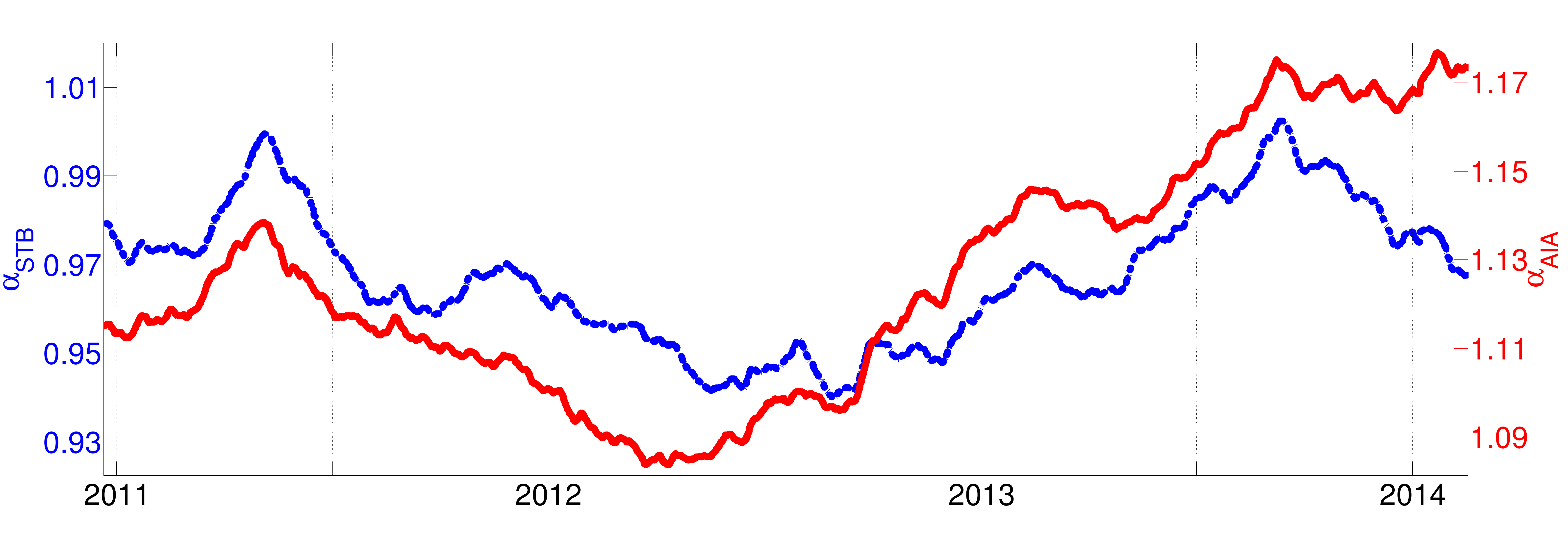}
\\
\includegraphics[width=0.75\textwidth]{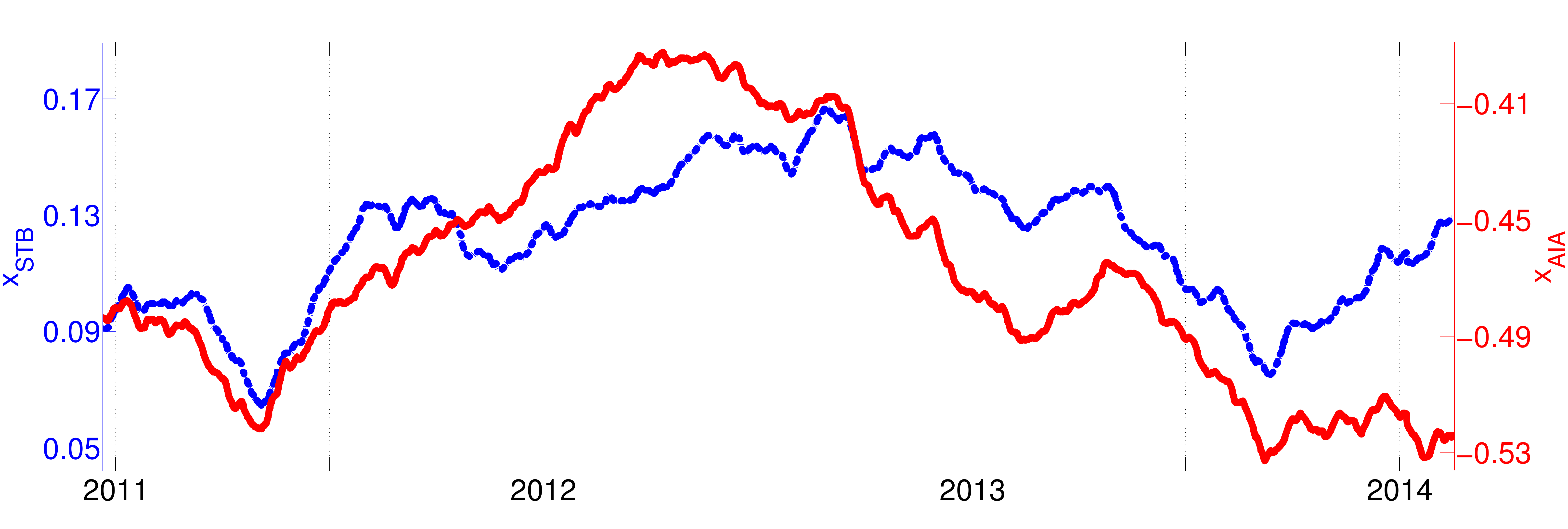}
\caption{Double-$y$ axis plots of the computed inter-instrument transformation factors $\alpha$ and $x$ of Eq.~(\ref{eq_scale_alpha_x}) for transforming STB images to STA (blue dashed line) and AIA images to STA (red solid line) from 12/21/2010 to 02/17/2014.  The computational parameters are the same as in Fig.~\ref{fig_iit_hist}.\label{fig_iit}}
\end{figure}

We have computed the IIT factors for transforming STB to STA and AIA to STA from 12/21/2010 to 02/17/2014 at a 6-hour cadence using a one-year averaging window.  The factors are computed using the EUV images after the application of the limb-brightening correction of Eq.~(\ref{eq_lbcc_beta_y}) with the computed LBCC factors from Fig.~\ref{fig_lbcc_alldata}.  The results are shown in Fig.~\ref{fig_iit}.
We see that, like the LBCC factors in Fig.~\ref{fig_lbcc_alldata}, the inter-instrument transformation factors vary over time.  Although the changes are small, they are within the sensitivity range of the threshold values used in our CH detection algorithm, re-enforcing the need for a data-driven approach.  

\subsection{Validation of preprocessing results}
\label{sec_preproc_valid}
The overarching goal of our preprocessing is to transform multi-instrument EUV intensity measurements to a common vantage point and unit scaling as best we can (Secs.~\ref{sec_lbcc} and \ref{sec_iit}). This is not perfectly achievable because of inherent geometric and observational limitations such as projection effects, time synchronization differences, obscuration, variable apparent coronal radius, etc.  However, we can still characterize the performance of our preprocessing by examining overlapping observations of the spacecraft.

\begin{figure}[tbp]
\centering
\includegraphics[width=4in]{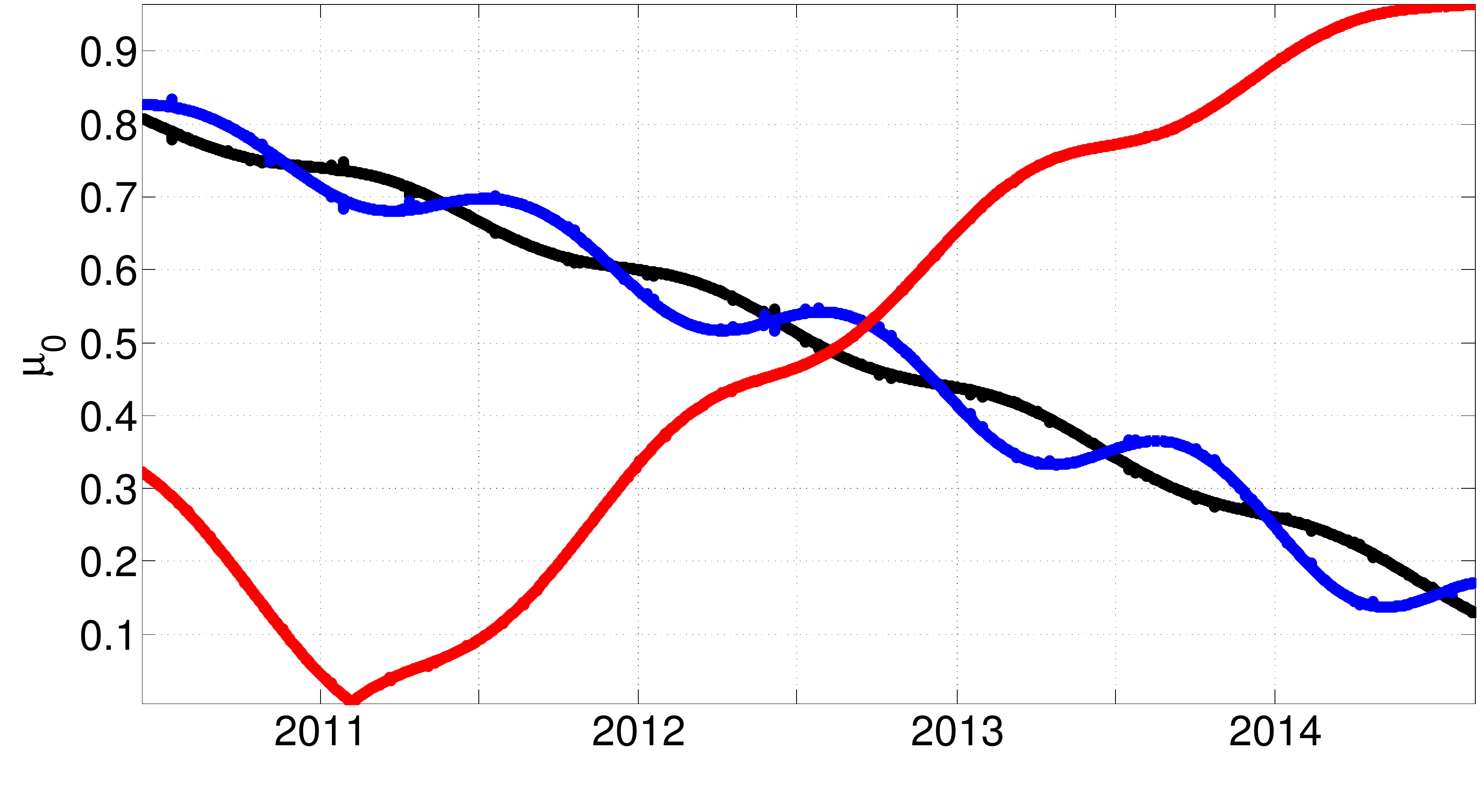}
\caption{The midpoint $\mu$ value of the overlap region, $\mu_0$, as a function of time for each instrument pair. The pairs are: STA--AIA (black), STB--AIA (blue), and STA--STB (red).
\label{fig_valid_mu0}}
\end{figure}

We start by taking EUV images from each instrument and mapping them to a common heliographic grid (Sec.~\ref{sec_mapping}). Then we determine the midpoint $\mu$ value of the overlap region between two instruments, $\mu_0$, and select regions for inter-comparison by extracting overlapping data within a small strip defined by $\mu_0\pm\Delta\mu$.  Choosing a small $\Delta\mu$ restricts the variation from the mean line-of-sight angle and helps limit projection effects (we choose $\Delta\mu=0.05$). The value of $\mu_0$ changes naturally over time due to the variation of the orbital position and orientation of the spacecraft (see Fig.~\ref{fig_valid_mu0}).  These variations affect our analysis by changing the mean line-of-sight angle in the overlap.

Denoting the mapped intensity data in each $\Delta \mu$-limited overlap region as $J$, we first explore how average intensity values change before and after our final IIT preprocessing step. Defining the percent difference of the mean intensity as: 
\begin{equation}
D_{\mbox{\tiny pdm}}=100\times\frac{\overline{J_a}-\overline{J_b}}{\overline{J_b}},\qquad \overline{J}=\frac{1}{N}\sum_i^N J(i),
\end{equation}
we compute $D_{\mbox{\tiny pdm}}$ for every map and overlap pair in our dataset. The average of $D_{\mbox{\tiny pdm}}$ before and after applying the IIT for all pairs with $\mu_0>0.4$ is shown in Table~\ref{table_valid}. The corresponding standard deviations for these data are also shown.
%\begin{table}[tbp] 
%\centering 
%\begin{tabular}{lrr} 
%\multicolumn{3}{c}{{\bf $D_{\mbox{\tiny pdm}}$ of $J$}}\\
%\hline
%\rowcolor[HTML]{FFFFBB} {\bf Before IIT} & Mean & $\sigma_{\mbox{\tiny std}}$ \\
%\hline
%\rowcolor[HTML]{DDDDDD} AIA--STA & $-10.81\%$ & $5.25\%$ \\
%\rowcolor[HTML]{DDDDFF} AIA--STB & $-13.80\%$ & $5.60\%$ \\
%\rowcolor[HTML]{FFDDDD} STA--STB & $4.03\%$ & $6.13\%$ \\
%\hline
%\hline
%\rowcolor[HTML]{FFFFBB} {\bf After IIT} & Mean & $\sigma_{\mbox{\tiny std}}$ \\
%\hline
%\rowcolor[HTML]{DDDDDD} AIA--STA & $0.05\%$ & $5.87\%$ \\
%\rowcolor[HTML]{DDDDFF} AIA--STB & $-0.31\%$ & $5.84\%$ \\
%\rowcolor[HTML]{FFDDDD} STA--STB & $0.67\%$ & $5.59\%$ \\
%\hline
%\end{tabular}
%\qquad
%\begin{tabular}{lrr} 
%\multicolumn{3}{c}{{\bf NRMSD of $J$}}\\
%\hline
%\rowcolor[HTML]{FFFFBB} {\bf Before IIT} & Mean &  $\sigma_{\mbox{\tiny std}}$ \\
%\hline
%\rowcolor[HTML]{DDDDDD} AIA--STA & $0.15$ & $0.042$ \\
%\rowcolor[HTML]{DDDDFF} AIA--STB & $0.18$ & $0.043$ \\
%\rowcolor[HTML]{FFDDDD} STA--STB & $0.11$ & $0.042$ \\
%\hline
%\hline
%\rowcolor[HTML]{FFFFBB} {\bf After IIT} & Mean & $\sigma_{\mbox{\tiny std}}$ \\
%\hline
%\rowcolor[HTML]{DDDDDD} AIA--STA & $0.11$ & $0.039$ \\
%\rowcolor[HTML]{DDDDFF} AIA--STB & $0.11$ & $0.037$ \\
%\rowcolor[HTML]{FFDDDD} STA--STB & $0.10$ & $0.040$ \\
%\hline
%\end{tabular}
%\caption{The mean and standard deviation ($\sigma_{\mbox{\tiny std}}$) of $D_{\mbox{\tiny pdm}}$ and NRMSD computed in the $J$ overlap regions for each instrument pair with $\mu_0 > 0.4$ over the entire data set before and after applying the IIT. \label{table_valid}}
%\end{table}
\begin{table}[tbp] 
\centering 
\begin{tabular}{lrr} 
\multicolumn{3}{c}{{\bf $D_{\mbox{\tiny pdm}}$ of $J$}}\\
\hline
\hline
{\bf Before IIT} & Mean & $\sigma_{\mbox{\tiny std}}$ \\
\hline
AIA--STA & $-10.81\%$ & $5.25\%$ \\
AIA--STB & $-13.80\%$ & $5.60\%$ \\
STA--STB & $4.03\%$ & $6.13\%$ \\
\hline
\hline
{\bf After IIT} & Mean & $\sigma_{\mbox{\tiny std}}$ \\
\hline
AIA--STA & $0.05\%$ & $5.87\%$ \\
AIA--STB & $-0.31\%$ & $5.84\%$ \\
STA--STB & $0.67\%$ & $5.59\%$ \\
\hline
\end{tabular}
\qquad
\begin{tabular}{lrr} 
\multicolumn{3}{c}{{\bf NRMSD of $J$}}\\
\hline
\hline
{\bf Before IIT} & Mean &  $\sigma_{\mbox{\tiny std}}$ \\
\hline
AIA--STA & $0.15$ & $0.042$ \\
AIA--STB & $0.18$ & $0.043$ \\
STA--STB & $0.11$ & $0.042$ \\
\hline
\hline
{\bf After IIT} & Mean & $\sigma_{\mbox{\tiny std}}$ \\
\hline
AIA--STA & $0.11$ & $0.039$ \\
AIA--STB & $0.11$ & $0.037$ \\
STA--STB & $0.10$ & $0.040$ \\
\hline
\end{tabular}
\caption{The mean and standard deviation ($\sigma_{\mbox{\tiny std}}$) of $D_{\mbox{\tiny pdm}}$ and NRMSD computed in the $J$ overlap regions for each instrument pair with $\mu_0 > 0.4$ over the entire data set before and after applying the IIT. \label{table_valid}}
\end{table}

We find that the IIT dramatically improves the average value of $D_{\mbox{\tiny pdm}}$, reducing the percent difference from over $10\%$ to less than $1\%$ for the STA--AIA and STB--AIA pairs.  For STA--STB this difference drops from $4\%$ to less than $1\%$. This is not surprising because the mean is mostly sensitive to the relative scaling between the instruments, but it does confirm that the IIT is serving its purpose of normalizing intensities between the spacecraft. The standard deviations of $D_{\mbox{\tiny pdm}}$ do not change substantially but remain relatively low within $\pm6\%$. This indicates that a baseline level of variation is present in the data. The presence of variations is reasonable because the corona is an extended optically thin medium, and each spacecraft will naturally have an oppositely directed perspective through it in these overlap regions.

\begin{figure}[htbp]
\centering
$\begin{array}{cc}
\mbox{Before IIT} & \mbox{After IIT}
\\
\includegraphics[width=3.25in]{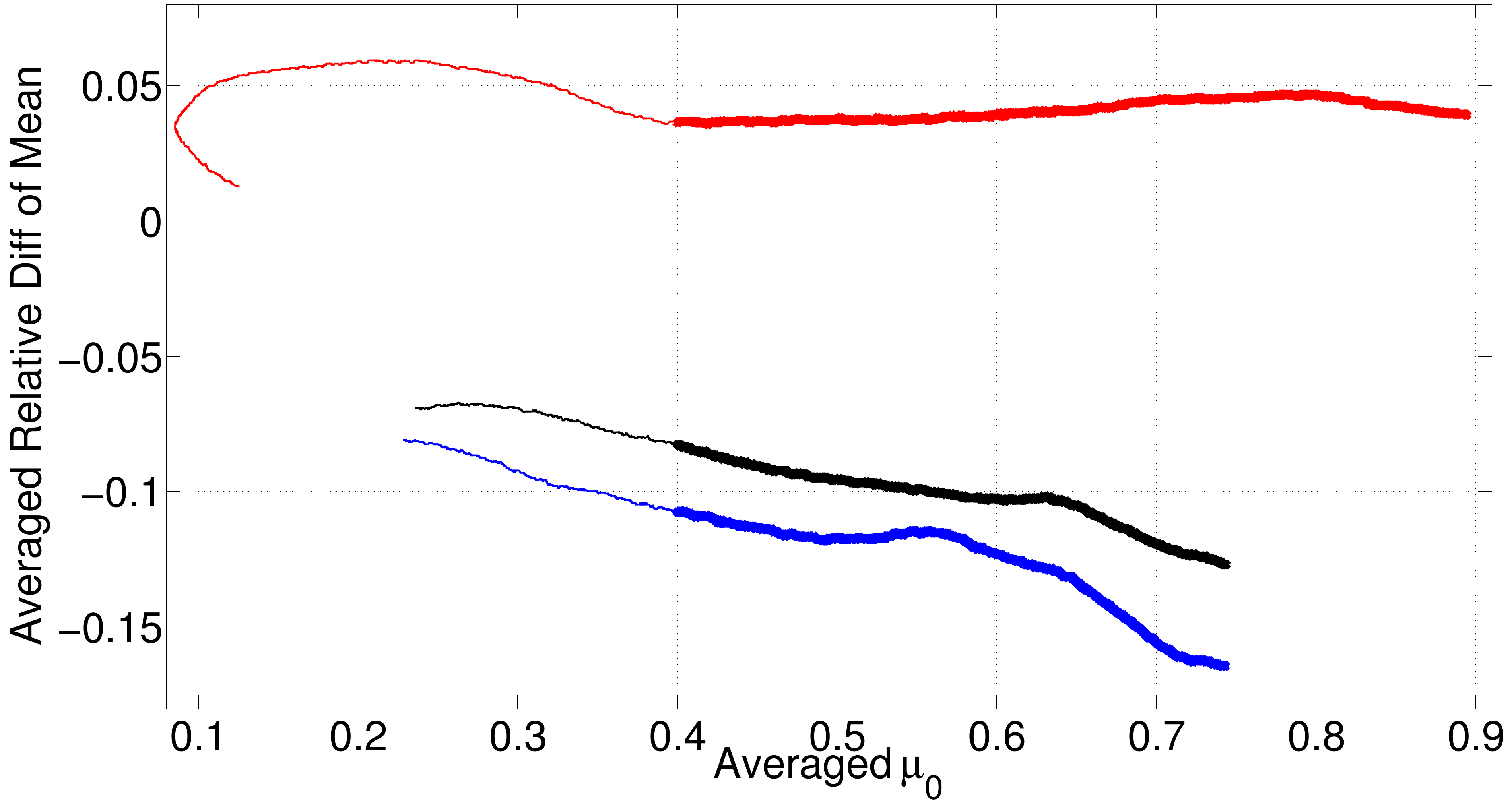} &
\includegraphics[width=3.25in]{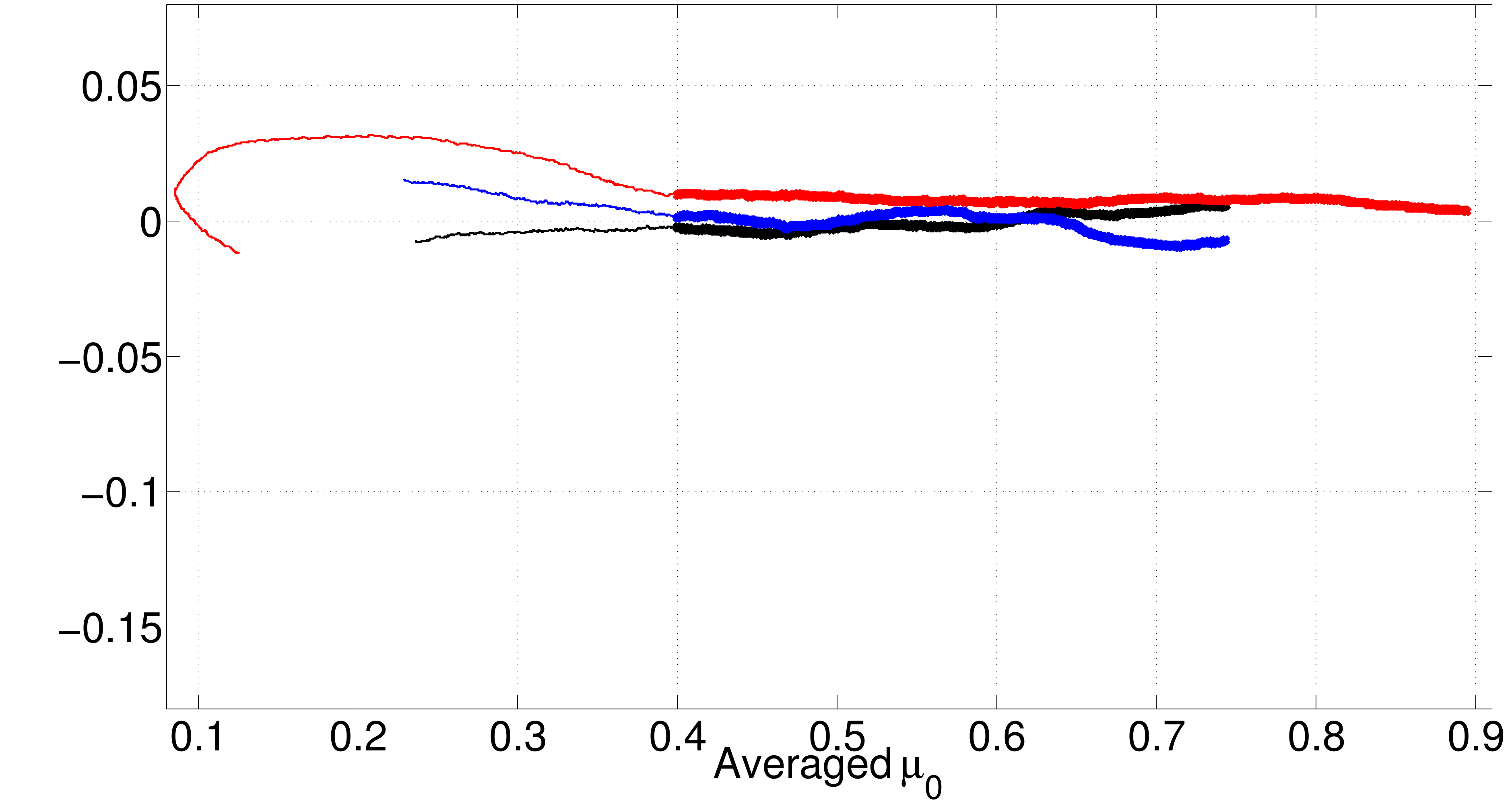}
\\
\includegraphics[width=3.25in]{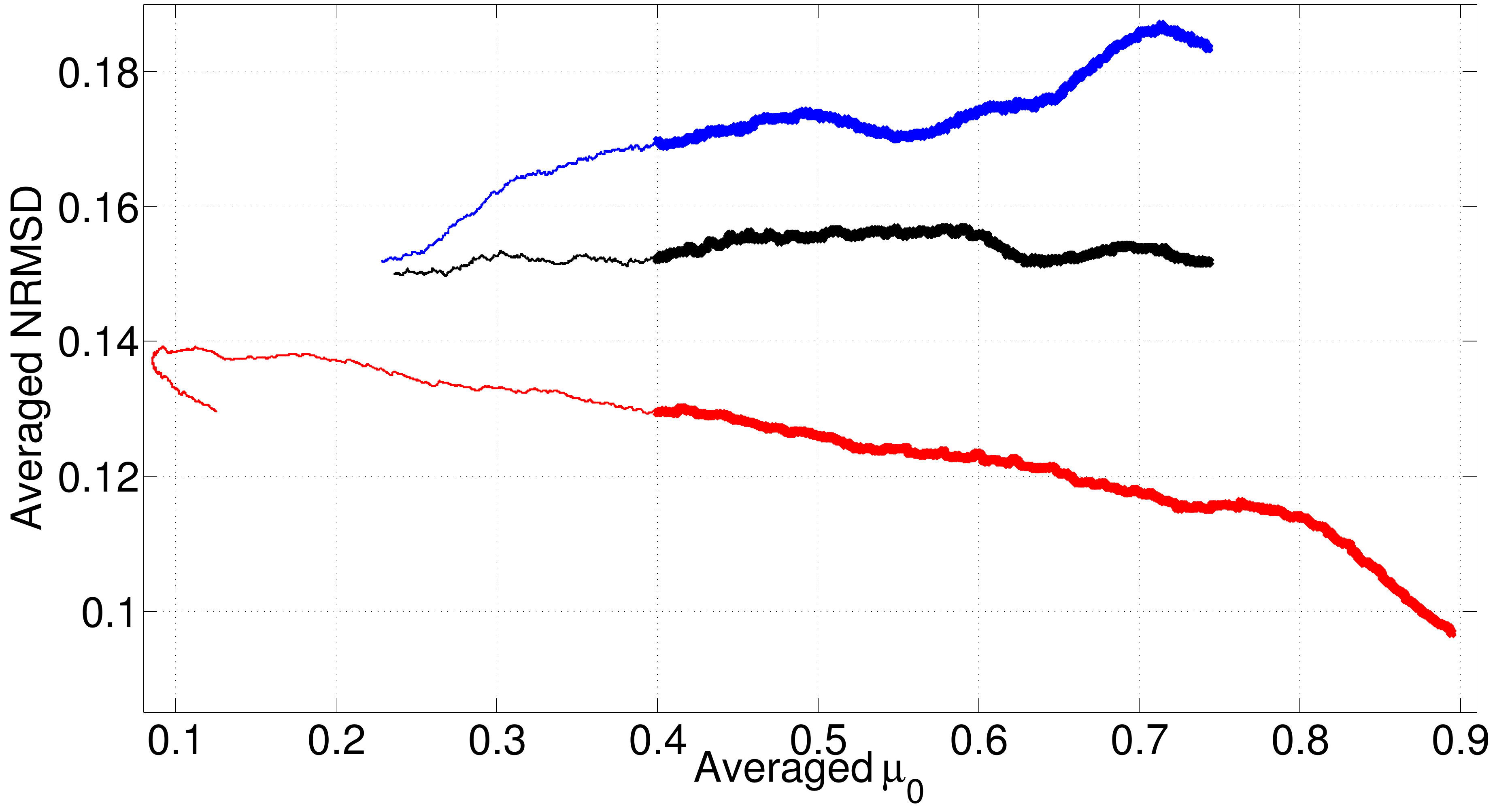} & 
\includegraphics[width=3.25in]{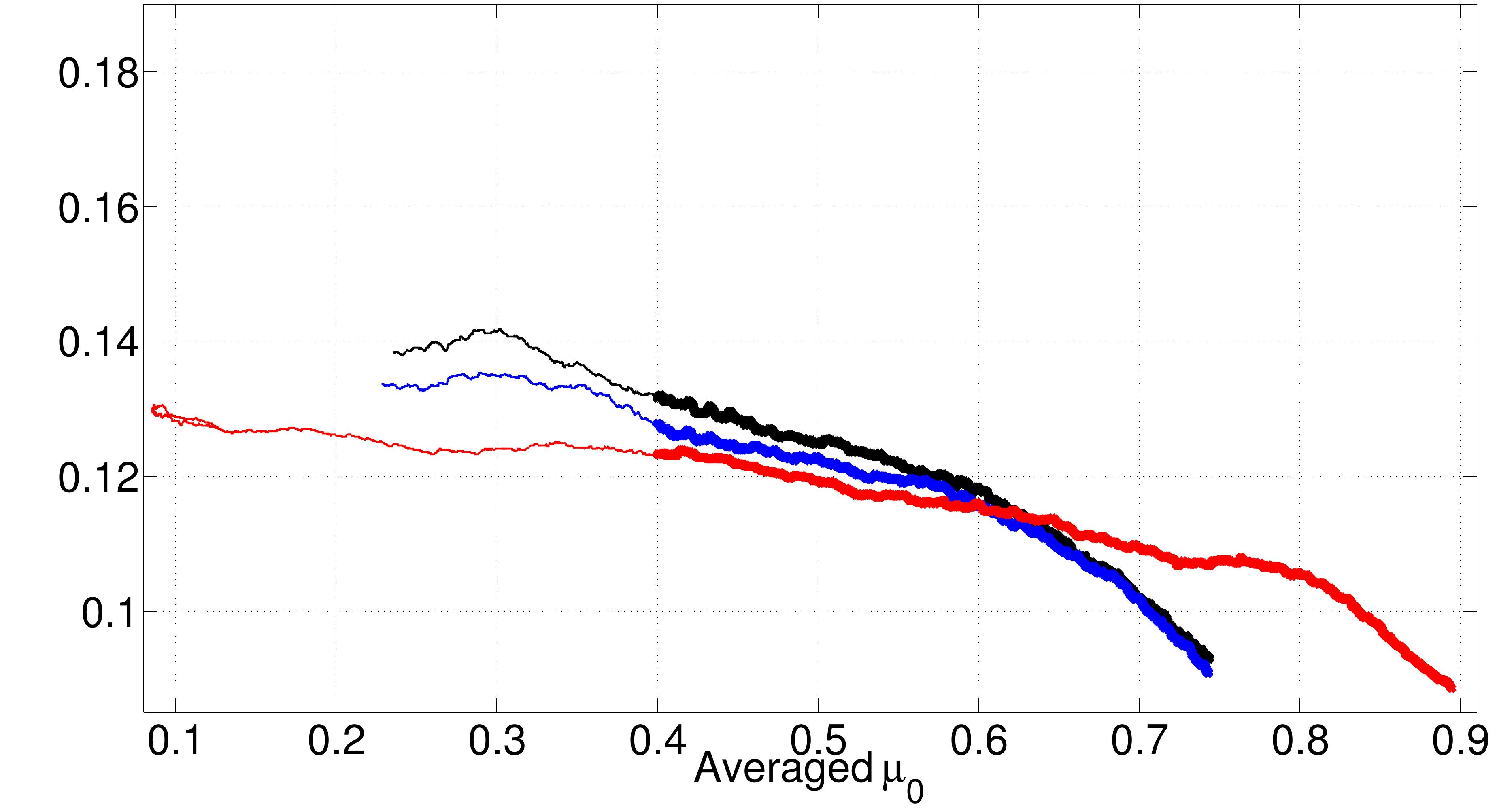}
\end{array}$
\caption{One-year running averages of $D_{\mbox{\tiny pdm}}$ and NRMSD for each spacecraft pair before and after IIT has been applied as a function of $\mu_0$ (refer also Fig.~\ref{fig_valid_mu0}).  The pairs are colored as follows: STA--AIA (black), STA--STB (blue), and STA--STB (red).  The dashed lines represent data for $\mu_0<0.4$ (poor viewing geometry).
\label{fig_valid_results}}
\end{figure}

We also compute one-year running averages of $D_{\mbox{\tiny pdm}}$ over the entire data set and see a similar story. These curves are shown for each instrument pair in the top row of Fig.~\ref{fig_valid_results}, with the dashed lines indicating where $\mu_0<0.4$ (poor viewing). 
Before applying the IIT, the mean curves all have their own non-zero offset and these offsets are removed afterwards. Even more importantly, we see a non-zero slope in some of the curves before applying the IIT (particularly relative to AIA) and find that these slopes are substantially reduced afterwards. This indicates that the time-dependent calculation of the IIT is warranted and that the transformation is working properly.

Next we examine the normalized root-mean-square-difference (NRMSD) between all pixels inside an overlap region:
\begin{equation}
\mbox{NRMSD}=\frac{1}{J_{\mbox{\tiny max}}-J_{\mbox{\tiny min}}}\sqrt{\frac{1}{N}\sum_i^N (J_a(i)-J_b(i))^2}.
\end{equation} 
The NRMSD can be thought of as a general quality indicator of the of the pixel-to-pixel match between instruments. For consistent normalization, we choose $J_{\mbox{\tiny max}}-J_{\mbox{\tiny min}}=4$ to be the approximate total intensity range for all images.

Intrinsic differences due to opposed viewing projections through the corona are expected to contribute significantly to the NRMSD. This means that even with perfect limb brightening and intensity corrections, we do not expect the NRMSD to trend to zero. Despite this fact, the NRMSD can be a useful indicator of preprocessing quality by comparing the values of the three overlap pairs to each other (STA--STB, STA--AIA, STB--AIA), mainly because we should expect consistent results between pairs for a common $\mu_0$.

We give the complete time average and standard deviation of the NRMSD before and after applying the IIT in Table~\ref{table_valid} (again for $\mu_0>0.4$), and show the corresponding one-year running averages in the bottom row of Fig.~\ref{fig_valid_results}. As with $D_{\mbox{\tiny pdm}}$, the average NRMSD values are reduced with the application of the IIT. Most importantly we see that the mean of the NRMSD for all three instrument pairs becomes much more consistent with one another, suggesting that their variations also become similarly distributed. 

This consistency can be explored further by looking at the one-year averages of NRMSD vs. $\mu_0$. Before applying the IIT, the curves do not share a common value or slope, but afterwards they converge to one another, especially between $\mu_0=0.4\mbox{--}0.6$. A slope is expected even after applying the IIT in this case because the change in mean projection angle with $\mu_0$ will influence the pixel-to-pixel variation even if the overall intensity scaling is identical. Unsurprisingly, the NRMSD is worse with a poor viewing angle and improves as $\mu_0$ increases. Lastly, the difference between the STA--STB pair (red) and the STA/B--AIA pairs (blue/black) around $\mu_0=0.7\mbox{--}1.0$ is not unreasonable considering that these favorable elongations occur at different times, once near solar maximum for STA--STB (2013 onwards) and once near solar minimum for the pairs with AIA (2010--2011).

Despite the systematic limitations of the problem, we have shown with these diagnostics that our preprocessing pipeline produces a common intensity scaling for all spacecraft pairs, and that the relative scaling and measurement variance is small and consistent between them (at least in a statistical or average sense).  With the preprocessing of the images completed, we can now move on to the next step in our data pipeline.

%%%%%%%%%%%%%%%%%%%%%%%%%%%%%%%%%%%%%%%%%%%%%%%%%%%%%%%%%%%%%%%%%%%%%%%%%%%%%%%%%%%%%%%%%%%%%%%%%%%%%%%%%%%%%%%%%%%%%

\section{Coronal hole map generation}
\label{sec_alg}
After the preprocessing steps of Sec.~\ref{sec_preproc} are completed, the images are ready to be used with a coronal hole detection algorithm.  Although there are many segmentation algorithms available in the image processing community (as discussed in Sec.~\ref{sec_intro}), they are often not suited for CH detection, or are difficult to obtain and/or use.  We therefore have written our own segmentation algorithm modular code called {\tt ezseg} that utilizes a dual-threshold region-growing style algorithm with variable connectivity written in FORTRAN-OpenMP and C-CUDA.  We also provide MATLAB interfaces to {\tt ezseg} for fast and easy use.  The resulting maps for each instrument are then merged to form the synchronic CH maps using MATLAB scripts.

%%%%%%%%%%%%%%%%%%%%%%%%%%%%%%%%%%%%%%%%%%%%%%%%%%%%%%%%%%%%%%%%%%%%%%%%%%%%%%%%%%%%%%%%%%%%%%%%%%%%%%%%%%%%%%%%%%%%%

\subsection{Coronal hole detection algorithm: {\tt ezseg}}
\label{sec_ezseg}
As discussed in Sec.~\ref{sec_intro}, the simplest technique for image segmentation is intensity thresholding.  Due to the transient nature of structures on the sun in both intensity levels and abundances, automatic threshold selection algorithms such as the popular ``graythresh'' \citep{1979_Otsu_Grayscale_auto-thresh} algorithm are not useful in this context.  Additionally, simple thresholding for CH detection itself is problematic due to the overlap of the tails of the intensity distributions of CHs with other features such as dark filament channels and dark regions of the quiet sun.  An example of this is shown in \citet{2009_Kirk_PCHD_good-pole-thresh}, where thresholding is used to capture polar CHs, and in the process, detects much of the quiet sun (in that study the authors were only concerned with polar holes and used a unique tracing algorithm so quiet sun detection was not a problem).  Some have relied on using magnetic field data to exclude such features \citep{2014_Lowder_CHD_AIA_EUVI}, but it is important that our technique not rely on such data because photospheric magnetic field observations are not made by the STEREO spacecraft, and a key requirement is that our technique takes advantage of the instantaneous (near) full-sun EUV coverage provided by these spacecraft. We have found that the combination of the preprocessing steps of Sec.~\ref{sec_preproc} with a region growing style algorithm (described below) is typically sufficient to distinguish CHs from other structures without the need of magnetic data.

Our algorithm is a simplified region growing algorithm using one threshold ($t_1$) for seed placement and a second value ($t_2$) for the stopping criteria.  This is similar in spirit to the popular ``flood-fill'' or ``magic-wand'' tools that are common in image processing software. The use of region growing avoids false detection of disconnected dark areas of quiet sun by only growing regions that are seeded with intensities corresponding to the darkest regions of coronal holes.   

%A previously developed solar image segmentation software called SWAMIS \citep{2007_DeForest_SWAMIS} uses a similar algorithm and is currently used to detect structures in magnetic field data for the HEK database.  The code can in principle also be used for CH detection in EUV images as well, and it is freely available for download and use.  However, due to its automatic per-image method for choosing threshold values, it may not be well suited for time-dependent synchronic CH evolution studies.  Additionally, its setup and use can be somewhat involved, as it is more of a collection of tools, not a ready-to-use code.  We therefore produce our own code that is tailored for fast, simple CH detection.

In our simplified implementation of region growing, we do not track/identify individual regions that then have to be merged, rather we only use local information to grow and stop growing the region.  Therefore, the region average and detection history do not need to be stored in specialized data structures, and there is no order dependence allowing the algorithm to be formulated in an embarrassingly parallel manner.  Although for some cases, the form of the algorithm may be less efficient than standard methods, the parallelism allows us to implement the algorithm on both shared-memory processors and GPUs, allowing it to perform well for our needs.  For example, a CH detection on the STA image (2048x2048) on 06/10/2010 required 275 iterations, which took a total of $2.6$ seconds using two threads on a Core i3 3.3GHz CPU, and only $0.3$ seconds using a GTX970 GPU (including data transfer time).

An important feature of the algorithm is to check the connectivity of a pixel in all 8 neighboring pixels and enforce a connectivity condition that requires $n$ consecutive marked pixels for inclusion in the region.  This helps to avoid false connectivity due to noise or ``bleed-out'' from small adjacent dark quiet sun pixel regions.  Through testing, we have found that $n=3$ yields the best detection results overall.

\begin{figure}[tbp]
\centering
\includegraphics[height=1.5in]{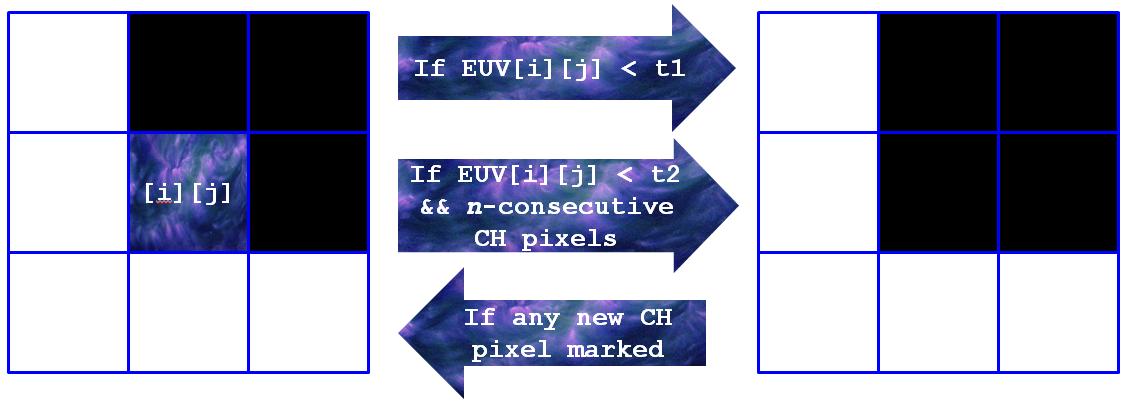}
\caption{Schematic of the iterative procedure in the {\tt ezseg} coronal hole detection algorithm.\label{fig_chdalg}}
\end{figure}

The algorithm starts with a blank CH map with pixel locations corresponding to missing/unused EUV data flagged.  Each viable pixel location is independently checked to see if the EUV image at that location is below $t_1$, and if it is, the pixel location in the CH map is marked.  Then, in each subsequent iteration, each pixel location is checked and is marked as a CH pixel if it is above $t_1$, below $t_2$, and has the requisite number of consecutive neighboring CH pixels.  This iterative procedure continues until no additional pixels are added to the CH map;   Fig.~\ref{fig_chdalg} shows a schematic depiction.

\begin{figure}[tbp]
\centering
\includegraphics[width=0.45\textwidth]{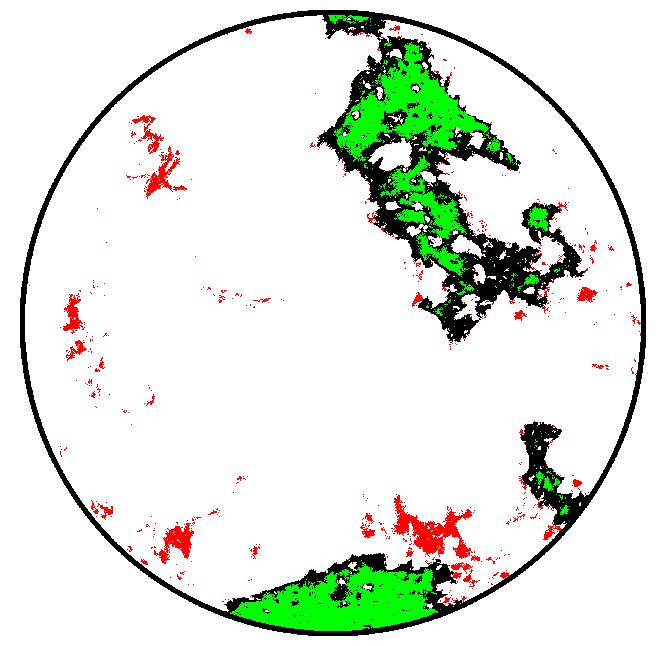}
\includegraphics[width=0.45\textwidth]{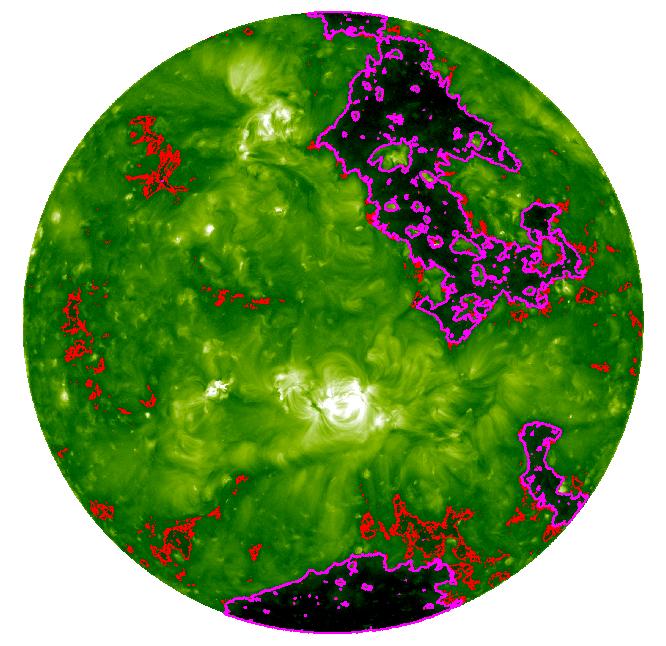}
\caption{Left:  Coronal hole map for AIA on 02/03/2011 produced by thresholding of $t_1$ (green), thresholding of $t_2$ (red) and the {\tt ezseg} algorithm (black).  Right:  Corresponding preprocessed AIA EUV image.  The {\tt ezseg} detection result is shown as the magenta contour, while the additional false detections by only using a threshold of $t_2$ is shown as red contours. The $t_1$ and $t_2$ values used are $0.95$ and $1.35$ respectively.\label{fig_t1t2alg}}
\end{figure}

Fig.~\ref{fig_t1t2alg} shows the advantage of using a region growing approach versus single-value thresholding.
The green areas are the result of using a threshold of $t_1$, the red areas of using only $t_2$, and the black areas (which include green areas) are the result of the two-threshold region growing algorithm.  We see that the region growing algorithm captures the CH boundaries but avoids adding the darker regions of quiet sun that would be detected when using a single threshold of $t_2$.

The main challenge in using the algorithm is choosing $t_1$ and $t_2$ such that a desired CH map is produced over a large number of sequential frames.  It is not clear if/how the CH intensities vary over the solar cycle (in which case the thresholds would need to be modified over time), and the choice of $t_2$ is somewhat subjective (indeed the use of different color tables to display the images can change the perceived detection quality).  There are different strategies for choosing $t_1$ and $t_2$.  For example, the SWAMIS code uses preselected sigma values of the histogram distribution to select the values.  However, since there may be no CHs in an image and/or large bright active regions, such a selection is not ideal for automatic CH detection over longer periods of time.  In order to address this issue in the simplest way possible, we have manually selected optimal values for $t_1$ and $t_2$ for multiple times in the range of interest.  This subjective process yielded no distinct trends in the way $t_1$ and $t_2$ changed over time, and the results were approximately constant over the five year period.  We therefore have chosen to use a single average value for each threshold ($t_1=0.95$ and $t_2=1.35$) for use with the entire time period.  These values apply to preprocessed images (as described in Sec.~\ref{sec_preproc}) and yield good detection results overall.  They can easily be modified as needed for improving detection for a specific data set, time, or time-range of interest.  Also, we note that the {\tt ezseg} algorithm is written in a way such that the user can manually (or though other automatic means) specify seed points in the input array, eliminating the need to choose a $t_1$ intensity value.

%%%%%%%%%%%%%%%%%%%%%%%%%%%%%%%%%%%%%%%%%%%%%%%%%%%%%%%%%%%%%%%%%%%%%%%%%%%%%%%%%%%%%%%%%%%%%%%%%%%%%%%%%%%%%%%%%%%%%

\subsection{Mapping}
\label{sec_mapping}
In order to avoid possible detection errors introduced by mapping the disk images \citep{2004_Deforest_Optimized_Coord_image_transforms}, as well as allowing the detection algorithm to work on smaller arrays, we have computed the CH maps for each instrument's line-of-sight images separately.  We now map the resulting CH disk-maps (and the images themselves) to a full-sun grid using simple linear interpolation.  We use a $\sin(\zeta)\mbox{--}\phi$ grid (where $\zeta=\pi/2-\theta$ is latitude) whose size is determined by setting the number of $\sin(\zeta)$ points to the maximum number of pixels in the disk images along a vertical centered cut.  This avoids excessive under- or over-sampling.

\begin{figure}[tbp]
\centering
$\vcenter{\hbox{\includegraphics[width=3in]{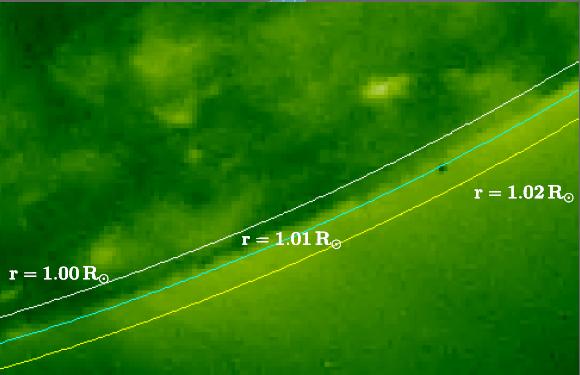}}}$\qquad
$\begin{array}{rcc}
\; & R_0=R_{\odot} & R_0=1.01\;R_{\odot}
\\
\rotatebox{90}{\;\;\;\;\;\;\;\;AIA} & \hbox{\includegraphics[width=1.25in]{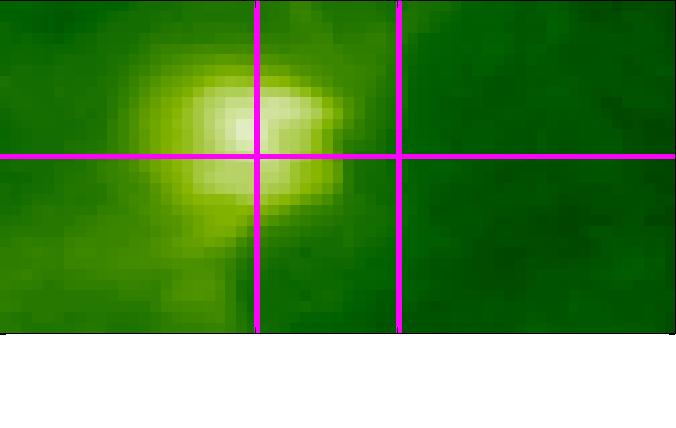}} &
\hbox{\includegraphics[width=1.25in]{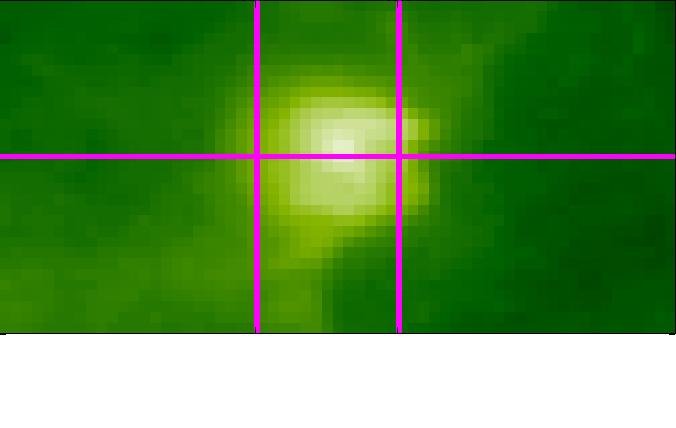}}
\\
\rotatebox{90}{\;\;\;\;\;\;\;\;STB} \; &\hbox{\includegraphics[width=1.25in]{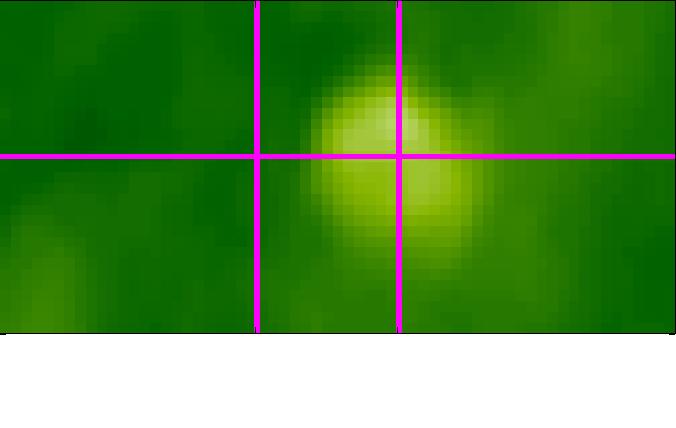}} &
\hbox{\includegraphics[width=1.25in]{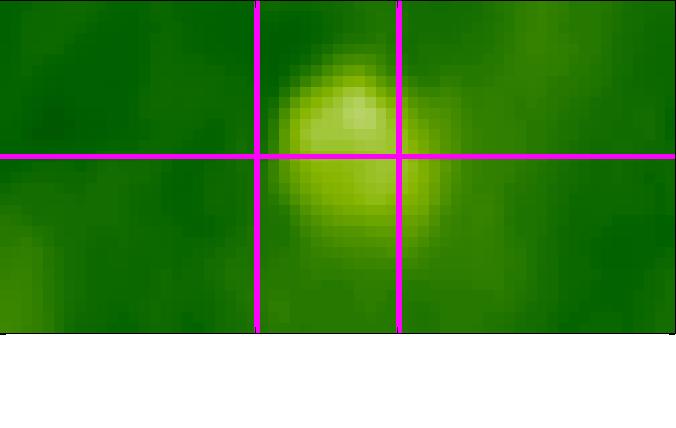}}
\\
\rotatebox{90}{\;\;\;\;\;\tiny $\frac{\mbox{AIA}-\mbox{STB}}{\mbox{STB}}$} \; &\hbox{\includegraphics[width=1.25in]{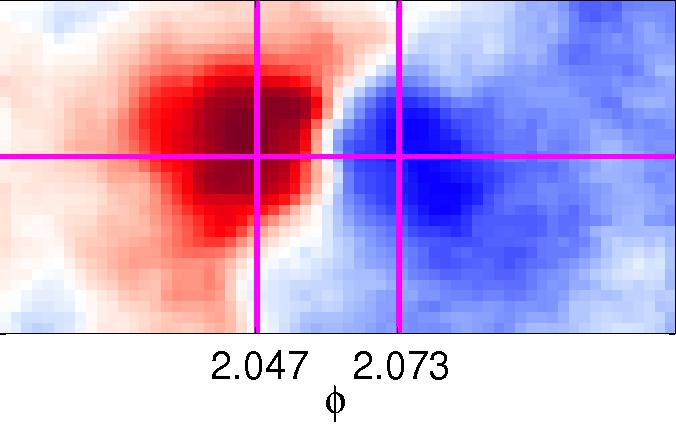}} &
\hbox{\includegraphics[width=1.25in]{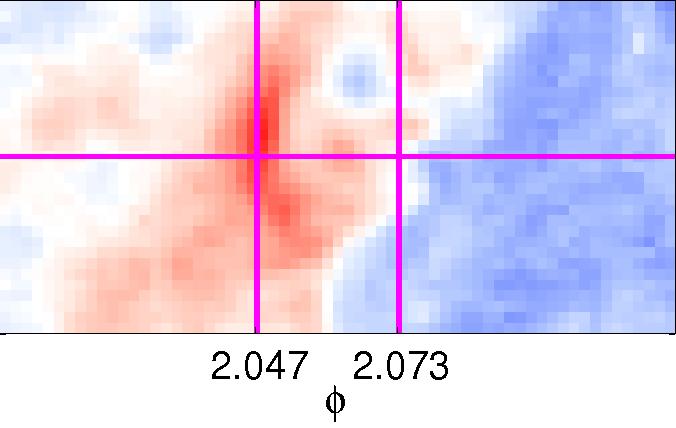}}
\end{array}$
\caption{Left: Example EUV image with radial contours displaying that the effective lower coronal boundary ($R_0$) is noticeably distinct from the photospheric radius.  Right:  Example of the effects of the assumed $R_0$ value on mapping overlapping EUV images.  We show close-ups ($\Delta \phi=0.12$, $\Delta \sin(\zeta)=0.04$) of EUV mapped images of a bright structure taken from AIA and STB in their overlapping data region.  The image from each instrument as well as the relative difference between them are shown for $R_0=R_{\odot}$ (left column) and $R_0=1.01\,R_{\odot}$ (right column).  The shift of the bright structure's position between instruments when using $R_0=1.01\,R_{\odot}$ is seen to be much less than that when using $R_0=R_{\odot}$. \label{fig_r101}}
\end{figure}

Typically, disk images of the sun are mapped assuming a photospheric radius ($R_{\odot}$).  However, in the case of EUV images such as those used here, the apparent radius of the line-of-sight disk image is slightly larger than $R_{\odot}$ due to the bulk of the emitting plasma residing in the lower corona (as shown in Fig.~\ref{fig_r101}).    This is an important effect to keep in mind, especially when projecting multiple disk images to the same map, as we have found that the difference between the photospheric and lower coronal radii is large enough to cause overlapping mapped images to display a noticeable shift relative to each other (see Fig.~\ref{fig_r101}).
Therefore, we use an estimate of the effective lower coronal boundary ($R_0=1.01\,R_{\odot}$) for all preprocessing and mapping steps.

\begin{figure}[tbp]
\centering
\subfigure[ ]{\includegraphics[width=0.2\textwidth]{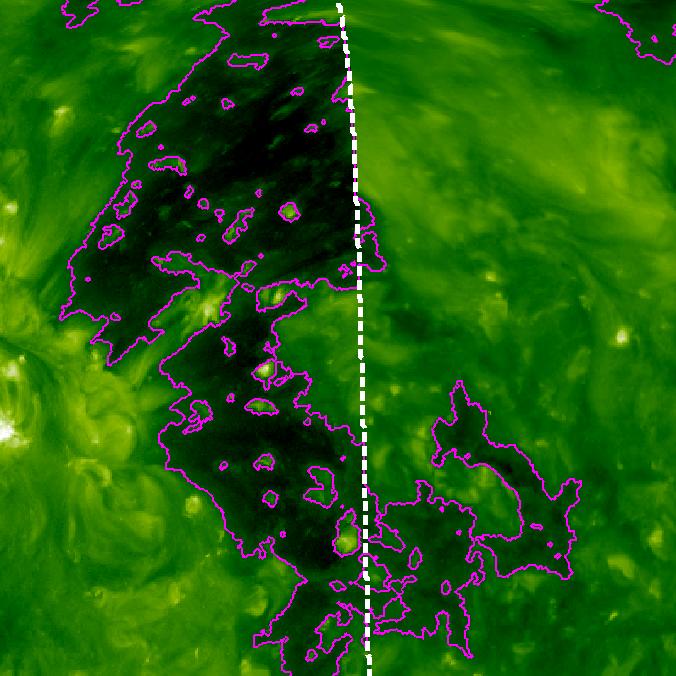}}
\subfigure[ ]{\includegraphics[width=0.2\textwidth]{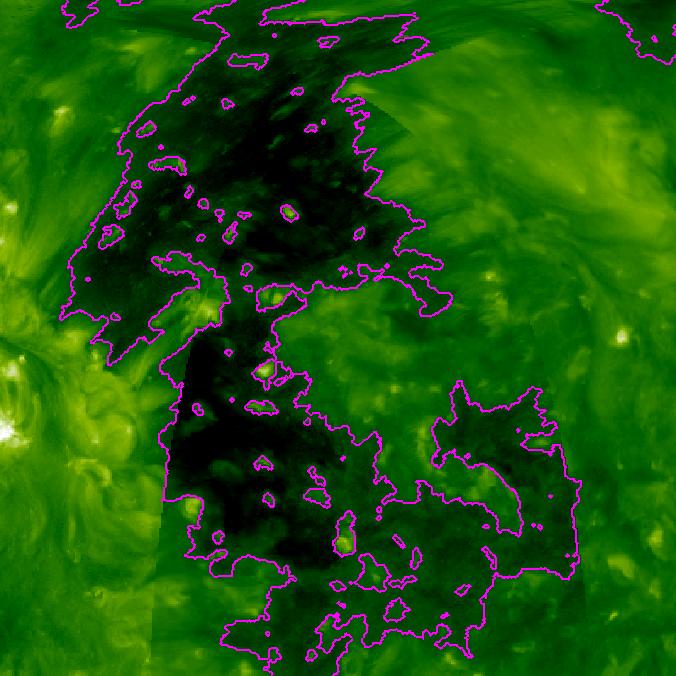}}
\caption{Example of merging EUV images and CH maps from different instruments in overlapping regions.  The merged preprocessed images of AIA and STA data from 02/04/2011 are shown with (a) maximum-$\mu$ merging (with the white line denoting the maximum $\mu$ edge of each data) and (b) minimum intensity merging. The parameters for the CH detection and preprocessing steps are the same as described in the text.\label{fig_merge}}
\end{figure}

Since it is often the case that images from two (or more) instruments will overlap in the map, it is necessary to select or merge the data to be used. We have implemented two ways of doing this.  The first is simply to use the data that has the highest $\mu$-value associated with it, since the higher the $\mu$, the more direct the viewing angle and the less distortion there will be from the coordinate transformation.  This can sometimes produce ``cut-off'' lines through the images and coronal hole maps that sweep through the structures over time.  This may happen when the CH line-of-sight becomes obscured by nearby quiet-sun or active regions, in which case, another instrument might have a better view of the CH  even though its line-of-sight is closer to the limb.  Therefore, we have implemented a merge mapping where each disk image is first mapped to its own Carrington map, and then the three maps are merged by taking the lowest intensity values of the overlap.  The coronal hole map data is then set based on which data was used for the EUV map.  We find that this method produces EUV and CH maps that tend to be more continuous for dark structures at the seams and work well for the present context. An example showing the difference between the two types of map merging is shown in Fig.~\ref{fig_merge}.
Of course, there are instances where a CH becomes obscured by intervening structures from all viewpoints and therefore has portions drop out from the detection; this is an unavoidable systematic error inherent to mapping the EUV corona from limited viewpoints. We note that the methods of merging the overlap regions discussed here are only a {\em selection} criteria of which data to use, i.e. they do not modify any of the image or CH disk data.

One problem with either of the two mapping methods described above is that data that is very close to the limb is often quite distorted, and it may be desirable to not include it in the map.  This can be avoided by setting a minimum cut-off for $\mu$, denoted $\mu_{\mbox{\scriptsize cut}}$, which only allows data with $\mu \ge \mu_{\mbox{\scriptsize cut}}$ to be used.  We find that a value of $\mu_{\mbox{\scriptsize cut}}=0.4$ is a good choice for avoiding distortions while still retaining most of the data.  For cases where one is willing to tolerate distorted data past the $\mu$-cutoff if it is the only data available, we can use two $\mu$ cutoffs: one for the merging regions with overlap, and the other for non-merging regions.  We choose to use the maximum amount of available data (using a merge cutoff of $\mu=0.4$ and a non-merge of $\mu=0$) so that we can create EUV and CH maps with the best merged data in the overlapped regions, while still allowing the use of disk data that extends all the way to the limb/poles (the merged map in Fig.~\ref{fig_merge} used these parameters).  

%%%%%%%%%%%%%%%%%%%%%%%%%%%%%%%%%%%%%%%%%%%%%%%%%%%%%%%%%%%%%%%%%%%%%%%%%%%%%%%%%%%%%%%%%%%%%%%%%%%%%%%%%%%%%%%%%%%%%

\section{Sample results}
\label{sec_results}
Here we display several examples of CH maps generated by {\tt ezseg} using STA, STB, and AIA data, after preprocessing with the techniques described in Sec.~\ref{sec_preproc}.   

\begin{figure}[tbp]
\centering
\includegraphics[height=0.2\textheight]{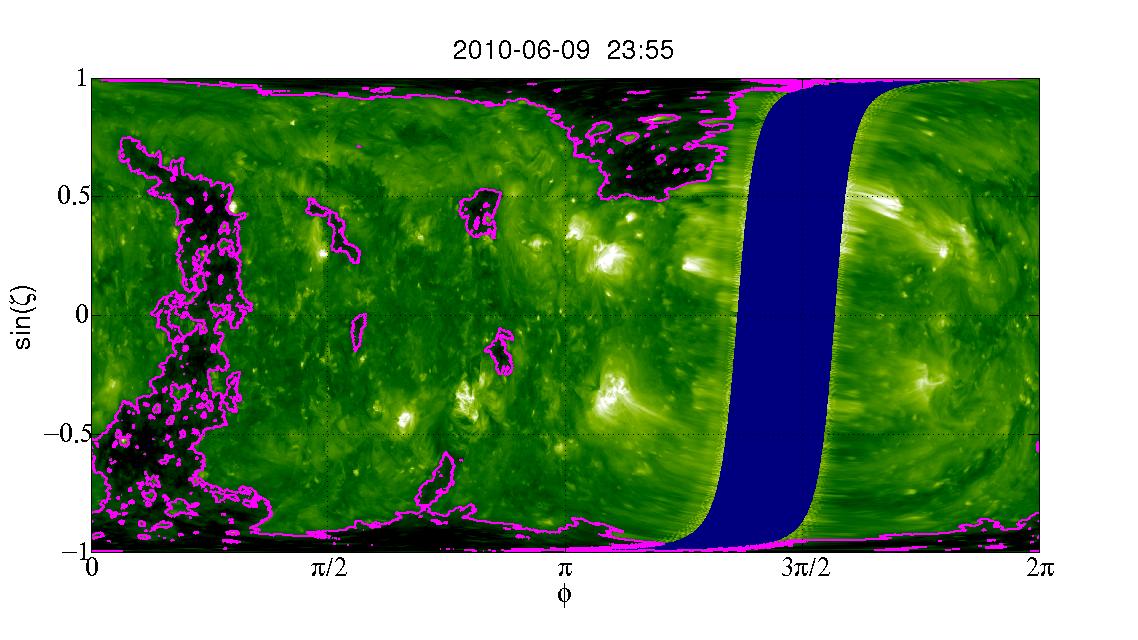}
\includegraphics[height=0.2\textheight]{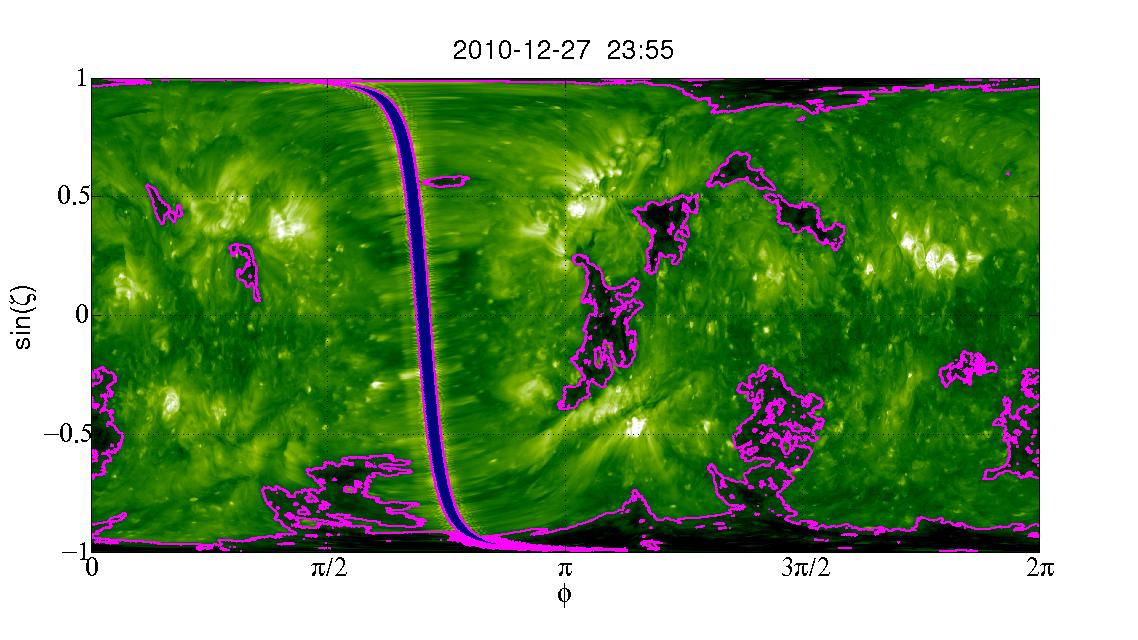}
\\
\includegraphics[height=0.2\textheight]{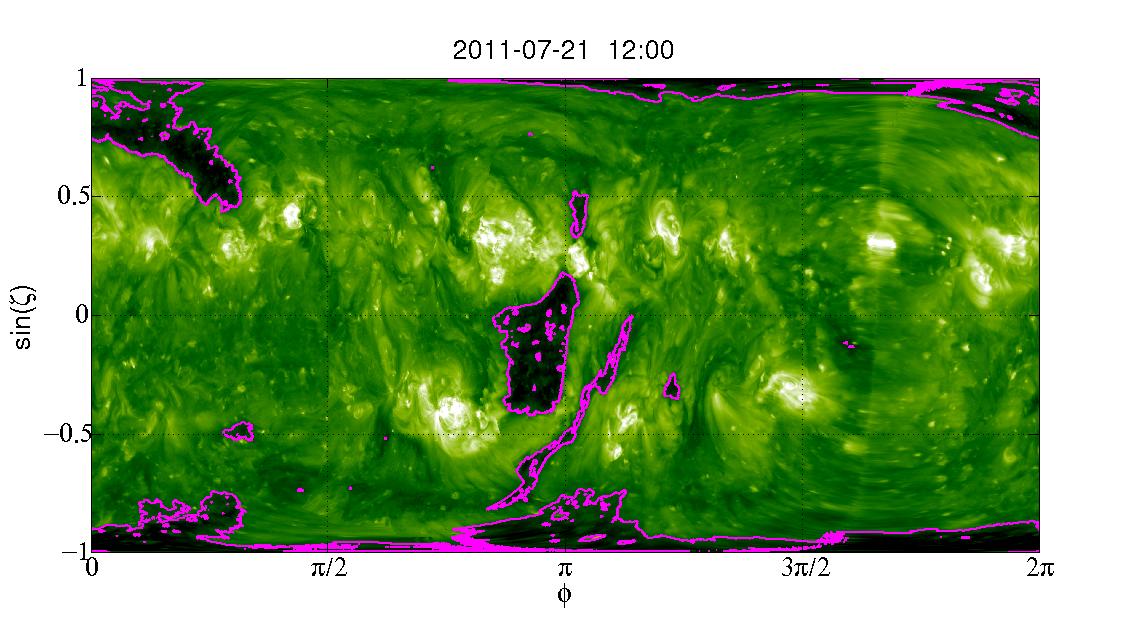}
\includegraphics[height=0.2\textheight]{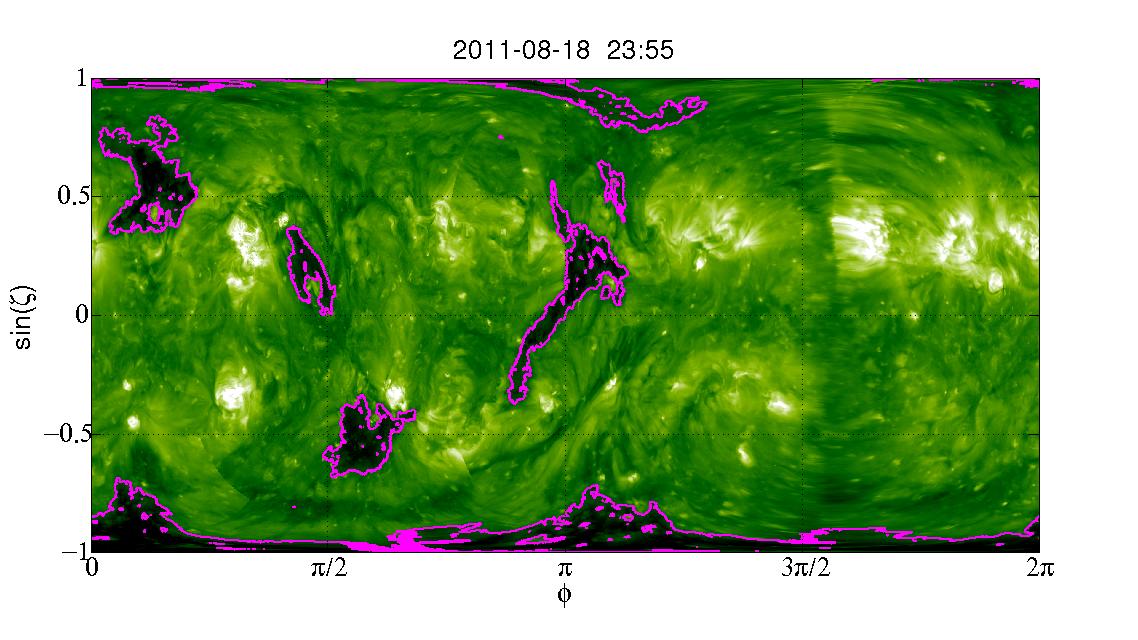}
\\
\includegraphics[height=0.2\textheight]{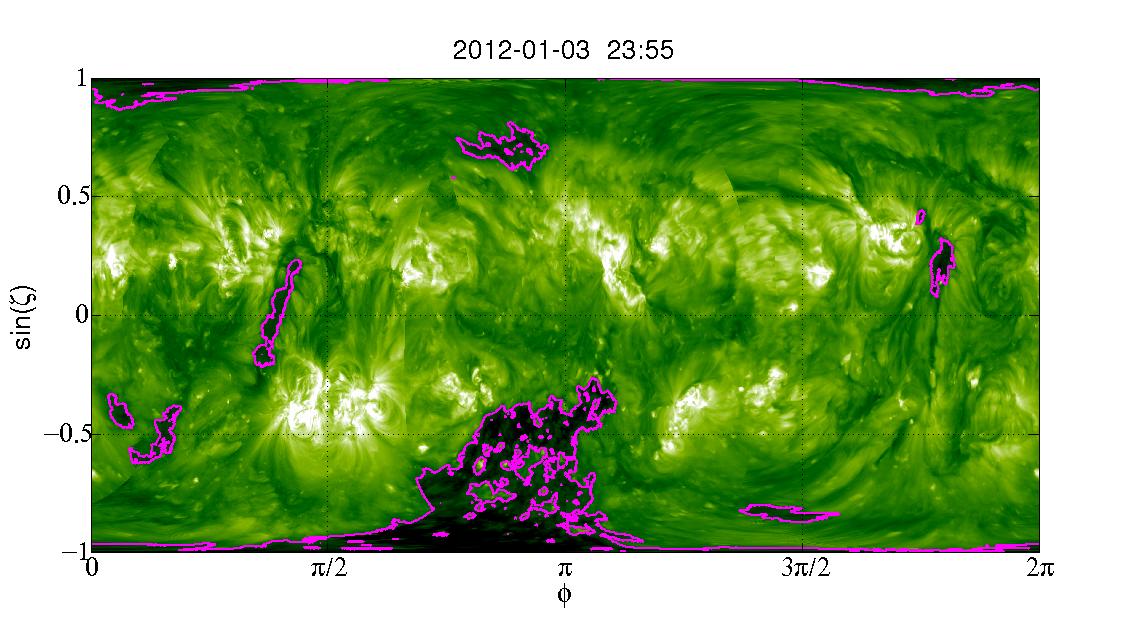}
\includegraphics[height=0.2\textheight]{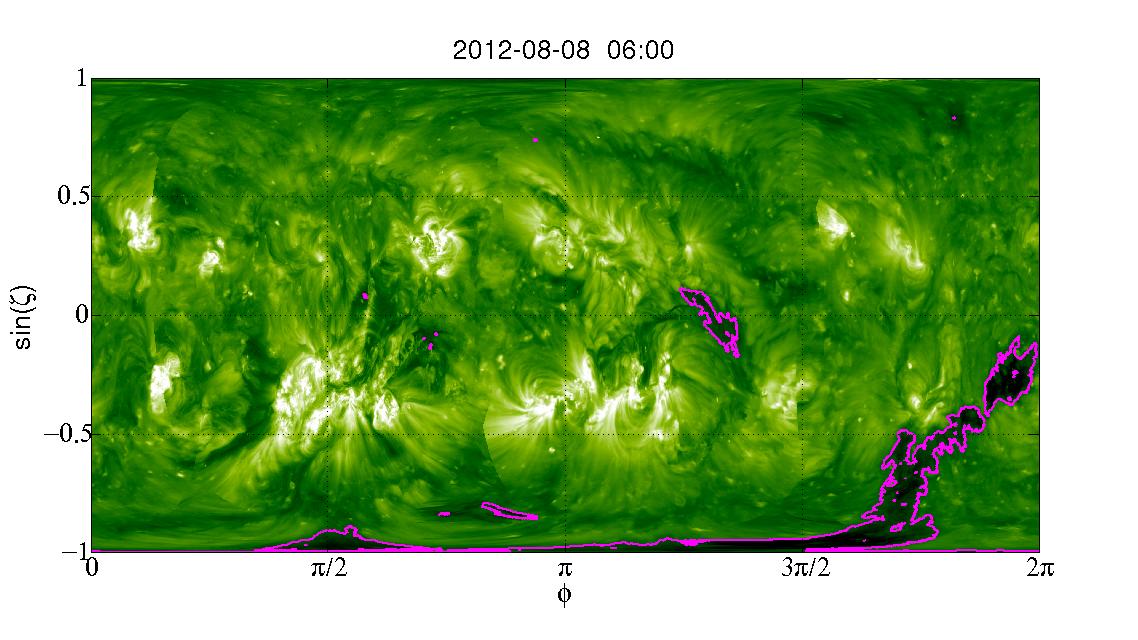}
\\
\includegraphics[height=0.2\textheight]{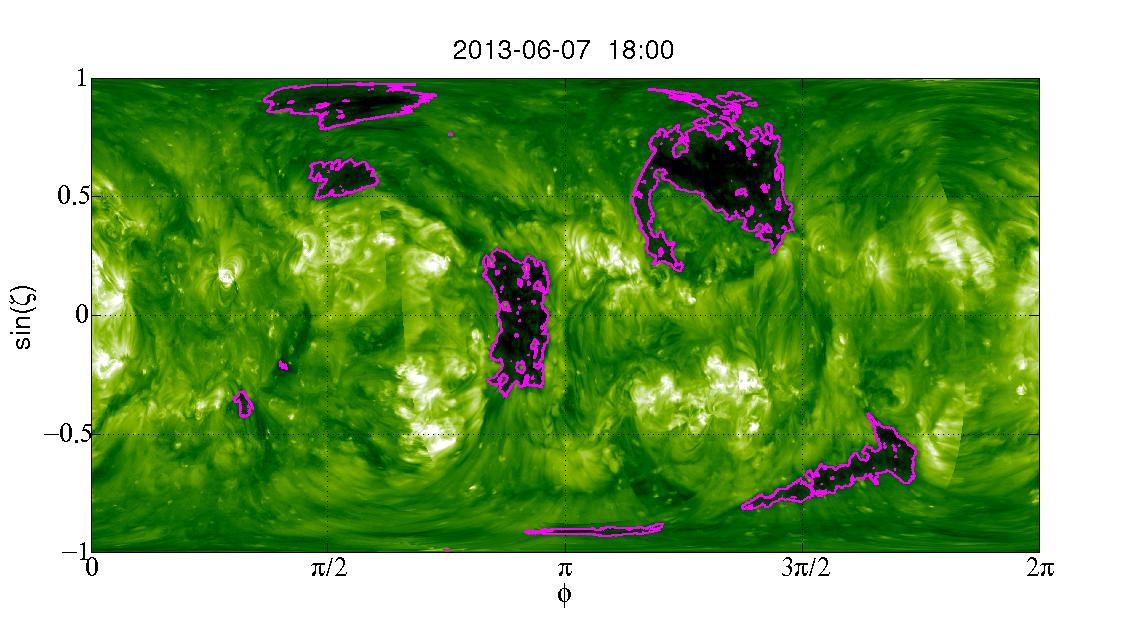}
\includegraphics[height=0.2\textheight]{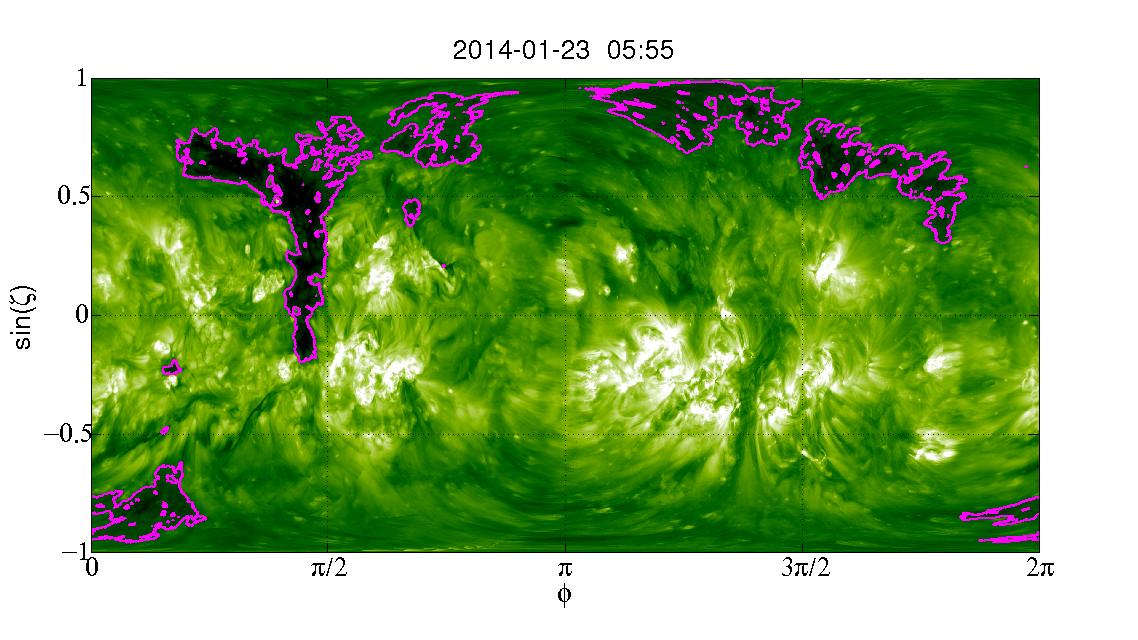}
\caption{Examples of synchronic EUV maps with the detected CH map overlaid as a contour.  All detections use the threshold values $t_1=0.95$ and $t_2=1.35$. For time periods where no limb-brightening correction or inter-instrument transformation factors have been computed, the nearest computed values in time are used.  The date and time for the STA data for each frame is indicated, while the time for the STB and AIA data is no more than $\pm10$ minutes from the STA time. \label{fig_samples}}
\end{figure}

In Fig.~\ref{fig_samples} we show an assortment of synchronic EUV maps with an overlay contour of the detected CH maps within the 5-year time period where STEREO-A/B and AIA were all active (a full set of such maps at 6-hour cadence from 06/10/2010 to 08/18/2014 are made available online (see Sec.~\ref{sec_conclusion})).
The data-driven preprocessing allows for the production of nearly seamless full-sun synchronic maps, and the CH detection performs well throughout the data set, even with the use of constant values for the threshold parameters, $t_1$ and $t_2$. 

\begin{figure}[tbp]
\centering
\subfigure[01/22/2011 to 02/09/2011]{\includegraphics[width=\textwidth]{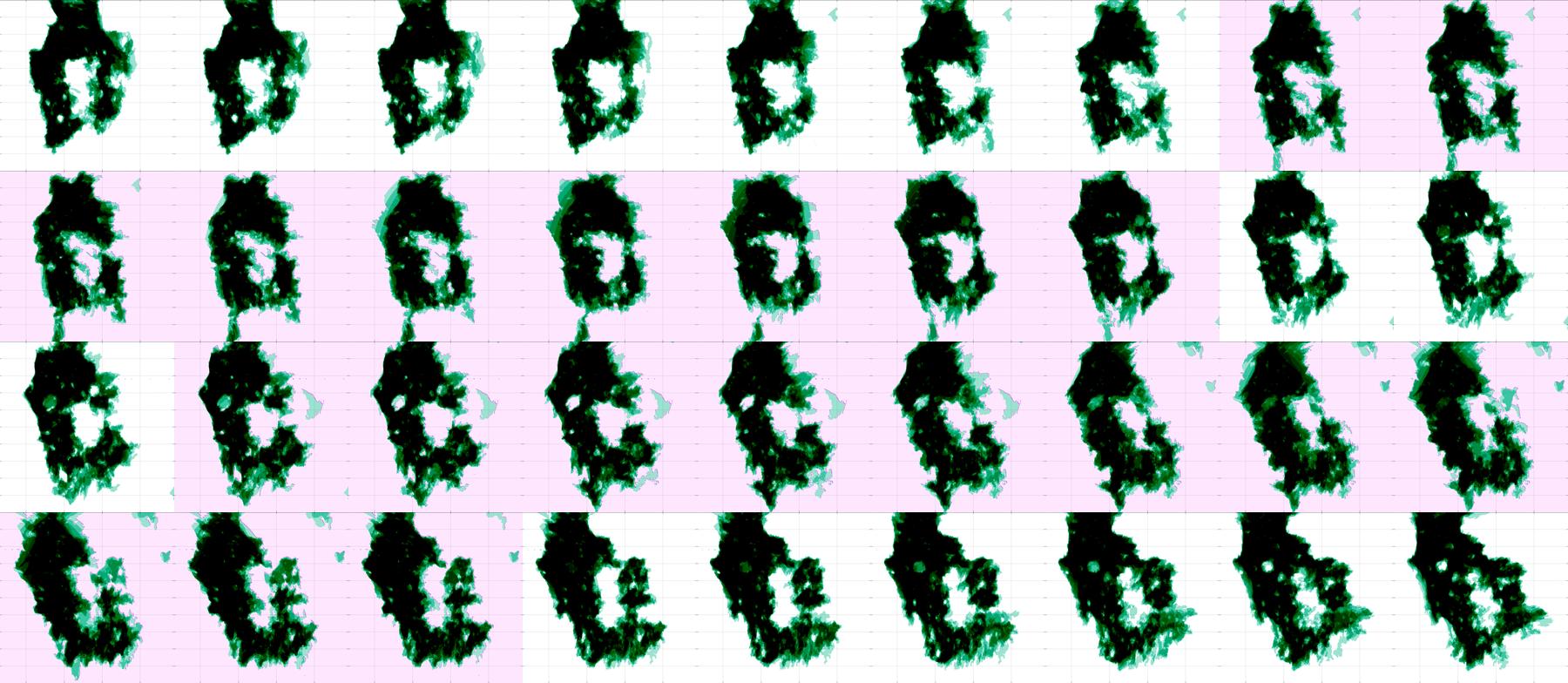}}\\
\subfigure[05/27/2013 to 06/18/2013]{\includegraphics[width=\textwidth]{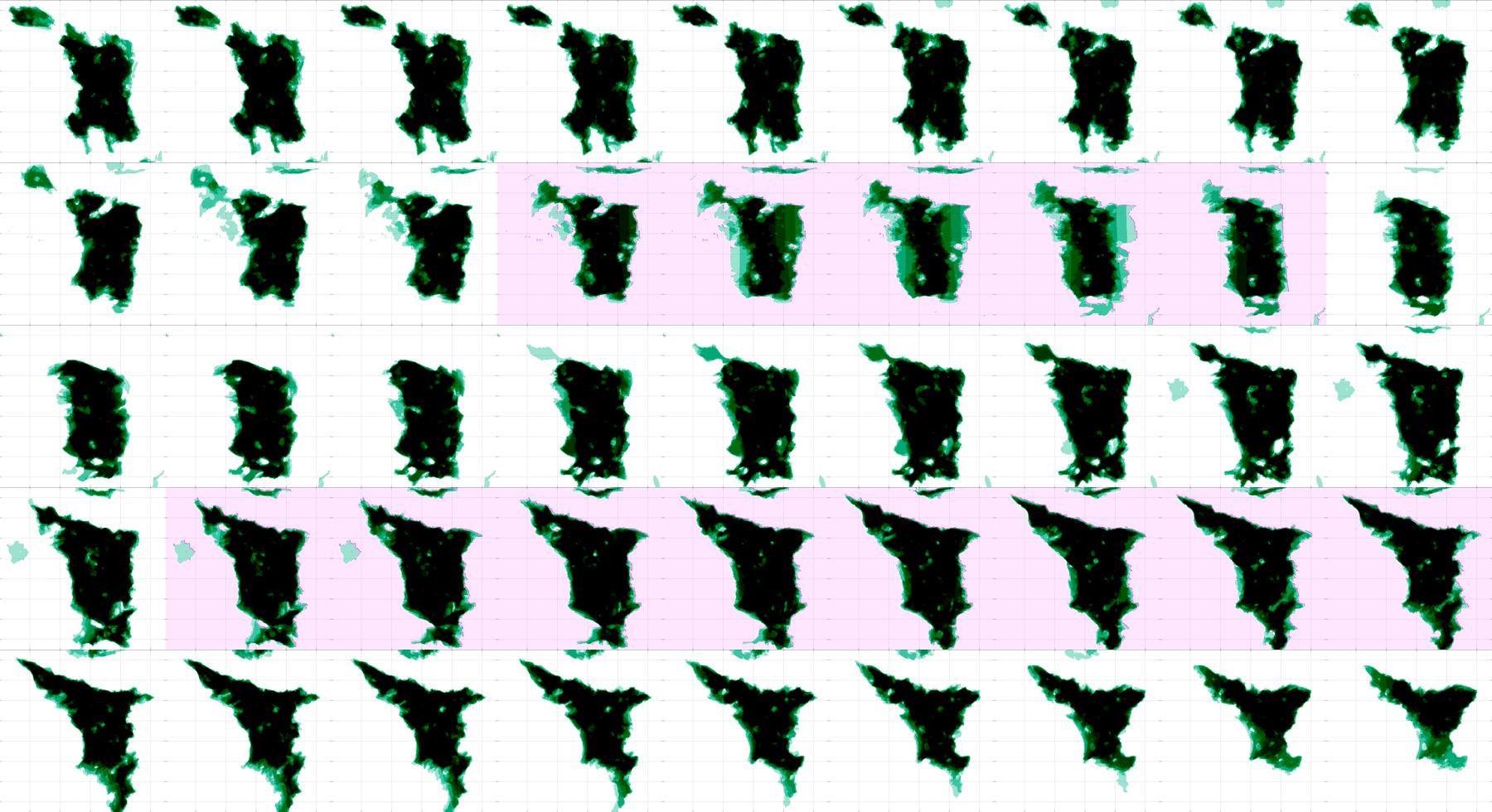}} 
\caption{(a) Depiction of coronal hole evolution from 01/22/2011 to 02/09/2011 for the CH in the region defined by $\theta\in\{37^{\circ},102^{\circ}\}$ and $\phi\in\{166^{\circ},258^{\circ}\}$.  (b) Depiction of coronal hole evolution from 05/27/2013 to 06/18/2013 for the CH in the region defined by $\theta\in\{63^{\circ},110^{\circ}\}$ and $\phi\in\{130^{\circ},185^{\circ}\}$.  Each evolution is shown as a series of 66-hour centered persistence maps at 12-hour cadence. The pink shaded frames indicate that the given map contains overlapping data at the central time.\label{fig_chd_time_dept}}
\end{figure}

Our automated pipeline allows us to observe the evolution of CHs over time for longer periods of time than a single instrument's images would allow.  In Fig.~\ref{fig_chd_time_dept}, we show the evolution of two CHs; one from a quiet sun time (01/22/2011 to 02/09/2011) and the other from a more active time (05/27/2013 to 06/18/2013).  
Since the current form of {\tt ezseg} occasionally finds minor CH areas at the borderline of $t_1$ that flicker in and out, we display smoother 66-hour persistence maps. A persistence map is a floating point mean of a time-series of CH maps \citep[see][for details]{2014_Lowder_CHD_AIA_EUVI}, and just like a running average, a time-centered persistence map can be created using neighboring data for any time in our data set. In this case, both CHs evolve over the time periods shown, with the CH from 05/27/2013 to 06/18/2013 showing a great deal of shape transformation and the CH from 01/22/2011 to 02/09/2011 appearing to rotate.  Note that both evolutions were detected with the same coronal hole detection parameters even though the time periods are over 2 years apart.  Once again, this is only possible due to the data-driven preprocessing. 

\begin{figure}[htbp]
\centering
\includegraphics[width=0.45\textwidth]{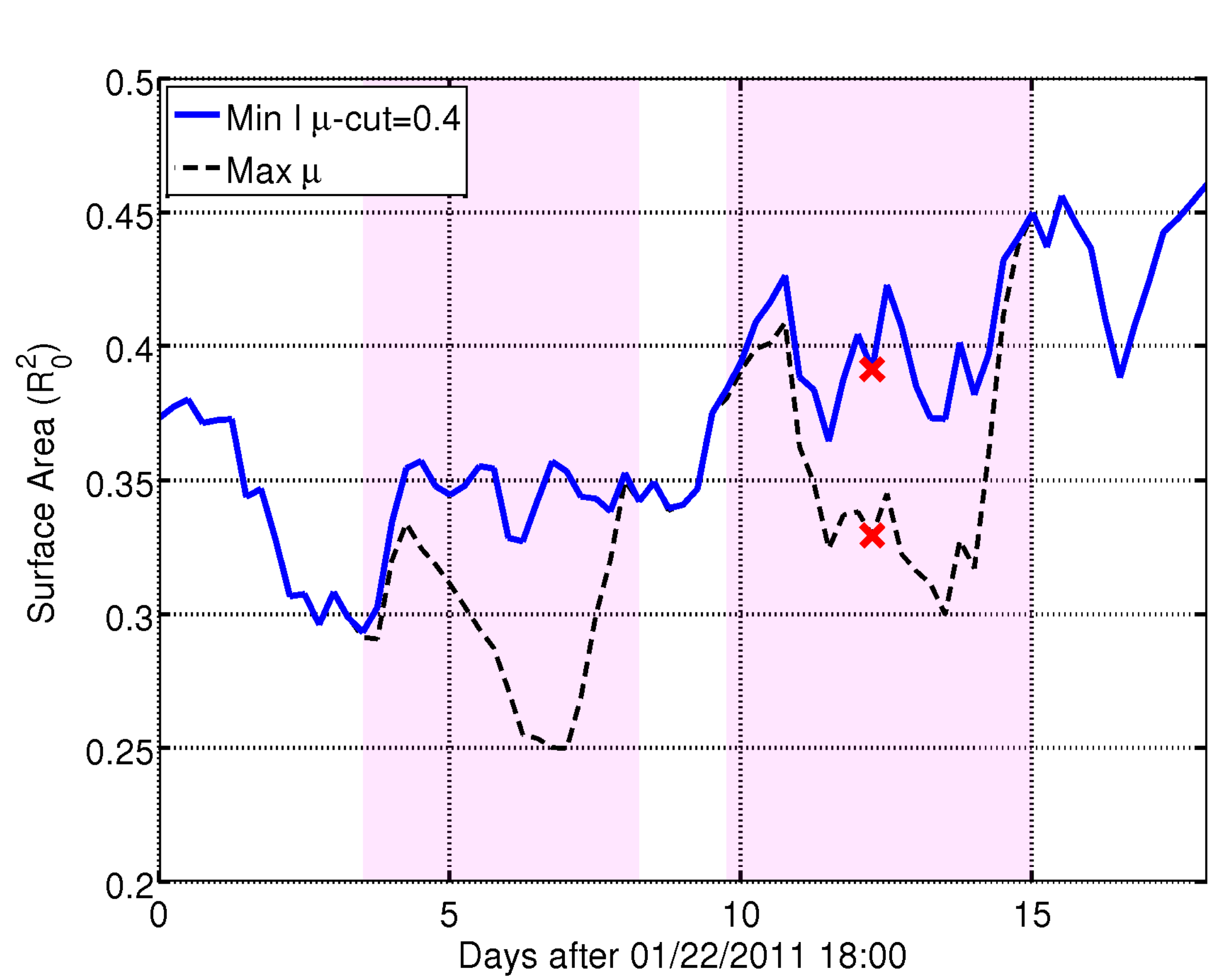}
\includegraphics[width=0.45\textwidth]{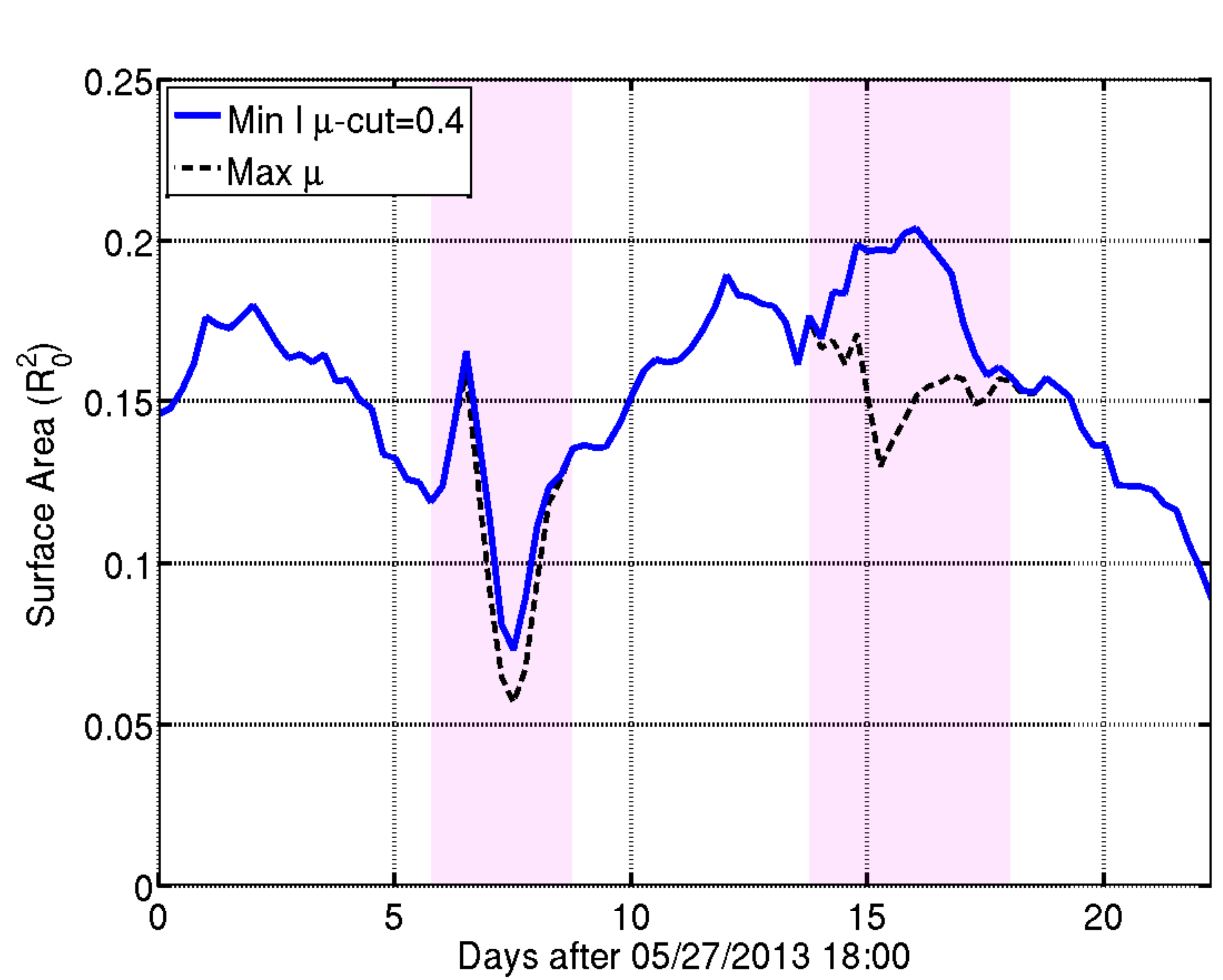}
\caption{Computed coronal hole area vs. time for the two CHs shown in Fig.~\ref{fig_chd_time_dept} (a) and (b).  The areas are shown using the two overlap-data merging methods from Sec.~\ref{sec_mapping}: a minimum-intensity overlap merging method with $\mu_{\mbox{\tiny cut}}=0.4$ (blue solid line) and a maximum $\mu$ merge (black dashed line).  The pink shaded regions represent times when overlap occurs and these times correspond to the pink shaded frames in Fig.~\ref{fig_chd_time_dept}.  The red markers in the left plot correspond to the merge comparison images shown in Fig.~\ref{fig_merge}.\label{fig_charea}}
\end{figure}

Time-dependent analysis of these maps allows one to track the properties of the CHs over time.  An example is given in Fig.~\ref{fig_charea} where we show the computed surface area of the CHs over time.  The area of each CH, in units of $R_0^2$, is computed by isolating the region containing the CH and multiplying the area factor $\Delta \sin(\zeta)\Delta \phi$ by the total number of CH pixels.
The CH area for 01/22/2011 to 02/09/2011 seems to be fairly constant with only a slow overall rise in surface area, even though the CH shape itself is evolving.  It also seems that the minimum intensity form of merging the CH maps does a better job of tracking the CH area through instrument transitions than the maximum $\mu$ approach (see Sec.~\ref{sec_mapping}).  This conclusion is reasonable because it is unlikely for the CH area to exhibit sudden jumps only during overlap transitions.  In contrast, both merging methods appear to have trouble tracking the CH area for the other time period. This can be seen in the right area plot during the first overlap, which has dramatic up and down spikes in the trend.  Observing the evolution (shaded frames of Fig.~\ref{fig_chd_time_dept} (b)), this oscillation in area does not appear to be true CH evolution, but rather is due to geometric obscuration by surrounding quiet sun structures, first from one side and then the other. This issue is compounded by the very small overlap of the instruments at this time, which worsens the mean projection angle and increases the likelihood of obscuration from both sides.

\begin{figure}[tbp]
\centering
\subfigure[]{\includegraphics[width=0.32\textwidth]{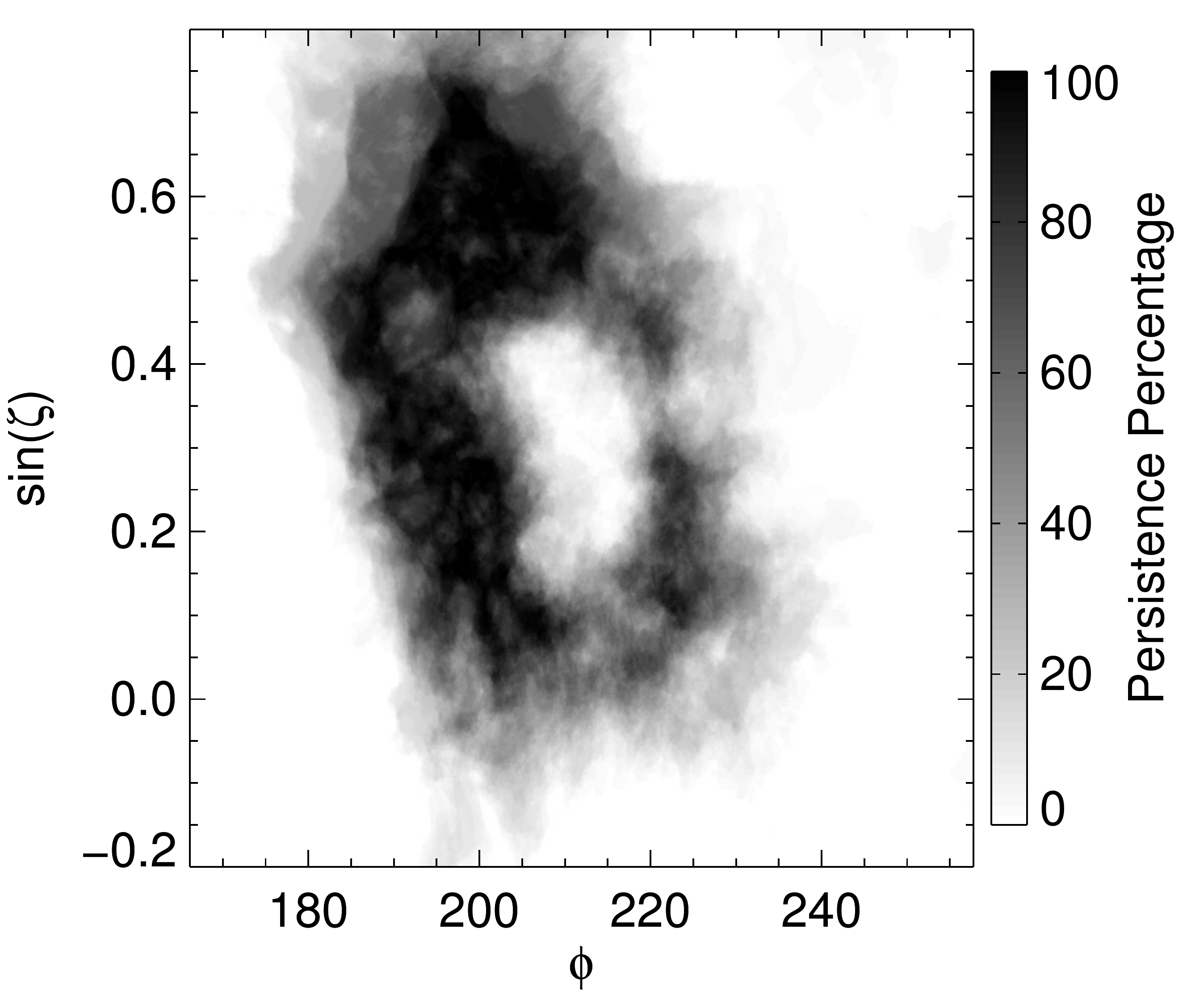}}
\subfigure[]{\includegraphics[width=0.32\textwidth]{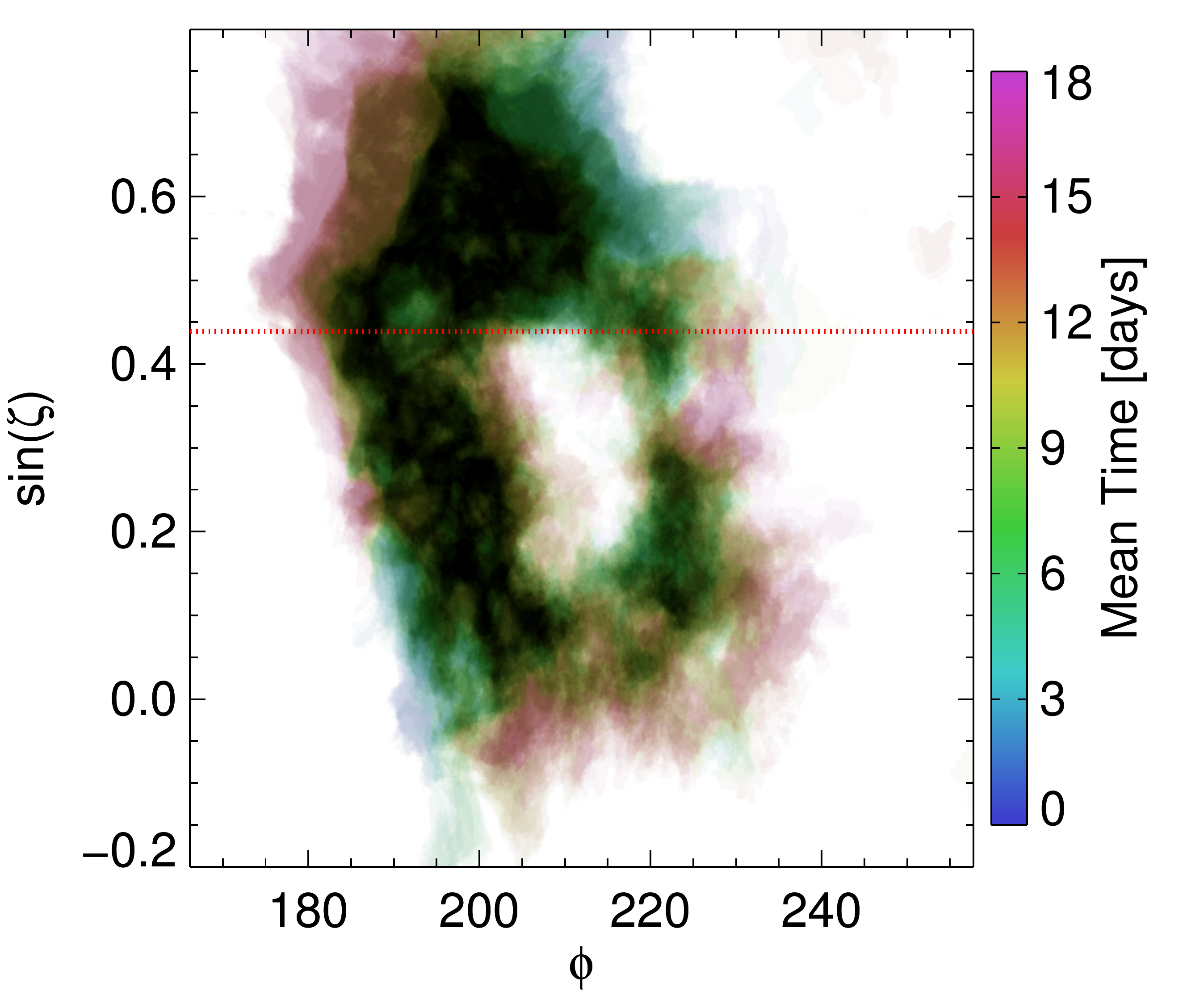}}
\subfigure[]{\includegraphics[width=0.32\textwidth]{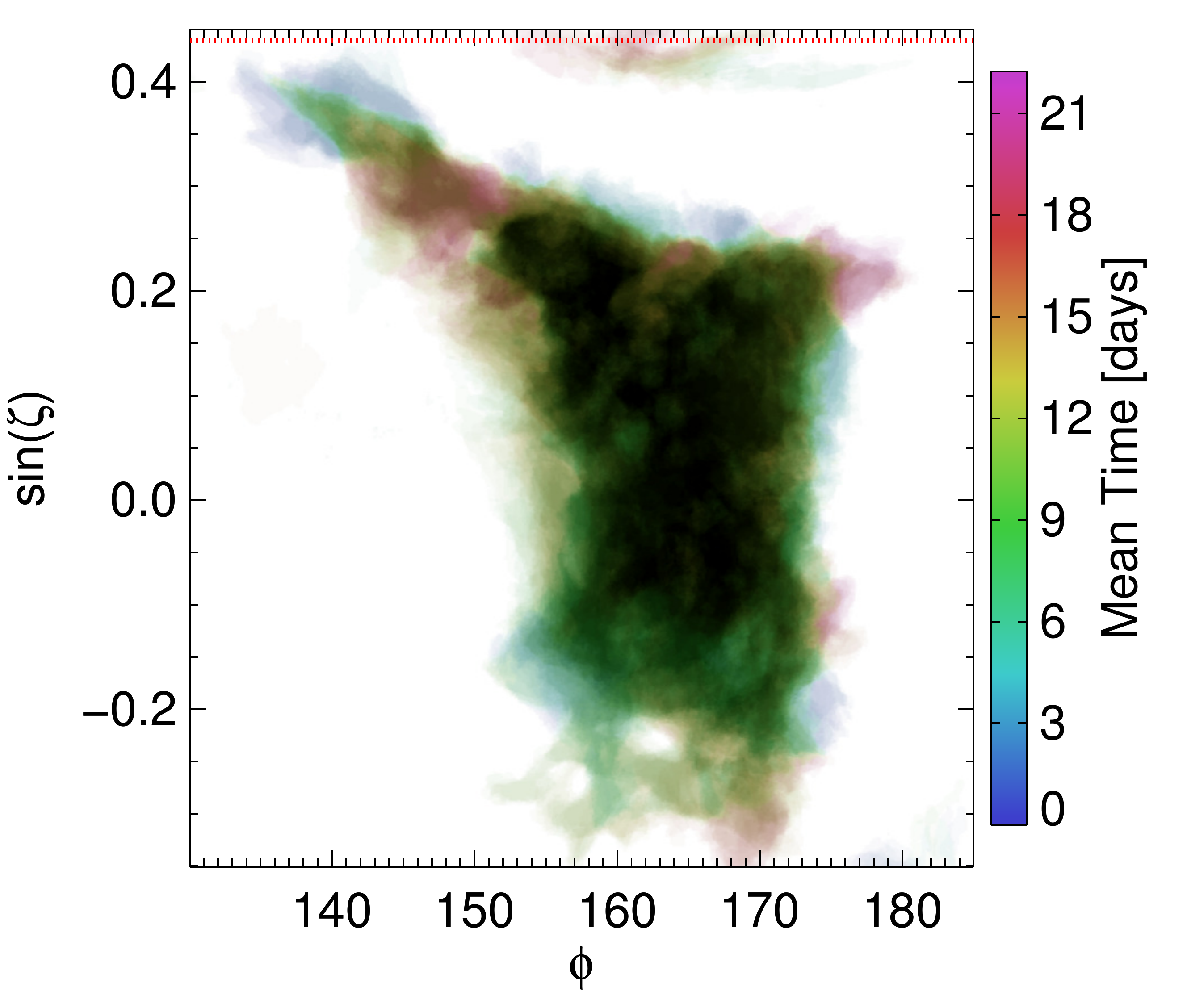}}
\caption{Persistence maps for the coronal hole evolutions of Fig.~\ref{fig_chd_time_dept}. (a) Grayscale persistence map ($P$) for the CH evolution from 01/22/2011 to 02/09/2011. (b) Time-colored evolution map (Eq.~(\ref{eq_timecolormap})) for the same CH evolution in (a).  (c) Time-colored evolution map of the CH evolution from 05/27/2013 to 06/18/2013.  The red lines in (b) and (c) show the location the Carrington latitude of $\sim 26^{\circ}$.\label{fig_ch_per}}
\end{figure}

Lastly, in an attempt to illustrate CH evolution with a single visualization, we introduce a method for constructing time-colored evolution maps based on persistence data.  Given the series of CH maps $C(x,y,t)$, where pixels identified as belonging to a CH are set to $C=1$ and otherwise $C=0$, each pixel in the time-color evolution map is colored using an HSV color table with 
\begin{alignat}{2}
\label{eq_timecolormap}
H^{\prime} &= \mean{t}/\Delta t, \\
S &= P, \notag \\
V &= 1-P, \notag 
\end{alignat}
where $\Delta t=t_{\mbox{\tiny stop}}-t_{\mbox{\tiny start}}$ is the length of time of the CH evolution, $\mean{t}$ is the integrated mean time of the detection, and $P(x,y)\in[0,1]$ is the normalized persistence map. $\mean{t}$ and $P$ are computed as: 
\[
\mean{t} = \frac{1}{P}\,\int_{t_{\mbox{\tiny start}}}^{t_{\mbox{\tiny stop}}} (t-t_{\mbox{\tiny start}})\,C(x,y,t)\,dt,
\]
\[
P(x,y) = \frac{1}{\Delta t}\int_{t_{\mbox{\tiny start}}}^{t_{\mbox{\tiny stop}}} C(x,y,t)\,dt,
\]
and $H^{\prime}\in[0,1]$ corresponds the chosen range of hue values ($H$).  Since hue is cyclical, we truncate the full range of hue values to start at $H=242$ (blue) backwards through to $H=297$ (pink) leaving out the pink-to-blue transition values.  Thus, blue color in the map represents CH area near the start of the evolution time while pink represents CH area near the end time.  The map becomes darker and more color saturated in areas of high persistence, and is brighter and has less color saturation in areas of low persistence.  For example, a CH traveling from left to right in the time period of the persistence map would appear as a streak of color ranging in blue on the left to pink  on the right.  In Fig.~\ref{fig_ch_per}, we show examples of time-color evolution maps for both CH evolutions from Fig.~\ref{fig_chd_time_dept}.  We also show a standard gray-scale persistence map for the CH from 01/22/2011 to 02/09/2011 to illustrate the advantage of the new colorized map.
One can see that the time-colorized evolution map allows one to visualize the time-evolution of the CHs in a single glance.  For example, the map for 01/22/2011 to 02/09/2011 seems to indicate that the central core of the CH remains steady throughout the evolution, while the outer areas are undergoing differential rotation.  This can be seen qualitatively by plotting the latitude where the Sun's rotation rate matches the Carrington rotation rate ($\sim\!26^{\circ}$) and noting that the CH clearly shows left-sided (slower) motion above the Carrington latitude line and right-sided motion (faster) below.  In contrast, the CH from 05/27/2013 to 06/18/2013 does not show any discernible signal of differential rotation. However, this equatorial hole was present with nearby active regions emerging, evolving, and sometimes partially obscuring the coronal hole. Further analysis of CH evolutions are underway and plan to be discussed in future publications.

%%%%%%%%%%%%%%%%%%%%%%%%%%%%%%%%%%%%%%%%%%%%%%%%%%%%%%%%%%%%%%%%%%%%%%%%%%%%%%%%%%%%%%%%%%%%%%%%%%%%%%%%%%%%%%%%%%%%%

\section{Summary, products, and future work}
\label{sec_conclusion}
In this paper we have described a process to generate synchronic coronal hole evolution maps from multiple EUV images.  We applied the method to EUV images from STEREO-A/B EUVI 195\AA\, and SDO AIA 193\AA\, from 06/10/10 to 08/18/14, which represents the majority of the time span when STEREO-A/B and AIA were all simultaneously functioning.  

The ability to produce automatic synchronic coronal hole maps using the techniques described here allows for the analysis of coronal hole evolution. The surface areas of coronal holes were tracked over a longer period of time than the images from a single instrument could provide. We then produced persistence maps of the CHs using a new time-colored method allowing for the trends in a CH evolution to be observed in a single generated image.  For a CH in the beginning of 2011, we observed a compelling signal of differential rotation in the CH.  Further study of CH evolution using the detection methods presented in this paper are planned.

An important element in this work is the development and application of several preprocessing steps. These include PSF deconvolution, limb-brightening correction, and inter-instrument transformations that help to sharpen the images and reduce scattered light, equalize intensity values from disk center to limb, and equalize intensity values across instruments respectively. In addition to their practical image processing utility, the limb-brightening correction results illustrate their sensitivity to the temperature structure of the corona, which we plan to explore further. Although we have only shown preprocessing parameter data for specific wavelengths and instruments, the same processing steps can be applied to any available wavelength/instrument combination.

The preprocessing steps applied here allow the use of a simple segmentation algorithm to produce acceptable coronal hole maps.  The algorithm (denoted {\tt ezseg}) is a dual-threshold based region growing algorithm with variable connectivity constraints.  It seeds CH map points based on an intensity threshold value, and then grows the regions iteratively until a growth intensity threshold value is reached or the connectivity falls below the required consecutive neighbor criteria.   The choice of the seed and growth thresholds are currently not determined automatically, and require manual selection.  However, we have found that (due to the new preprocessing) a single set of values work surprisingly well over a long span of time.  The coronal hole maps for each instrument are then combined into a single $\sin(\zeta)$--$\phi$ map where any overlapping regions of the images are merged by selecting the minimum intensity values to bring out the coronal holes.

We have shown that the current {\tt ezseg} algorithm produces very useful CH maps over time.  However, the algorithm has some limitations whose improvement could enhance more detailed CH evolution studies.  The time-dependent CH maps can suffer from a flickering effect where CHs (usually small ones or ones near instrument borders) will pop in and out of the produced maps.  This effect is caused by using singular threshold values over time. Another common problem is ``bleed-out'' where dark quiet sun happens to be near a CH, causing the algorithm to sometimes follow the path of the quiet sun and mistakenly detect it as being attached to the CH.  Our use of a connectivity requirement reduces this problem significantly, but more improvement is desired.  These issues may be mitigated by the use of a more advanced detection algorithm, such as the seeded region growing algorithm (SRG) of \citet{1997_Menhert_Improved_SRG} that, given seed locations, segments an image without the need of any tuning parameters.  Further along those lines would be the SRG algorithm of \citet{2005_Fan_SRG_autoseed_autotrack} that automatically chooses the seed locations, and auto-tracks the segmentation through time, which would be useful for studying coronal hole evolution.  To our knowledge, the SRG algorithm has not yet been utilized in any form in the context of solar images, and looks to be a promising area to pursue.  

Several data products are publicly provided to the community at a dedicated site\footnote{URL: {\tt www.predsci.com/chd}}. These include:
\begin{itemize}
\item High-resolution synchronic EUV and CH maps from 06/10/2010 to 08/18/2014 at 6-hour cadence.
\item New preprocessing data for STEREO-A/B EUVI 195\AA\, and SDO AIA 193\AA\ for performing limb-brightening correction and inter-instrument intensity normalization for the time period 12/21/2010 to 02/17/2014.
\item A modular region-growing image segmentation algorithm {\tt ezseg} (that can be linked into any desired driver code) implemented in both an OpenMP-enabled FORTRAN and GPU-accelerated CUDA-C version.
\item A ready-to-use implementation of our CH detection pipeline in the form of a MATLAB driver script ({\tt euv2chm}) that reads in EUV images, applies our new preprocessing steps, performs coronal hole detection using MATLAB's MEX interface to call {\tt ezseg}, and generates the merged EUV and CH maps.  An SSW/IDL script to prepare image data for use in {\tt euv2chm} from \emph{fits} images is also included ({\tt euv2chm\_data\_prep.pro}).
\item The MATLAB scripts used to generate the LBCC and IIT preprocessing data.  These are included for advanced users who wish to use the procedure described in this paper with alternative data sets (such as data with alternate PSFs and/or other EUV channels).
\end{itemize}
Ultimately, we hope that the method and tools presented here can be useful for those wishing to study coronal hole evolution as well as studying the full-Sun EUV corona and/or limb brightening in general. These methods are particularly relevant for looking at global EUV evolution in time and for applications where a careful treatment of center-to-limb and inter-instrument intensity variations is required.
 
%%%%%%%%%%%%%%%%%%%%%%%%%%%%%%%%%%%%%%%%%%%%%%%%%%%%%%%%%%%%%%%%%%%%%%%%%%%%%%%%%%%%%%%%%%%%%%%%%%%%%%%%%%%%%%%%%%%%%

\acknowledgements
The authors would like to thank the anonymous reviewer for his/her constructive comments and suggestions, which served to greatly improve the manuscript and led to important additions to the methodology.  This work was supported by AFOSR, AFRL, NASA's LWS TR\&T program, and NSF's FESD program.

\appendix

\section{Theoretical limb-brightening correction curves}
\label{sec_lbcc_theory}
A theoretical limb-brightening correction curve (LBCC) can be computed by calculating the observed intensity as a function of $\mu$ by assuming an idealized hydrostatic atmosphere at a given coronal temperature.  For example, \citet{2003_Andretta_analytic_limb_curve} used a constant-gravity approximation for generating such a curve where it was used to estimate photon fluxes in EUV emission. Here we illustrate the effect of limb brightening using a model that incorporates a $1/r^2$ gravitational fall-off for completeness.

In order to compute the emission intensities, the electron density as a function of height above the base of the corona ($R_0$) is required.  Hydrostatic equilibrium for $r\ge R_0$ in the solar corona can be described as
\begin{equation}
\label{eq_lbcc_anay1}
\frac{dP}{dr}=-g_0\,\rho\,\left(\frac{R_0}{r}\right)^2,
\end{equation}
where $g_0=G\,M_{\odot}/R_{0}^2$, and $P$ is the pressure given by $P=n\,k_B\,T$, where $n=\rho/\tilde{\mu}\,m_p$ is the number density, where $\tilde{\mu}$ is the mean molecular weight (we use the coronal value of $\tilde{\mu}=0.6$ \citep{2014_Priest_BOOK}), $k_B$ is the Boltzmann constant, and $m_p$ is the proton mass.  Using an isothermal approximation, Eq.~(\ref{eq_lbcc_anay1}) can be written as
\begin{equation}
\label{eq_lbcc_anay2}
\frac{d\rho}{dr}=-\frac{\rho}{\lambda(T)}\,\left(\frac{R_0}{r}\right)^2,
\end{equation}
where we define $\lambda(T)$ as
\begin{equation}
\label{eq_lbcc_lambda}
\lambda(T) = \frac{k_B\,T}{g_0\,\tilde{\mu}\,m_p}.
\end{equation}
Assuming all particle species are perfectly coupled, the solution to Eq.~(\ref{eq_lbcc_anay2}) yields the following form for the electron density:
\begin{equation}
\label{eq_lbcc_anay3}
n_e=n_{e0}\,\mbox{exp}\left[-\frac{R_0}{r}\,\frac{r-R_0}{\lambda(T_e)}\right],
\end{equation}
where $n_{e0}$ is the electron density at the base of the corona ($r=R_0$).
\begin{figure}[tbp]
\centering
\includegraphics[width=2.5in]{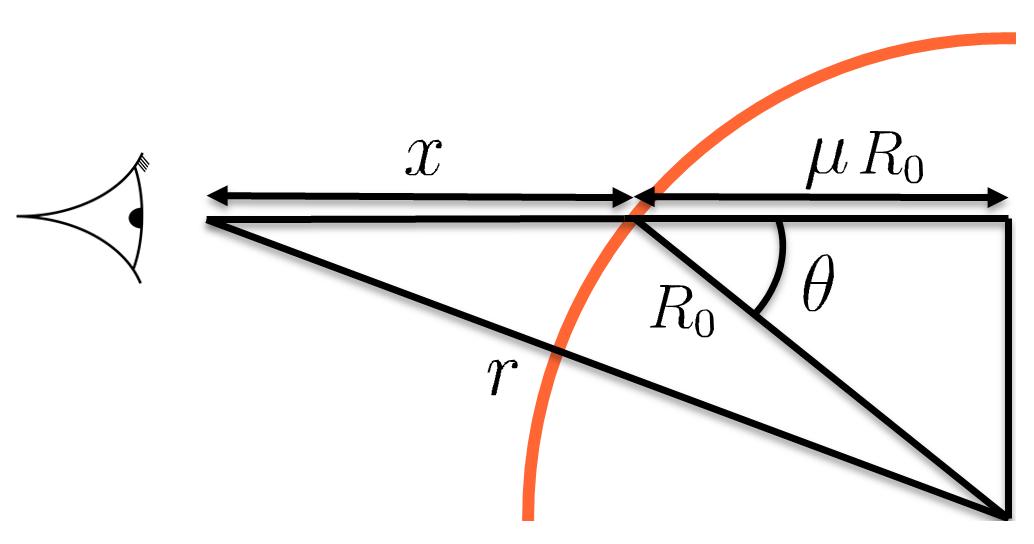}
\caption{Coordinate system for computing limb-brightening correction curves. The center-to-limb angle $\theta$ is shown in a congruent angle position.\label{fig_lbcc_coords}}
\end{figure}
The emissivity at any point is found by multiplying the instrument-dependent response function $R(T_e,n_e)$ (which includes the area factor, etc.) by the number density squared. To obtain the observed emission ($F$) for an observer at infinity as a function of $\mu=\cos\theta$ where $\theta$ is the center-to-limb angle, we integrate over a line-of-sight as shown in Fig.~\ref{fig_lbcc_coords} described as
\begin{alignat}{2}
\label{eq_limb_int}
F(\mu;T_e)&=\int_{0}^{\infty} R(T_e,n_e)\,n_e^2(r)\,dx, 
\end{alignat}
where $r$ is a function of $x$ and $\mu$, parametrized by $R_0$, defined as
\begin{equation}
\label{eq_lbcc_r_def}
r(x;\mu)=\sqrt{x^2+2\,x\,\mu\,R_0+R_0^2}.
\end{equation}

We ignore the weak number density dependence of $R(T_e,n_e)$ for the relevant ranges of density and temperature observed by STA, STB, and AIA, which allows us to choose a typical density $\tilde n_e$ and pull $R(T_e,\tilde n_e)$ out of the integral to yield
\begin{alignat}{2}
\label{eq_limb_int2}
F(\mu;T_e)&\approx R(T_e,\tilde n_e)\,n_{e0}^2\,\int_{0}^{\infty} \mbox{exp}\left[-2\,\frac{R_0}{r}\,\frac{r-R_0}{\lambda(T_e)}\right]\,dx.
\end{alignat}
Unlike the case of constant gravity, the integrand of Eq.~(\ref{eq_limb_int2}) converges to a constant value as $x\rightarrow\infty$, causing the integral to diverge \citep[a classical proof of the existence of the solar wind, e.g. ][]{2009_Lang_BOOK}.  However, we have found numerically using typical coronal temperatures and a base coronal radius of $R_0=1.01\,R_\odot$, that the percent difference in the computed integral using an integration limit of $x_{\mbox{\tiny max}}=10\,R_\odot$ versus $x_{\mbox{\tiny max}}=10,000\,R_\odot$ is $\ll 0.1\%$.  Thus, the divergence of the integral of Eq.~(\ref{eq_limb_int2}) is quite slow and since most of the plasma that emits in the EUV bands used in this paper is very near the Sun, truncated computations of Eq.~(\ref{eq_limb_int2}) are a reliable means of obtaining approximate emission values.  The LBCC is then defined (in log-space)  as
\[
L(\mu;T_e)=\mbox{log10}\left[\frac{F(\mu;T_e)}{F(1;T_e)}\right].
\]

It is interesting to note that at disk center ($\mu=1$), the equivalent integral of  Eq.~(\ref{eq_limb_int2}) for a constant gravity model has an analytic solution given by
\begin{equation}
\label{eq_lbcc_mu1}
F_{g_{\mbox{\tiny const}}}(1,T_e) = \frac{n_{e0}^2\,k_B}{2\,g_0\,\tilde{\mu}\,m_p}\,R(T_e)\,\,T_e,
\end{equation}
which shows that under such an approximation, for a given density at the surface, the intensity is directly proportional to the temperature multiplied by the response function at that temperature.  This implies a correspondence between image intensity and temperature within an EUV image.  While such a direct correspondence is somewhat suspect, this could be justified (at least on average) by a heuristic physics based argument. 
Drawing from a long history of idealized coronal heating scaling laws \citep[e.g.][and references therein]{2002_Aschwanden+Schrijver_HeatingScalingLaws}, if one assumes that there is on average some general power law correspondence between number density and temperature (no matter how weak) then the strong dependence of emissivity on density implies that the intensity of a given structure will be positively correlated with temperature.

\def\myitemsep{5pt}
\bibliographystyle{apj}
\bibliography{CHD}

%To get equal-spaced weight bins: Take circle of radius 1.  
%\[
%A_{1,2}=\pi\,(l_2^2-l_1^2)=\pi\,((1-\mu_2^2)-(1-\mu_1^2))=\pi(\mu_1^2-\mu_2^2).
%\]
%Want equal-area annuli of area $A=A_{tot}/(n-1)$.
%We have $\mu_1=1$, so we need $\mu_2=\sqrt{1-A_{1,2}/\pi}=\sqrt{1-1/(n-1)}$.
%Now for $\mu_2$ to $\mu_3$ we have $A/\pi=\mu_2^2-\mu_3^2=1-A/\pi - \mu_3^2$, so $\mu_3=\sqrt{1-2A/\pi}=\sqrt{1-2/(n-1)}$.
%Therefore we have:
%\[
%\mu_i=\sqrt{1-\frac{i-1}{n-1}}, \qquad l_i=\sqrt{\frac{i-1}{n-1}}, \qquad i={1\rightarrow n}.
%\]

\end{document}